\documentclass[useAMS,usenatbib]{mn2e}
\usepackage[draft]{hyperref}

\usepackage{graphicx}
\usepackage{times}
\usepackage{caption}
\usepackage{color}
\usepackage{float}
\usepackage{verbatim}
\usepackage{fixltx2e}
\usepackage{comment}
\usepackage{amsmath}

\title[Environmental dependence of the galaxy stellar mass function in the Dark Energy Survey]{Environmental dependence of the galaxy stellar mass function in the Dark Energy Survey Science Verification Data}

\author[J.~Etherington et al.]{\parbox[t]{\textwidth}{J.~Etherington$^{1}$,\textsuperscript{\thanks{E-mail: j.etherington@gmail.com}}~D.~Thomas$^{1,2}$, C.~Maraston$^{1,2}$, I.~Sevilla-Noarbe$^{3}$, K.~Bechtol$^{4,5}$, J.~Pforr$^{6}$,  P.~Pellegrini$^{7,8}$, J.~Gschwend$^{7,8}$, A.~Carnero~Rosell$^{7,8}$, M.~A.~G.~Maia$^{7,8}$,  L.~N.~da Costa$^{7,8}$, A.~Benoit-L{\'e}vy$^{9,10}$, M.~E.~C.~Swanson$^{11}$, W.~G.~Hartley$^{12}$, T.~M.~C.~Abbott$^{13}$, F.~B.~Abdalla$^{14,15}$, S.~Allam$^{16}$, R.~A.~Bernstein$^{17}$, E.~Bertin$^{18,9}$, D.~Brooks$^{14}$, E.~Buckley-Geer$^{16}$, M.~Carrasco~Kind$^{19,20}$, J.~Carretero$^{21,22}$, F.~J.~Castander$^{21}$, M.~Crocce$^{21}$, C.~E.~Cunha$^{23}$, S.~Desai$^{44}$, P.~Doel$^{14}$, T.~F.~Eifler$^{26,27}$, A.~E.~Evrard$^{28,29}$, A.~Fausti~Neto$^{8}$, D.~A.~Finley$^{16}$, B.~Flaugher$^{16}$, P.~Fosalba$^{21}$, J.~Frieman$^{16,30}$, D.~W.~Gerdes$^{29}$, D.~Gruen$^{23,31}$, R.~A.~Gruendl$^{19,20}$, G.~Gutierrez$^{16}$, K.~Honscheid$^{32,33}$, D.~J.~James$^{13}$, K.~Kuehn$^{34}$, N.~Kuropatkin$^{16}$, O.~Lahav$^{14}$, M.~Lima$^{15,8}$, P.~Martini$^{32,36}$, P.~Melchior$^{37}$, R.~Miquel$^{38,22}$, J.~J.~Mohr$^{25,24,39}$, B.~Nord$^{16}$, R.~Ogando$^{8,7}$, A.~A.~Plazas$^{27}$, A.~K.~Romer$^{40}$, E.~S.~Rykoff$^{23,31}$, E.~Sanchez$^{41}$, V.~Scarpine$^{16}$, M.~Schubnell$^{28}$, R.~C.~Smith$^{13}$, M.~Soares-Santos$^{16}$, F.~Sobreira$^{42,8}$, G.~Tarle$^{29}$, V.~Vikram$^{43}$, A.~R.~Walker$^{13}$, Y.~Zhang$^{16}$}}

\begin{document}

\maketitle

\begin{abstract}
Measurements of the galaxy stellar mass function are crucial to understand the formation of galaxies in the Universe. In a hierarchical clustering paradigm it is plausible that there is a connection between the properties of galaxies and their environments. Evidence for environmental trends has been established in the local Universe. The Dark Energy Survey (DES) provides large photometric datasets that enable further investigation of the assembly of mass. In this study we use $\sim 3.2$ million galaxies from the (South Pole Telescope) SPT-East field in the DES science verification (SV) dataset. From grizY photometry we derive galaxy stellar masses and absolute magnitudes, and determine the errors on these properties using Monte-Carlo simulations using the full photometric redshift probability distributions. We compute galaxy environments using a fixed conical aperture for a range of scales. We construct galaxy environment probability distribution functions and investigate the dependence of the environment errors on the aperture parameters. We compute the environment components of the galaxy stellar mass function for the redshift range $0.15<z<1.05$. For $z<0.75$ we find that the fraction of massive galaxies is larger in high density environment than in low density environments. We show that the low density and high density components converge with increasing redshift up to $z\sim 1.0$ where the shapes of the mass function components are indistinguishable. Our study shows how high density structures build up around massive galaxies through cosmic time.  
\end{abstract}

\begin{keywords}
galaxies: evolution -- galaxies: formation -- galaxies: clusters: general -- galaxies: photometry -- galaxies: statistics
\end{keywords}

\section{Introduction} \label{sec:introduction}

Establishing an understanding of the assembly of mass in galaxies is a key goal in modern extragalactic physics and cosmology. Measurements of the galaxy stellar mass function through cosmic time for different populations of galaxies and as a function of environment are vital to inspect the nature of the assembly and also constrain models of the physical processes.  

It is widely assumed that dark matter accumulated and collapsed in a hierarchical fashion \citep{White1991}. Lambda Cold dark matter simulations \citep{Boylan-Kolchin2009} are able to emulate the clustering of structure observed in large scale galaxy surveys in the local Universe \citep{Springel2006}. However there are unresolved issues. The observed pattern of ``downsizing'' \citep{Cowie1996} remains a challenge even though some aspects of it can be understood in $\Lambda$CDM models assuming that there is some threshold of halo mass for efficient star formation \citep{Conroy2009}. Several flavours \citep{Fontanot2009} of downsizing have now been observed: chemo archaeological \citep{Worthey1992}, archaeological \citep{Thomas2005, Thomas2010}, downsizing in the star formation rate \citep{Conselice2007}, stellar mass \citep{Pozzetti2007, Maraston2013}, metallicity \citep{Maiolino2008}, nuclear activity \citep{Cristiani2004, Hasinger2005} and most recently black hole growth \citep{Hirschmann2012}.

The question of galaxy assembly is undoubtedly tied to the old adage: ``nature vs nurture''. Disentangling the internal and external physical processes involved is a taxing business. Modern works are providing clues; some by examining the environmental dependence of central and satellite galaxies separately \cite[e.g. for groups-][]{Carollo2013,Cibinel2013,Pipino2014}. The mass, either the stellar mass or the total mass of the dark matter halo, is widely thought to be the primary driver of galaxy evolution. This is a secular channel of evolution championing the `nature' argument. In the hierarchal clustering paradigm where a tree of halos merge and accrete forming larger structures it may follow that a galaxy's environment also has a role to play. 

Galaxies are subject to several external physical processes including ram pressure stripping \citep{Boselli2014, Fumagalli2014} galaxy harassment \citep{Farouki1981}, strangulation \citep{Larson1980} and cannibalism \citep{Nipoti2003}. These processes are certainly capable of stripping gas, shutting down star formation and transforming a galaxy's morphology. But to what degree are these processes ubiquitous in the Universe?    

In a hierarchical scenario it is clear that a galaxy's environment is not constant through cosmic time. It is in fact changing. The key parameter may therefore be the galaxy's integrated environment through time. Devising an observational proxy for this is challenging, perhaps even intractable with the snapshot observations we capture with modern galaxy surveys. The problem is not merely one of data collection but also of definition. In simulations for example, how do you define or even identify a galaxy’s environment when its constituent parts have not yet assembled? 

Galaxy surveys such as the SDSS have revolutionised the study of the galaxy population at low redshift \citep{Blanton2009}. Surveying large areas provides large statistics and reduces the error associated with cosmic variance. The first measurements of the galaxy stellar mass function of the local Universe were obtained by converting luminosity functions by simple modelling of the M/L of galaxies \citep{Cole2001, Bell2003, Kodama2003}. There are now several measurements of the galaxy stellar mass function for the local Universe \citep{Baldry2008, Li2009, Baldry2012} based on stellar masses derived from galaxy photometry. The GAMA survey has augmented SDSS with additional spectroscopic data to obtain mass complete samples to {\raise.17ex\hbox{$\scriptstyle\sim$}}$10^8M_{\odot}$ yielding the current state of the art mass function measurements in the local Universe \citep{Baldry2012}.

Investigations of the redshift evolution of the stellar mass function have until recently been restricted to data collected from spectroscopic pencil beam surveys. However the BOSS survey enabled \cite{Maraston2013} to study the evolution of the massive end of the stellar mass function using a sample of $400,000$ LRGs to a redshift of {\raise.17ex\hbox{$\scriptstyle\sim$}}0.6. The massive end of the mass function was found to be consistent with passive evolution in agreement with other works \citep{Pozzetti2010, Ilbert2010, Ilbert2013}. 

Pencil beam surveys such as the DEEP2 \citep{Newman2013} and zCOSMOS \citep{Lilly2007} surveys have typically captured data for no more than a few square degrees of the sky but they are complete to relatively high redshifts. These analyses exploit the measurements of 1000s to several tens of 1000s of galaxies. Some studies focus on the mass functions for the total galaxy population whereas others have also investigated the contributions made by different galaxy types; split by morphology and colour. There are relatively few works that have examined the role of galaxy environment on the stellar mass function. The earliest of these studies split galaxies into two types: field or cluster \citep{Balogh2001, Kodama2003}. More recent studies have quantified the environments and then examined the mass function for different environment bins for the local Universe \citep{McNaught-Roberts2014}, intermediate redshifts \citep{Bundy2006, Bolzonella2010, Vulcani2011} and high redshifts \cite{Mortlock2015}. The main finding of these works is the massive end of the galaxy stellar mass function is dominated by galaxies that reside within high density environments at low and intermediate redshifts but at higher redshifts the mass function is independent of environment. \cite{Davidzon2016} show the weakening of the environmental dependence of the mass function between redshifts of $0.5$ and $0.9$. 

Future studies of the galaxy stellar mass function require surveys of larger cosmological volumes. This is currently only feasible with large scale photometric surveys as spectroscopic surveys with a similar volume to DES for example, would be too costly and slow. Large number statistics therefore come at the cost of redshift precision.  

The aim of this work is to study the contributions to the stellar mass function from different environments as a function of redshift exploiting the DES SV (science verification) data taken before the first season of the survey's operation. 

The paper consist of six sections. In Section \ref{sec:data_params} we describe the data and the galaxy parameters. In Section \ref{sec:galaxy_environment} we present the galaxy environment measurements. In Section \ref{sec:mass_function} we show the mass function analysis. In Section \ref{sec:discussion} we discuss the robustness of the main results and lastly in Section \ref{sec:conclusions} we present our conclusions.

In this work we have assumed a Salpeter \citep{Salpeter1955} initial mass function, a cosmology with $\Omega_{m}=0.286$, $\Omega_{\Lambda}=0.714$ and $H_{0} = 70\;$kms$^{-1}$Mpc$^{-1}$ and have adopted comoving coordinates \citep[e.g.][]{Cooper2006} to calculate the distances between galaxies.

\section{Data and galaxy parameters} \label{sec:data_params}

\begin{figure*}
   \centering 
   \includegraphics[width=.55\linewidth]{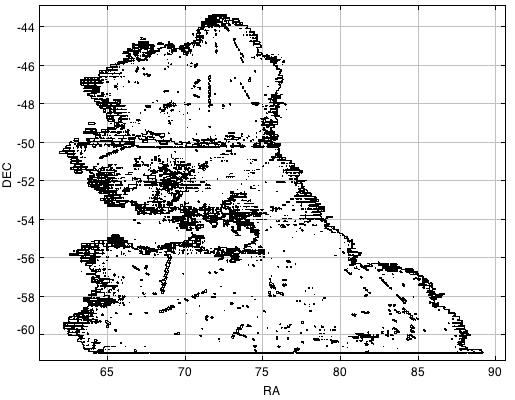}
\caption{Edges and holes of the SPT-E field. The field has a uniform depth of $23$ mag in i-band integrated apparent magnitude.}  
\label{fig:edges}
\end{figure*}

The DES is a multi-band ({\it g, r, i, z, Y}) photometric survey performed with the Dark Energy Camera (DECam, \citealp{Flaugher2015}) mounted on the 4-meter Blanco Telescope at Cerro Tololo Inter-American Observatory (CTIO), aimed at imaging {5000\  {\rm sq. deg}} of the southern sky out to redshift {\raise.17ex\hbox{$\scriptstyle\sim$}}$1.4$. The survey started in February 2013 but from November 2012 to February 2013, DES carried out a Science Verification (SV) survey. These observations provide science quality data for more than 250 deg$^{2}$ at close to the main survey's nominal depth ({\it i}-band 2-arcsec aperture magnitude$\simeq 24\ {\rm mag}$) for standard survey fields (e.g. SPT-E field) and/or deeper fields (used for calibration and/or supernovae studies). A number of analyses of the SV data have now been released \citep{DES2016, Palmese2016} including weak lensing shear measurements \citep{Jarvis2016}, CMB lensing tomography \citep{Giannantonio2016}, systematics \citep{Crocce2016, Leistedt2015}, LRG selection \citep{Rozo2016} and a study of the galaxy populations in massive clusters \citep{Hennig2016}.

The data stored in the catalogue of the SV coadded imaging created by the DES Data Management (DESDM) pipeline (\citealp{Mohr2012}, \citealp{Sevilla2011}, \citealp{Desai2012}) were then thoroughly tested and analysed by a team of DES scientists, who were able to construct a new photometric catalogue, called the \textit{SV Annual 1} (\textit{SVA1}) \textit{Gold} catalogue, containing 25,227,559 objects and extending over an area {\raise.17ex\hbox{$\scriptstyle\sim$}}250 deg$^{2}$. This catalogue is now publicly available\footnote{\url{http://des.ncsa.illinois.edu/releases/SVA1}.}.

In this work we use the Portsmouth COnstant \texttt{Mag\_MOdel} Depth Originated REgion (COMMODORE) galaxy catalogue (see Capozzi et al., in prep.), a subsample of the Gold SVA-1 catalogue. The COMMODORE catalogue was constructed to have homogeneous depth ({\it i}-band integrated magnitude $=23\ {\rm mag}$) so that it could be used for galaxy evolution studies and a bright end limit ({\it i}-band integrated magnitude $=16\ {\rm mag}$) to aid star-galaxy separation. The total area covered by this catalogue is {\raise.17ex\hbox{$\scriptstyle\sim$}}155 deg$^{2}$. However this area is not contiguous and for galaxy environment studies which require area contiguity, a subsample of the COMMODORE catalogue must be selected. In this paper we select from the COMMODORE catalogue only the galaxies in the SPT-E ({\raise.17ex\hbox{$\scriptstyle\sim$}}130 deg$^{2}$) field (which is the largest field in the SV dataset) to ensure the contiguity requirement is met. Fig \ref{fig:edges} shows the perimeter of the SPT-E field and the holes in the data (due to bright stars) after all of the processing steps have been applied.

\subsection{Photometric redshifts} \label{subsec:photometric_redshifts}

There are two types of methods used to measure photometric redshifts: template methods and training methods. \cite{Sanchez2014} performed extensive tests on the SV data using $13$ different photometric redshift codes. The codes were trained with {\raise.17ex\hbox{$\scriptstyle\sim$}}$6000$ spectroscopic redshifts obtained from existing datasets including VVDS Deep \citep{LeFevre2005, LeFevre2013}, VVDS Wide \citep{Garilli2008}, SDSS/BOSS \citep{Strauss2002, Eisenstein2001,Ahn2012}, ACES \citep{Cooper2012}, and 2dFGRS \citep{Colless2001} that matched the DES SV photometry. \cite{Sanchez2014} found that two of the training methods, one based on artificial neural networks (ANNz) \citep{Collister2004} and the other on prediction trees and random forests, called \textit{Trees for Photo-z} or TPZ \citep{Carrasco-Kind2013}, performed the best. The scatter of the difference between the photometric redshifts and the test set of spectroscopic redshifts for these methods was $\delta_{68}=0.08$. 

In this paper we utilise the output from the TPZ code because in addition to the best estimate of the redshift this code provides photometric redshift probability distribution functions (PDFs) for each galaxy. The photo-z PDFs consist of 200 bins spanning the redshift range: 0-1.8. Each bin has a width of 0.009 in redshift. 

We computed the 1 sigma width of each photo-z PDF and identified the galaxies at the $16^{\textrm{th}}$, $50^{\textrm{th}}$ and $84^{\textrm{th}}$ percentiles in the distribution of widths. The half-widths at these percentiles are: $0.041$, $0.055$ and $0.079$ respectively. We note that the 1 sigma width of the photo-z PDFs as measured here for all of the SPT-E galaxies is a different quantity from the scatter ($\delta_{68}$) measured between the peaks of the PDFs and the spectroscopic redshifts measured in \cite{Sanchez2014}.

In this paper we are interested in the propagation of the redshift errors into the derived galaxies properties: mass, absolute magnitude and galaxy environment. This in turn enables us to quantify errors on the environment components of the galaxy stellar mass function. 

\subsection{K-corrections, absolute magnitudes and stellar masses} \label{subsec:stellar_masses_fitting}

The galaxy properties used in our study are taken from the COMMODORE catalogue, which provides both galaxy physical properties (e.g. age, star formation history and stellar mass) and detectability related properties (e.g. k-correction, maximum accessible volumes and completeness factors as function of apparent magnitude and surface brightness). The physical properties are obtained via Spectral Energy Distribution (SED) fitting, using thirty two sets of theoretical templates (as in \citealp{Maraston2006} and \citealp{Capozzi2016}) constructed with the evolutionary stellar population synthesis models by \citet{Maraston2005} and \citet{Maraston2009} under the assumption of a Salpeter \citep{Salpeter1955} Initial Mass Function (IMF). These templates consist of four types of star formation histories including: single bursts (SSPs), exponentially-declining, truncated star formation rate (SFR) and constant SFR. The SED fitting was performed by means of the template fitting code \textsc{hyperz} \citep{Bolzonella2000}, using the DES photometry and fixing galaxy redshifts at the photometric values provided by the TPZ code. For each galaxy we then calculated detectability using the real observed-frame magnitudes calculated at given redshift for the galaxy's best fitting model found above. We use tables\footnote{C. Maraston, {\it in preparation}, available upon request} which provide for each of the 32 sets of templates, their observed-frame magnitudes (assuming a standard cosmology) for a fine grid of redshifts including $z=0$. The difference between the model magnitude at given $z$ and the same at $z=0$ is then the exact value of the fainting or brightening to be applied in the exact filter, without going through approximated $k$-corrections. These $\Delta$'s are then applied to the i-band absolute magnitudes. 

In evolutionary studies it is important to estimate the completeness of the galaxy sample in absolute magnitude and stellar mass as a function of redshift. To do this we follow a similar procedure as described in \cite{Pozzetti2010}. We obtain the completeness limits by computing the $90^{\textrm{th}}$ percentiles of the limiting stellar mass and absolute magnitudes within redshift bins for the faint and bright ends of the sample. Limits are constructed for the bright end in addition to the faint end because of the {\it i}-band apparent magnitude selection (see Section \ref{sec:data_params}) applied to construct the COMMODORE catalogue. We refer the reader to Capozzi et al. (in prep.), where the Portsmouth COMMODORE catalogue, the completeness limits and the galaxy property calculations are presented in detail.

\subsection{Error analysis} \label{subsec:error_analysis}

The main aim of this Section is to quantify the errors on the derived galaxy properties, i.e. the stellar masses and the i-band absolute magnitudes due to the errors on the photometric redshifts. We adopt a Monte-Carlo approach and generate many realizations of the SPT-E catalogue by drawing redshifts from the photo-z PDFs. An alternative is to start with the errors on the photometry itself and propagate them forwards \citep[e.g.][]{Taylor2009}. We use the photo-z PDFs as this makes our study easier to reproduce or modify as the SV photo-z PDFs are publicly available (see Section \ref{sec:data_params}). Since the TPZ photo-z PDFs are constructed using the photometric errors these approaches are comparable.

In Section \ref{subsubsec:sampling_tests} we present a series of tests on the redshift draws to: (i) verify that the draws are representative of the photo-z PDFs and (ii) quantify the difference between the statistics of the draws and the photo-z PDFs as a function of the number of catalogue realizations. We compute the stellar masses and i-band absolute magnitudes for the galaxies in each of the realizations of the SPT-E catalogues using \textsc{hyperz} as described in Section \ref{subsec:stellar_masses_fitting}. In Section \ref{subsubsec:property_errors} we present the distributions of the galaxy properties and their errors. 

\begin{figure*}
\begin{minipage}[t]{0.49\linewidth}
  \centering  
  \includegraphics[width=.85\linewidth]{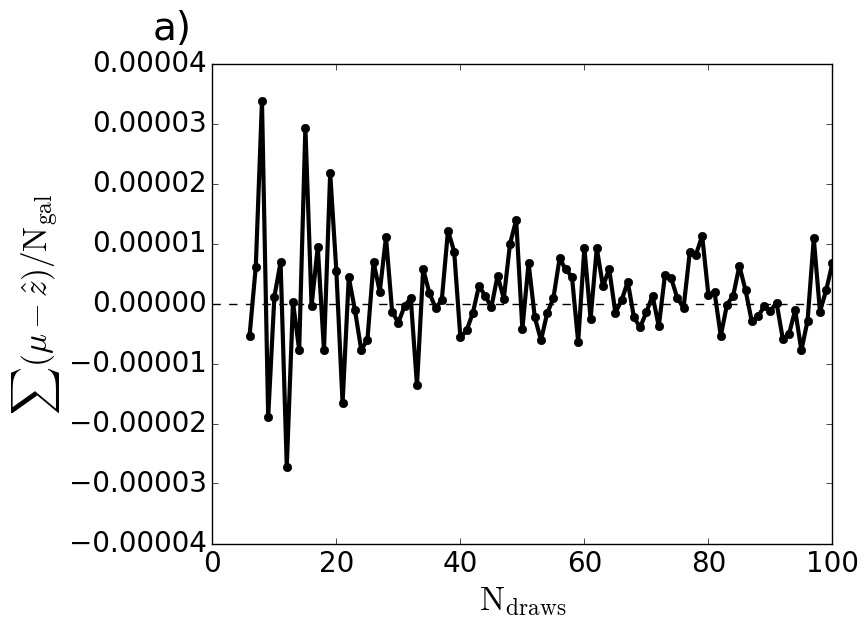}
\end{minipage}
\begin{minipage}[t]{0.49\linewidth}
  \centering  
  \includegraphics[width=.85\linewidth]{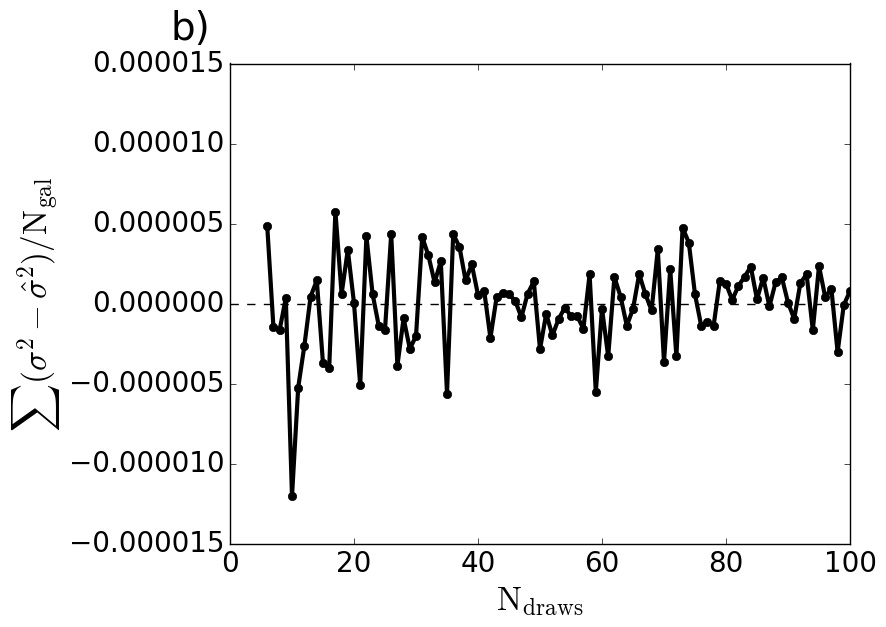}
\end{minipage}
\begin{minipage}[t]{0.49\linewidth}
   \centering 
   \includegraphics[width=.85\linewidth]{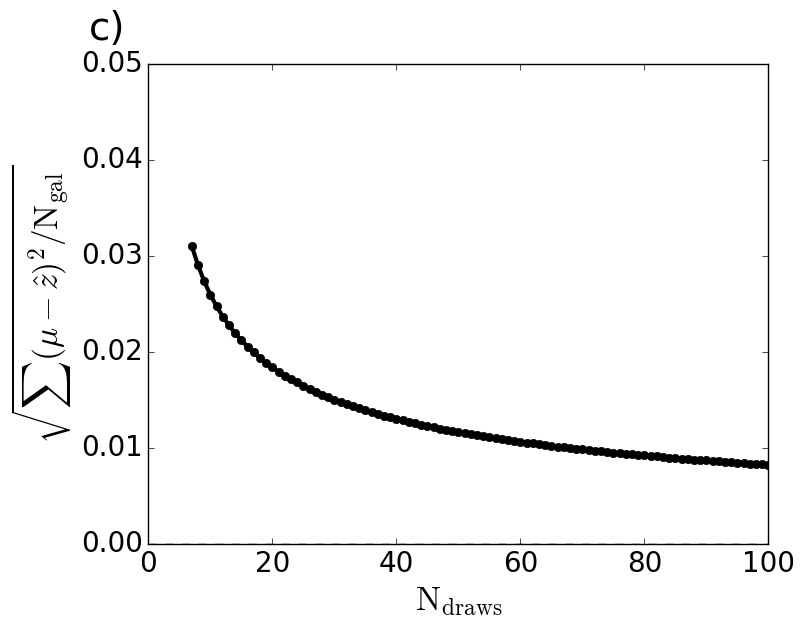}
\end{minipage}
\begin{minipage}[t]{0.49\linewidth}
  \centering  
  \includegraphics[width=.85\linewidth]{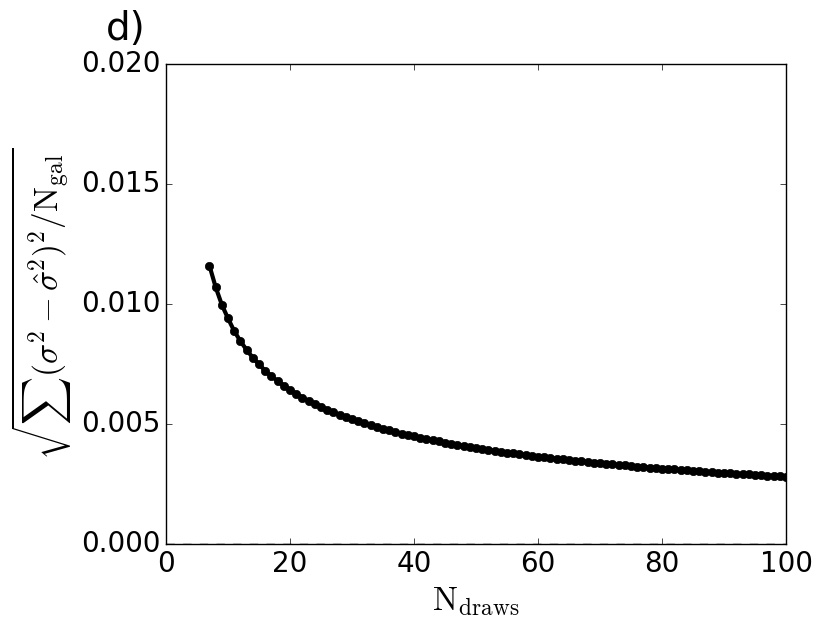}
\end{minipage}
\caption{Top: Bias for the mean (left) and variance (right) as a function of the number of draws. Bottom: RMSE for the mean (left) and variance (right) of the redshift draws as a function of the number of draws.}  
\label{fig:sampling_tests}
\end{figure*}

\subsubsection{Sampling Tests} \label{subsubsec:sampling_tests}

To investigate the variability due to the photometric redshift errors we generated $100$ realizations of the SPT-E catalogue drawing redshifts from the photo-z PDFs. To do this we constructed the photo-z cumulative distribution functions from the photo-z PDFs for each galaxy. The cumulative distribution function maps the range 0 to 1 to the possible redshifts for a galaxy. We drew random numbers from a uniform distribution spanning 0 to 1 and used the mappings to obtain redshift draws.

To verify that the sampled redshift draws were representative of the photo-z PDFs we calculated summary statistics. We obtained the ``true'' mean and variance of each photo-z PDF.  The mean is simply the expectation of the distribution: 

\begin{equation} \label{eq:pdf_mean}
\mu=E(z)=\Sigma z_{i}p_{i} \, .
\end{equation}

It is the sum of the redshift mid positions of the bars multiplied by the probabilities (heights) of the bars. The variance is the expectation of the squared distribution minus the expectation squared: 

\begin{equation} \label{eq:pdf_variance}
\sigma^{2}=E(z^{2}) - (E(z))^{2},\qquad E(z^{2})=\sum z_{i}^{2}p_{i} \, .
\end{equation}

The mean of the randomly generated redshift draws was calculated using:

\begin{equation} \label{eq:draws_mean}
\hat{z} = \frac{1}{N_{\textrm{draws}}} \sum z
\end{equation}

\noindent
and the variance of the redshift draws was calculated using:

\begin{equation} \label{eq:draws_variance}
\hat{\sigma}^{2} = \frac{1}{N_{\textrm{draws}}-1}\sum (z - \hat{z})^{2} \, .
\end{equation} 

\noindent
The range of 5-100 draws of each galaxy was investigated.

Using these summary statistics the biases and root mean squared errors (RMSE) between the ``true'' statistics (mean and variance) and the statistics from the samples as a function of the number of draws were calculated. The bias and RMSE of the mean are given by:

\begin{equation} \label{eq:mean_bias}
\textrm{Bias of the mean}=\frac{1}{N_{\textrm{gal}}} \sum (\mu - \hat{z}) \, ,
\end{equation}

\begin{equation} \label{eq:mean_rmse}
\textrm{RMSE of the mean}=\sqrt{\frac{1}{N_{\textrm{gal}}}\sum (\mu - \hat{z})^{2}} 
\end{equation}

\noindent
and the bias and RMSE of the variance are given by: 

\begin{equation} \label{eq:variance_bias}
\textrm{Bias of the variance}=\frac{1}{N_{\textrm{gal}}} \sum (\sigma^{2} - \hat{\sigma}^{2}) \, ,
\end{equation}

\begin{equation} \label{eq:variance_rmse}
\textrm{RMSE of the variance}=\sqrt{\frac{1}{N_{\textrm{gal}}}\sum (\sigma^{2} - \hat{\sigma}^{2})^{2}} \, .
\end{equation}

Fig \ref{fig:sampling_tests} shows the bias (top) and RMSE (bottom) for the mean (left) and variance (right) as a function of the number of redshift draws for each galaxy. Plots a) and b) show that the bias of the mean and the variance are approximately zero. This is expected as these estimators are unbiased. The plots show the random errors generated from the sampling process are of the order of $\sim 10^{-5}$ centered on zero. The RMSE for the mean decreases as approximately the square root of the number of redshift draws of each galaxy. The RMSE for the variance falls off slightly more rapidly with the number of draws, with a power of $-0.52$.

For $100$ random catalogues the RMSE of the mean is $0.0076$ and RMSE of the variance is $0.0028$ as shown in Fig \ref{fig:sampling_tests}. In addition we found that the RMSE of the standard deviation for the $100$ catalogues is $0.0082$. These numbers are between 5 and 10 times smaller than the typical width of the photo-z PDFs. Individual examples have been examined and with 100 draws minor offsets between the statistics of the samples and the PDFs can be introduced. PDFs that have multiple peaks, separated by relatively large redshifts can be sampled less effectively with only a small number of draws. Nevertheless, the precision quoted here, is sufficient for the purposes of this study. Since the errors decrease approximately as the square root of the number of draws $4$ times as many random catalogues (i.e. $400$ catalogues) would be required to halve the RMSEs.

\begin{figure*}
\begin{minipage}[t]{0.33\linewidth}
   \centering 
   \includegraphics[width=.99\linewidth]{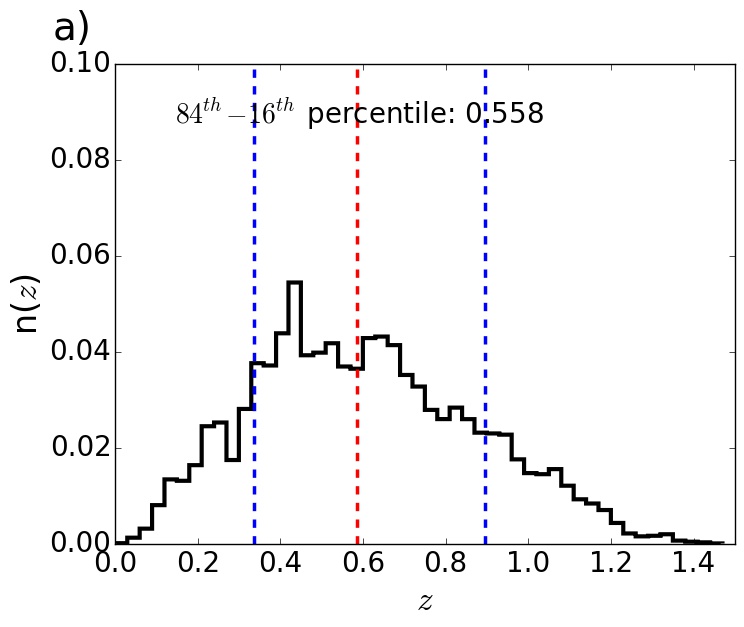}
\end{minipage}
\begin{minipage}[t]{0.33\linewidth}
  \centering  
  \includegraphics[width=.99\linewidth]{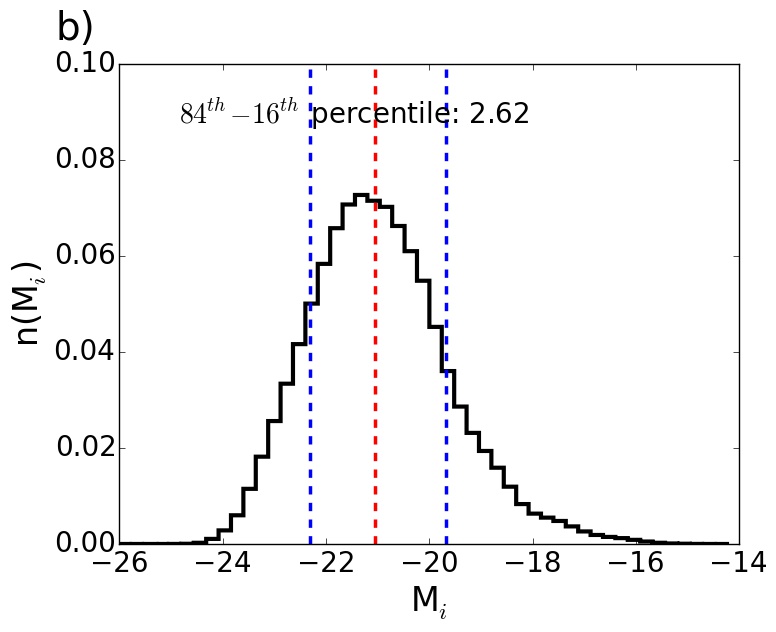}
\end{minipage}
\begin{minipage}[t]{0.33\linewidth}
  \centering  
  \includegraphics[width=.99\linewidth]{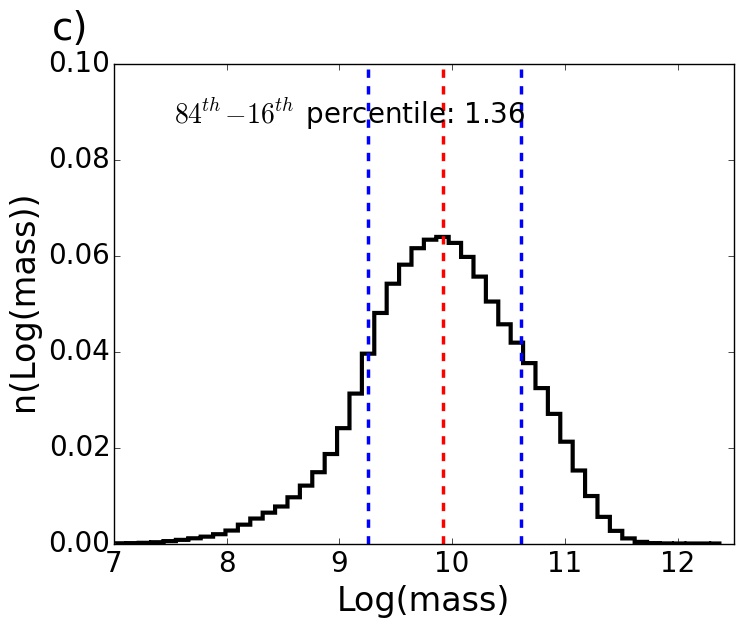}
\end{minipage} 
\begin{minipage}[t]{0.33\linewidth}
   \centering 
   \includegraphics[width=.99\linewidth]{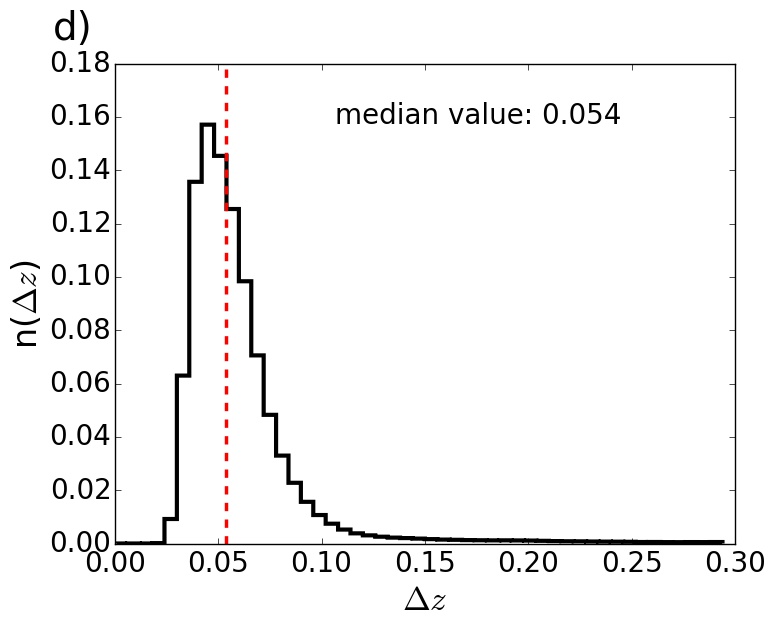}
\end{minipage}
\begin{minipage}[t]{0.33\linewidth}
  \centering  
  \includegraphics[width=.99\linewidth]{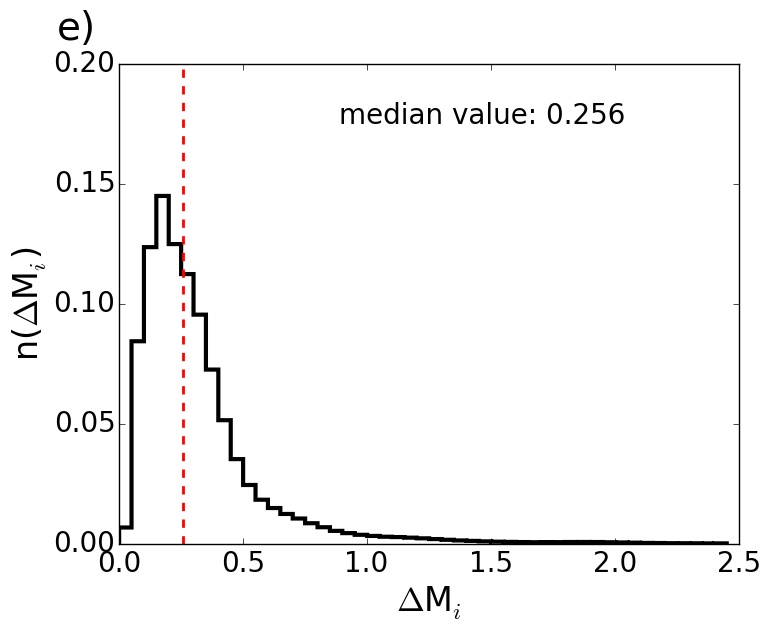}
\end{minipage}
\begin{minipage}[t]{0.33\linewidth}
   \centering 
   \includegraphics[width=.99\linewidth]{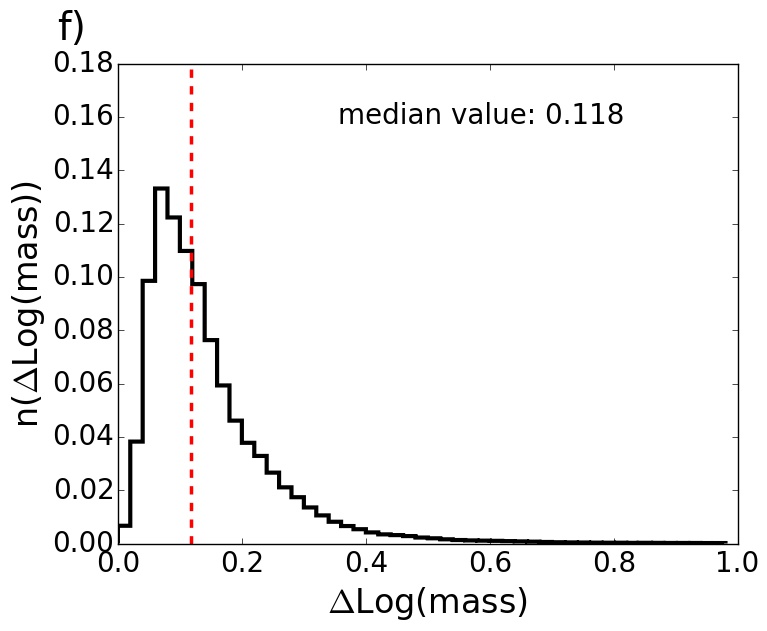}
\end{minipage} 
\begin{minipage}[t]{0.33\linewidth}
  \centering  
  \includegraphics[width=.99\linewidth]{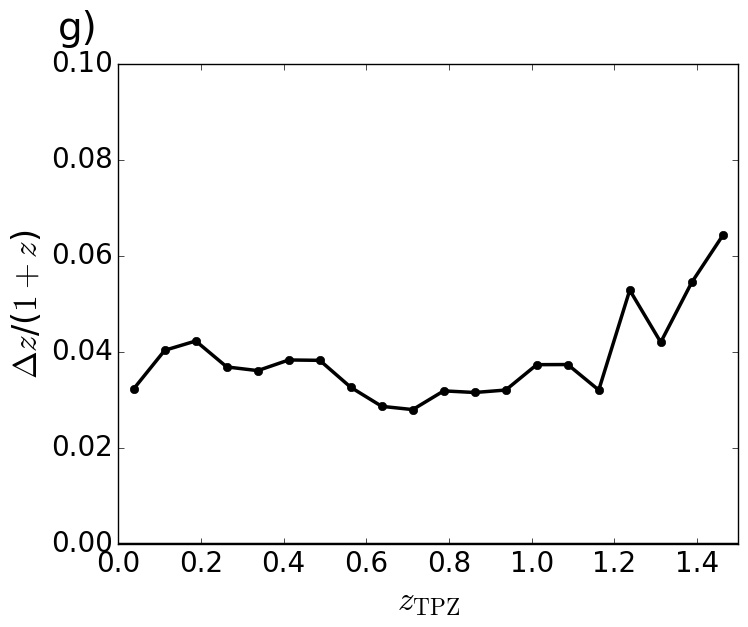}
\end{minipage} 
\begin{minipage}[t]{0.33\linewidth}
   \centering 
   \includegraphics[width=.99\linewidth]{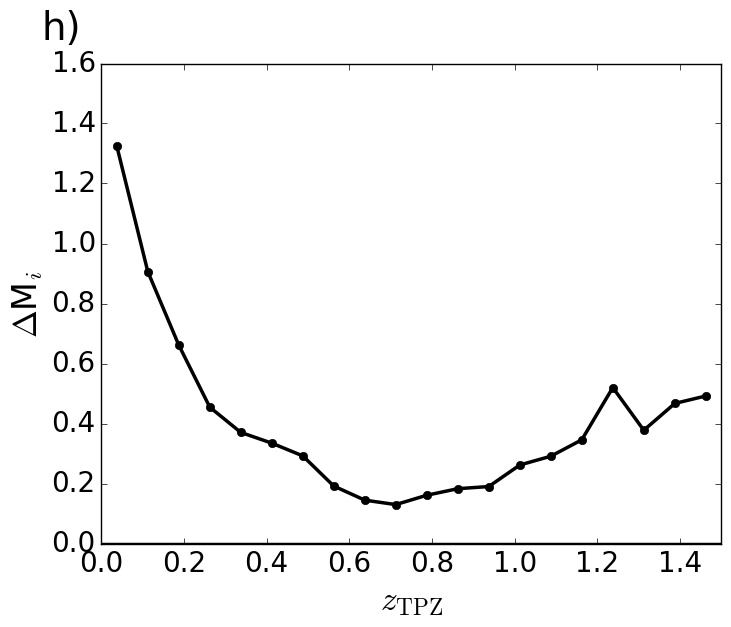}
\end{minipage}
\begin{minipage}[t]{0.33\linewidth}
  \centering  
  \includegraphics[width=.99\linewidth]{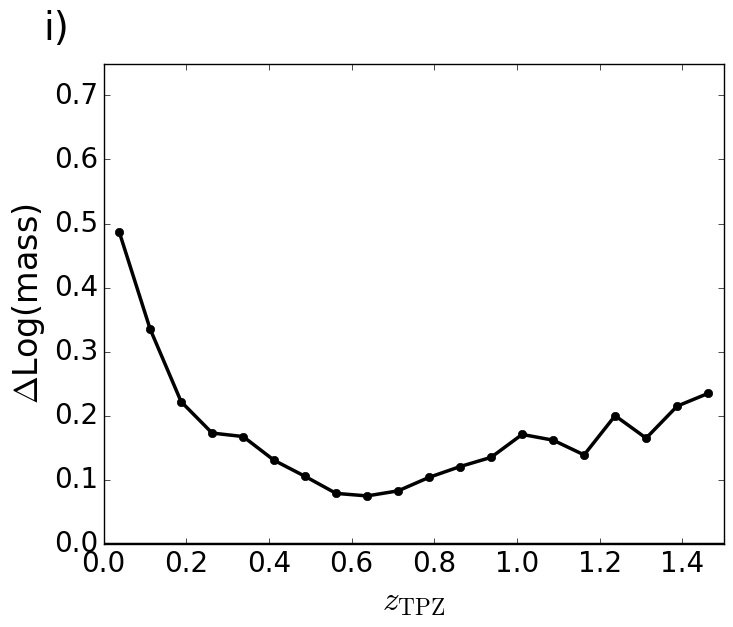}
\end{minipage} 
\begin{minipage}[t]{0.33\linewidth}
 \centering  
 \includegraphics[width=.99\linewidth]{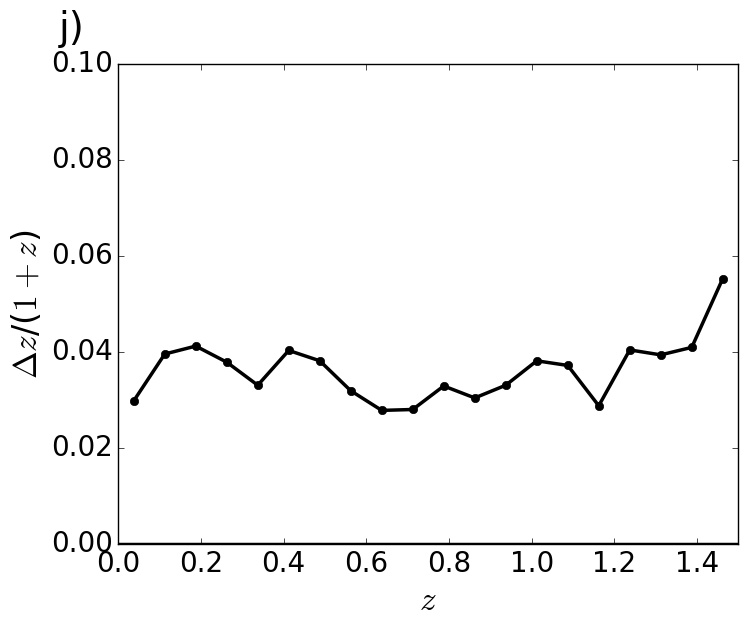}
\end{minipage}
\begin{minipage}[t]{0.33\linewidth}
  \centering  
  \includegraphics[width=.99\linewidth]{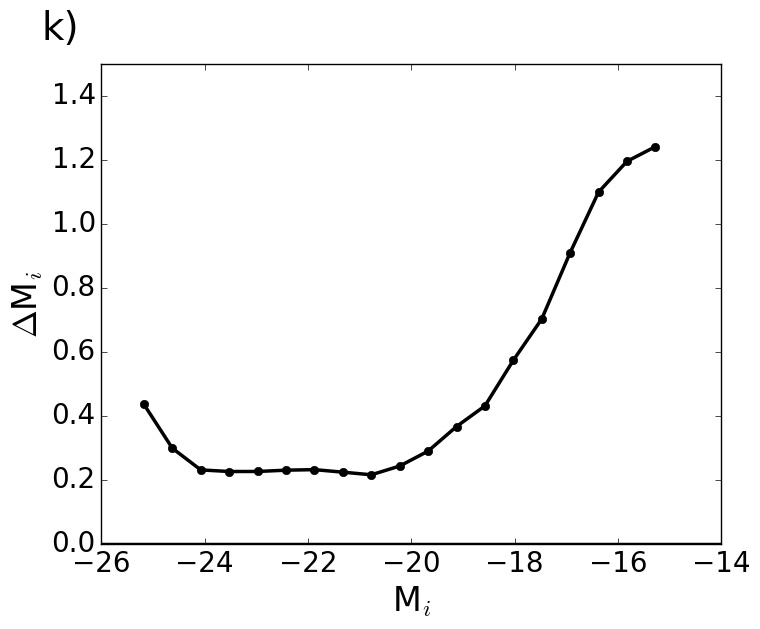}
\end{minipage} 
\begin{minipage}[t]{0.33\linewidth}
  \centering  
  \includegraphics[width=.99\linewidth]{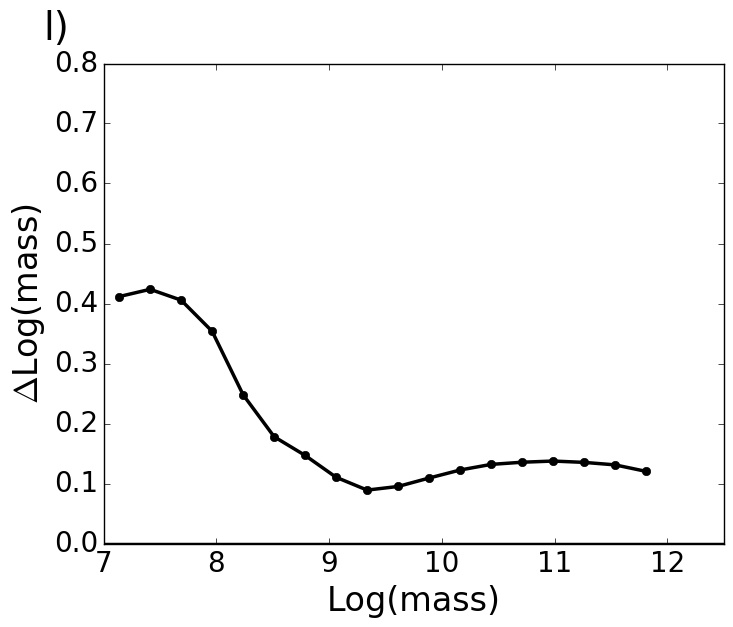}
\end{minipage} 
\caption{First row: Property distributions, second row: error distributions, third row: TPZ redshift dependence of the property errors and fourth row: property dependence of the errors. The first column is for redshift, the second column is for the i-band absolute magnitude and the third column is for the stellar masses. Stellar mass has units of solar masses. The red vertical dashed lines show the median values. The blue vertical dashed lines show the $16^{\textrm{th}}$ and $84^{\textrm{th}}$ percentiles.}
\label{fig:propagation_errors_properties}
\end{figure*}

\subsubsection{Galaxy properties: distributions and errors} \label{subsubsec:property_errors}

In this Section we present the collated results for the galaxy properties: redshifts, i-band absolute magnitudes and stellar masses derived from the 100 Monte-Carlo simulations. 

For each of the $3,207,756$ galaxies in the SPT-E field we computed the median, $16^{\textrm{th}}$ and $84^{\textrm{th}}$ percentiles (out of the 100 values) of the drawn redshifts, computed i-band absolute magnitudes and stellar masses. We quote the 1-sigma error (for each property) for each galaxy as half the difference between the $84^{\textrm{th}}$ and $16^{\textrm{th}}$ percentiles.

Fig \ref{fig:propagation_errors_properties} shows the (median) distributions of the redshifts, i-band absolute magnitudes and stellar masses in the first row, the distribution of errors on the properties in the second row, the TPZ redshift dependence of the property errors in the third row and the dependence of the property errors on the properties themselves in the fourth row. The left hand column shows the galaxy redshifts. The middle column shows i-band absolute magnitudes and the right hand column shows the galaxy stellar masses. Through out this paper the units of stellar mass is solar masses and stellar mass is plotted on a logarithmic scale. In the first row the vertical red dashed lines show the median values and the vertical blue dashed lines show the $16^{\textrm{th}}$ and $84^{\textrm{th}}$ percentiles of the redshift, i-band absolute magnitude and mass distributions. We quote the ranges of the property distributions in the plots as the difference between the $84^{\textrm{th}}$ and $16^{\textrm{th}}$ percentiles. In the second row the red vertical dashed lines mark the median property errors. We quote the median errors for comparison with the ranges of the distributions themselves quoted in the first row. 

The ranges of the distributions are: $0.56$, $2.62$ and $1.36$ for redshift, i-band absolute magnitude and stellar mass respectively and in the same order the median errors for these quantities are $0.054$, $0.256$ and $0.118$. The ratios of the ranges of the property distributions to the median property errors are: $10.3$, $10.2$ and $11.5$ for redshift, i-band absolute magnitude and mass respectively. This quantification suggests that all three of these properties can be studied, as the median error due to the photometric redshifts is an order of magnitude smaller than the ranges of the distributions of these properties. 

The third row of plots in Fig \ref{fig:propagation_errors_properties} shows the median errors (in redshift bins) on the properties as a function of the galaxies' TPZ redshifts. The redshift error corrected by a factor of $1+z$ is essentially constant at $\sim 0.035$ across the redshift range $0<z<1.0$. The error at $z\sim0.7$ is slightly smaller compared to the rest of the range. This is around the peak of the redshift distribution shown in plot a). The errors on the i-band absolute magnitudes and stellar masses behave in a similar way to each other. The errors are particularly large for $z<0.25$ and increase as the redshift is decreased. The errors stabilise at larger redshifts to values of $<0.4$ and $<0.2$ for the i-band absolute magnitude and stellar mass respectively. There is a ``sweet'' spot for both properties in the redshift range $0.6<z<0.7$ where the error is $<0.2$ for the i-band absolute magnitude and $<0.1$ for the stellar mass.

The fourth row of plots in Fig \ref{fig:propagation_errors_properties} shows the average $16^{\textrm{th}}$ and $84^{\textrm{th}}$ percentile property errors as a function of the median values of the properties themselves. As expected the redshift error dependence on the median redshift shown in plot j) is similar to the dependence on the TPZ redshift shown in plot g). The error dependence for the i-band absolute magnitude and masses also mirror each other. The errors are largest for the least luminous and least massive galaxies. This is in line with expectation as fainter galaxies are more difficult to measure. The i-band absolute magnitude error is relatively stable and $<0.25$ for galaxies with an absolute magnitude brighter than $-20.0$. Similarly the mass error is relatively constant at $<0.2$ for $\textrm{Log(M)}>8.5$.

\begin{figure*}
   \centering 
   \includegraphics[width=.75\linewidth]{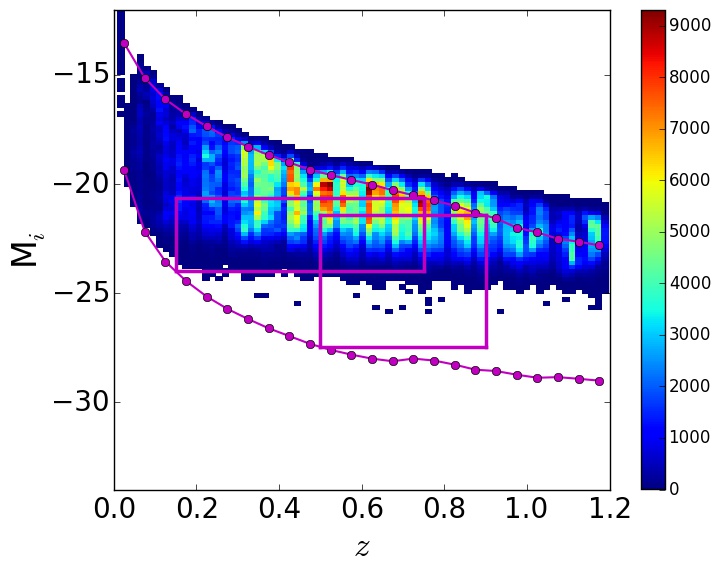}
\caption{Sample selection in the $M_{i}$-redshift plane. In this example the TPZ redshifts are used. The vertical magenta lines mark the redshift bounds. The number of galaxies in each 2-dimensional bin is represented with a colour as indicated in the colour bar. The magenta curves mark the lower and upper i-band absolute magnitude completeness limits. The magenta rectangular regions enclose the faint (low redshift) and bright (high redshift) density tracers. There is an overlapping redshift range between the faint and bright tracers that enables comparisons. Selection for each of the 100 Monte-Carlo simulations is performed in the same way using the redshifts drawn from the photo-z PDFs and the corresponding i-band absolute magnitudes.}
\label{fig:selection}
\end{figure*}

\section{Galaxy environment} \label{sec:galaxy_environment}

In this Section we present the galaxy environment measurements. We proceed with the Monte-Carlo approach and compute galaxy environments for all $100$ catalogue realizations to determine the error on the environment measurements of each galaxy. In Section \ref{subsec:environment_measurements} we describe the method we use to quantify galaxy environment. In Section \ref{subsec:environment_vs_parameters} we study the environment measurements and their errors as a function of the aperture parameters. In Section \ref{subsec:environment_characterization} we characterise the environment measurements that we employ later in the galaxy stellar mass function analysis. In Section \ref{subsec:environment_pdfs} we present the results of the Monte-Carlo simulations as environment PDFs.   

\subsection{Environment measurements} \label{subsec:environment_measurements}

In preparation for photometric surveys there have been a number of studies that have investigated the impact of redshift precision on measurements of galaxy environment. \cite{Cooper2005} concluded that for pencil beam surveys such as DEEP2 redshift measurements with errors $>0.02$ were unsuitable to measure \textit{local} galaxy densities. However more recently \cite{Fossati2015} showed using a semi-analytical model that it is possible to measure trends with galaxy environment. \cite{Etherington2015} examined the impact of redshift precision with a focus on large-scale surveys using SDSS data and found that the environmental signal in photometric data can be measured but needs to be optimised with a careful choice of aperture parameter values.

\subsubsection{Method} \label{subsubsec:method}

Galaxy environment has been measured with a variety of methods \cite[e.g][]{Carollo2013} including Voronoi Tessellation, fixed aperture, N$^{\textrm{th}}$ nearest neighbour methods and numerous variants of these. (see \cite{Muldrew2012} and \cite{Haas2012} for compilations).

In this work we employ a fixed aperture method. In this method a number density is calculated by counting the number of density tracing galaxies (see Section \ref{subsubsec:density_defining_pop}) found within an aperture centred on the target galaxy (the target galaxy is not included within the count) and dividing this by the volume of the aperture. There are three reasons for this choice of method: (i) fixed aperture methods are arguably easier to interpret because they compute densities over fixed scales, (ii) \cite{Shattow2013} showed that fixed aperture methods provide more robust measurements over cosmic time and (iii) fixed aperture methods are computationally less expensive compared to N$^{\textrm{th}}$ nearest neighbour methods. 

Several different aperture volumes have been used in previous studies including spheres \citep{Croton2005}, cylinders \citep{Gallazzi2009}, annuli \citep{Wilman2010}, cones \citep{Etherington2015} and ellipsoids \citep{Schawinski2007, Thomas2010}. In photometric surveys the errors in the redshift measurements along the line of sight are much greater than the errors in the angular measurements. The most appropriate aperture for a photometric survey is therefore conical in shape as this volume most effectively encompasses adjacent lines of sight. 

The aperture that we adopt is approximately a conical frustum, i.e. the volume that is left when you slice off the top of a cone. The volume is therefore calculated by taking the difference of two cones. We control the volume of the aperture with two parameters: the radius ($r$) of the cross section of the cone at the target galaxy and the ($\Delta z$) half-length of the aperture. In this study we investigate a range of radii: $0.1-3.0\;$Mpc and half-lengths: $0.08-0.3$ (in redshift). We count the number of density tracing galaxies (see Section \ref{subsubsec:density_defining_pop}) within the aperture and compute a density. Apertures that are found to be devoid of galaxies are assigned a nominal minimum density of $0.5$ galaxies per aperture. The density for each galaxy is then turned into a density contrast with respect to the mean density $\rho_{m}$ using the equation below:

\begin{equation} \label{eq:expectation}
\delta=(\rho-\rho_{m})/\rho_{m}
\end{equation}
 
The mean density $\rho_{m}$ is calculated within a redshift window centred on the target galaxy. We compute Log($1+\delta$) and refer to this quantity as the galaxy environment.

\subsubsection{Density Defining Populations} \label{subsubsec:density_defining_pop}

The distribution of stellar matter is a biased tracer \citep{Kaiser1984} of the large scale structure in the Universe. Maps of the total mass, baryonic and dark matter combined, have now been created using weak lensing measurements from the DES \citep{Vikram2015}. 

We follow the approach that has been adopted in previous galaxy environment studies \citep[e.g][]{Baldry2006,Thomas2010,Peng2010} and use only the galaxy distribution and assume this is an adequate proxy to trace the underlying density field. 

Populations of intrinsically faint galaxies are not detectable through the entire survey volume. To fairly trace the galaxy distribution we constructed volume limited samples. We employed two density defining populations which we called the faint and bright tracers because of the cuts on the absolute magnitudes of the galaxies. We employ luminosity cuts on the sample, rather than cuts in stellar mass because luminosities are more closely related to the observations and are less model dependent.

Fig \ref{fig:selection} shows an example of the sample selection of the galaxies from the SPT-E field using the i-band absolute magnitudes and in this example their TPZ redshifts. The colour bar indicates the number of galaxies in each grid cell. The upper and lower magenta curves mark the i-band absolute magnitude completeness limits. The faint and bright tracers consist of the galaxies within the magenta rectangles. The completeness limits were used to determine the extremities of the tracers. The faint tracer has a redshift range $0.15<z<0.75$ and an i-band absolute magnitude range $-24.0<M_{i}<-20.63$. The bright tracer has a redshift range $0.6<z<1.05$ and an i-band absolute magnitude range $-27.91<M_{i}<-22.37$. We purposely designed an overlap of the redshift ranges of the two tracers to enable comparisons between the two tracers. Table \ref{table:tracer_properties} summarizes the properties of the faint and bright tracers. Selection for each of the 100 Monte-Carlo simulations is performed in the same way using the redshifts drawn from the photo-z PDFs and the corresponding i-band absolute magnitudes.

\subsubsection{Survey edge and holes} \label{subsubsec:survey_edge}

Galaxy environment measurements require contiguous regions to ensure densities are not underestimated. The boundaries of the homogenised SPT-E field are not regular and there are holes in the data caused by bright stars. It is therefore important that the edges of the data are determined and appropriately managed. We used the HEALPix \citep{Gorski2005} software to identify the pixels that contained galaxies and those that did not. We populated the cells with random points. The density of random points was $>$10 times the average density of the galaxies. We then computed the angular distance between points that were inside the data footprint and the points outside of the footprint. We identified the set of points that were inside the footprint but were the closest to any point outside of the footprint. This set of random points defines the edge of the SPT-E data that we used in the scientific analysis. Fig \ref{fig:edges} shows the positions of the set of points that defines the edges of the footprint. To ensure the periphery of the data did not impact the environment measurements we applied a conservative cut and discarded galaxies that were less than $0.1$ degree away from an edge point but inside the footprint. After applying this cut the area of the footprint was $78.09$ deg$^{2}$.         

We manage the redshift boundaries by adjusting the depth of the aperture in cases where the aperture would cross over the boundary. We reduce the half depth of the aperture (of the infringing half) to be the comoving distance from the target to the boundary. This ensures the aperture fits inside the redshift range. The depth of the other half of the aperture that resides within the redshift range is not changed.

\begin{table}
\caption{Density defining population properties} \label{table:tracer_properties}
\begin{center}
    \begin{tabular}{@{}p{1.3cm}ll}
    \hline
  Property & Faint Tracer & Bright Tracer \\ 
    \hline
   $z$ & $0.15<z<0.75$ & $0.6<z<1.05$ \\
   $M_{i}$ & $-24.00<M_{i}<-20.63$ & $-27.91<M_{i}<-22.37$ \\ 
 Radius & $r=1.0\;$Mpc & $r=1.4\;$Mpc \\ 
 Half-depth & $\delta z$=0.1 & $\delta z$=0.2 \\
 Range Log(1+$\delta$) & $0.36$ & $0.42$ \\
 Median $\Delta$Log(1+$\delta$) & $0.096$ & $0.11$ \\
 Ratio & 3.8 & 3.8 \\
    \hline
    \end{tabular}
\end{center}
\end{table}

\begin{figure*}
\begin{minipage}[t]{0.32\linewidth}
   \centering 
   \includegraphics[width=.99\linewidth]{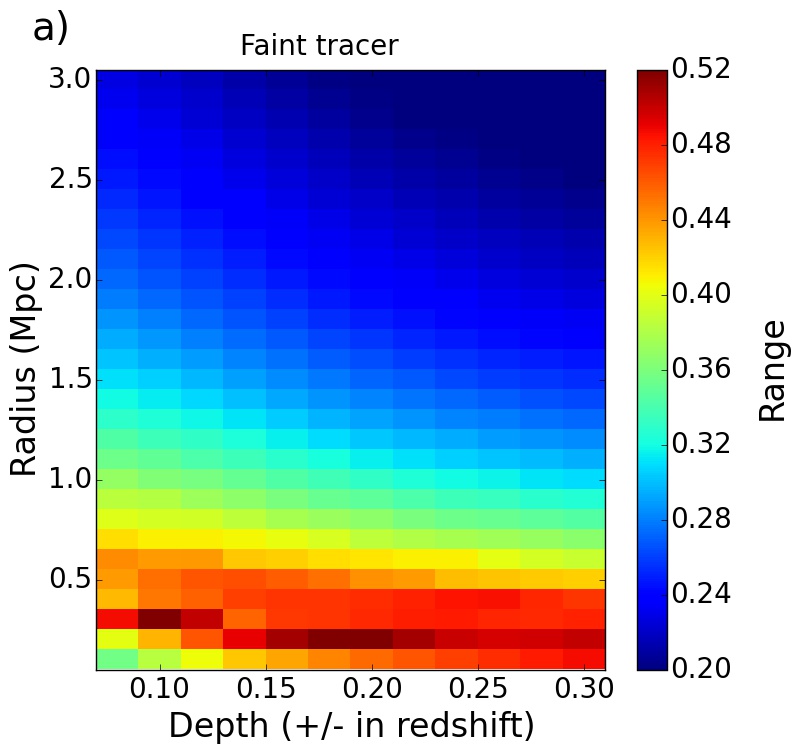}
\end{minipage}
\begin{minipage}[t]{0.32\linewidth}
  \centering  
  \includegraphics[width=.99\linewidth]{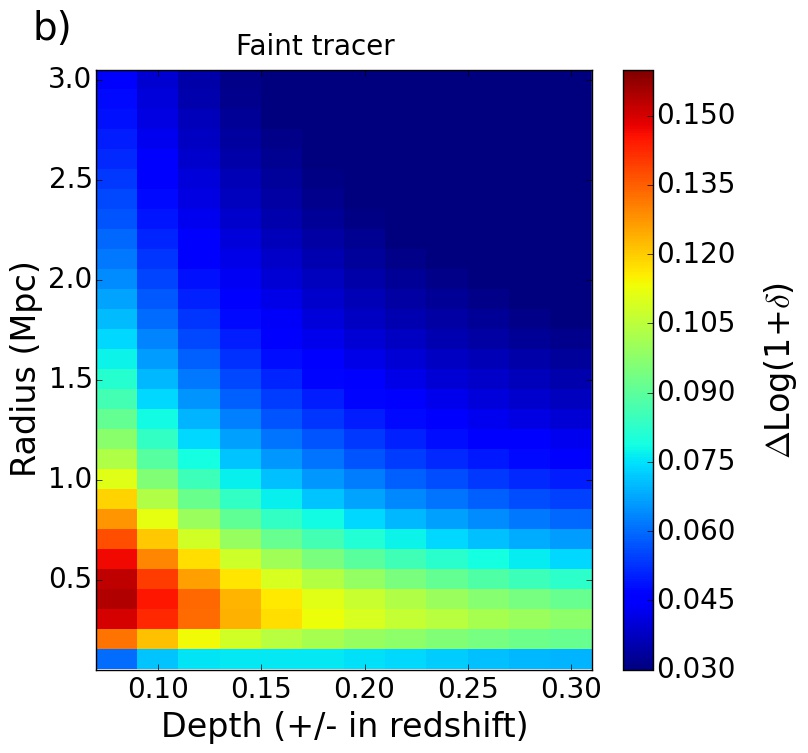}
\end{minipage}
\begin{minipage}[t]{0.32\linewidth}
  \centering  
  \includegraphics[width=.99\linewidth]{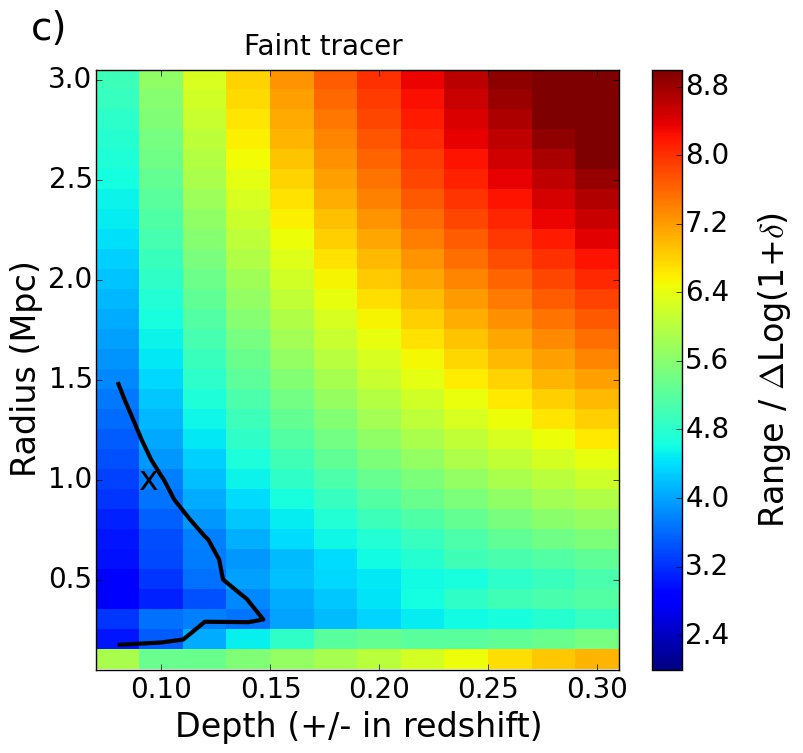}
\end{minipage}
\begin{minipage}[t]{0.32\linewidth}
   \centering 
   \includegraphics[width=.99\linewidth]{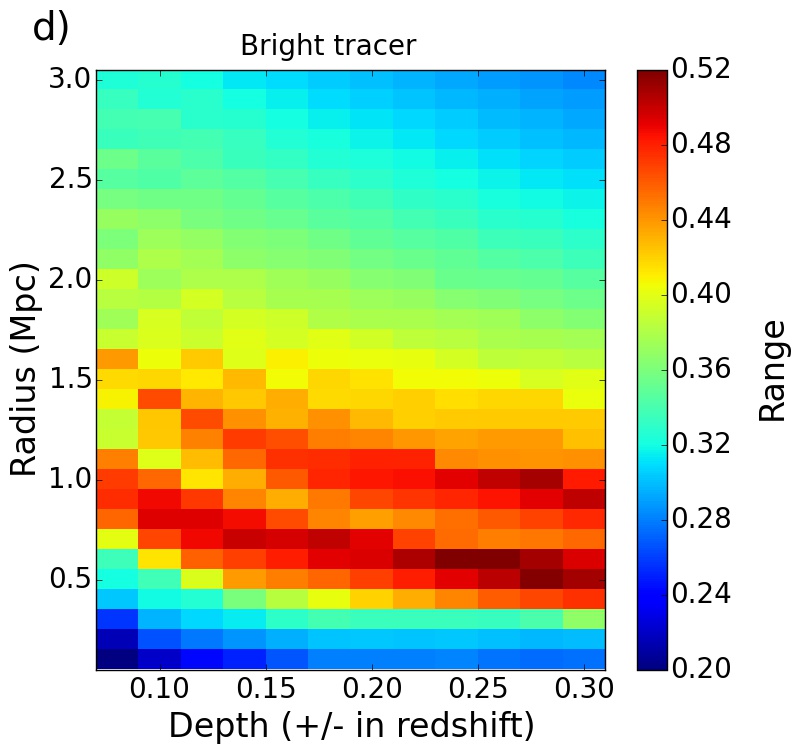}
\end{minipage}
\begin{minipage}[t]{0.32\linewidth}
  \centering  
  \includegraphics[width=.99\linewidth]{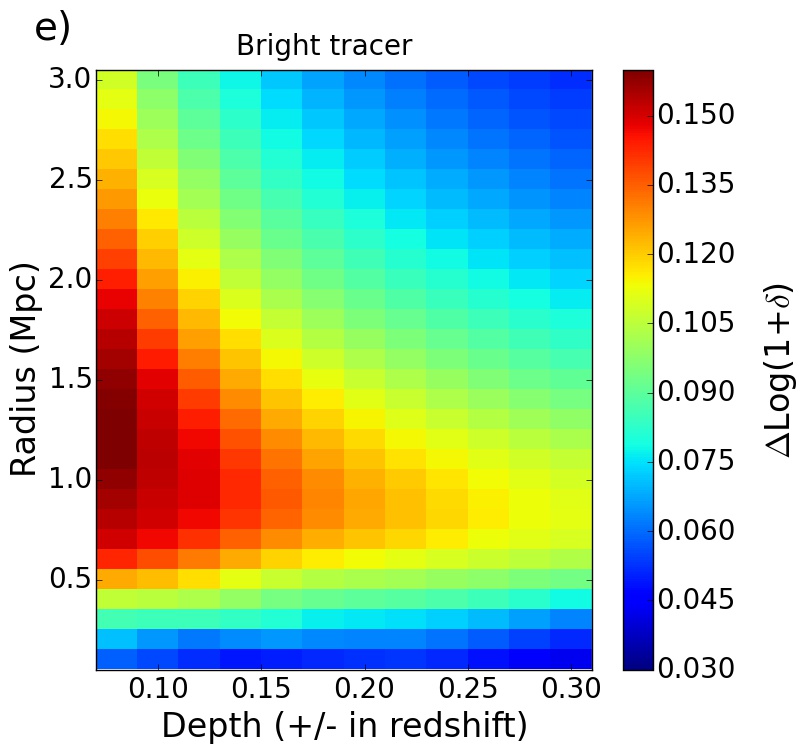}
\end{minipage}
\begin{minipage}[t]{0.32\linewidth}
  \centering  
  \includegraphics[width=.99\linewidth]{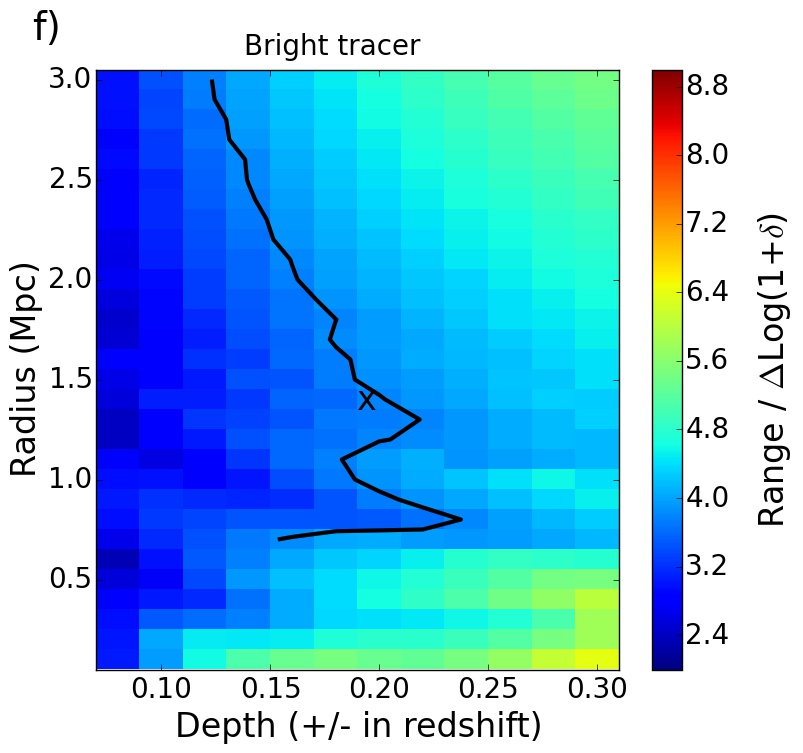}
\end{minipage}
\caption{Left column: range of the environment distributions (difference between the $84^{\textrm{th}}$ and $16^{\textrm{th}}$ percentiles) as a function of the radius and depth of the aperture. Middle column shows the median error of the environment measurements as a function of the radius and depth. Right column shows the ratio of the range of the environment distribution to the median error as a function of the radius and depth. Top row is for the faint tracer and the bottom row is for the bright tracer. The contour lines in plots c) and f) are for a ratio of $3.8$.}  
\label{fig:envs_ratio_by_param}
\end{figure*}

\begin{figure*}
\begin{minipage}[t]{0.49\linewidth}
   \centering 
   \includegraphics[width=.99\linewidth]{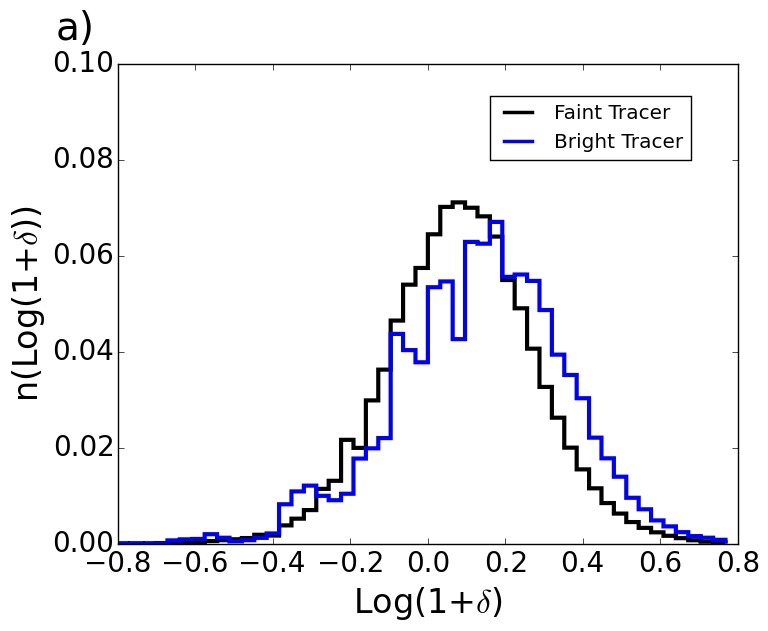}
\end{minipage}
\begin{minipage}[t]{0.49\linewidth}
  \centering  
  \includegraphics[width=.99\linewidth]{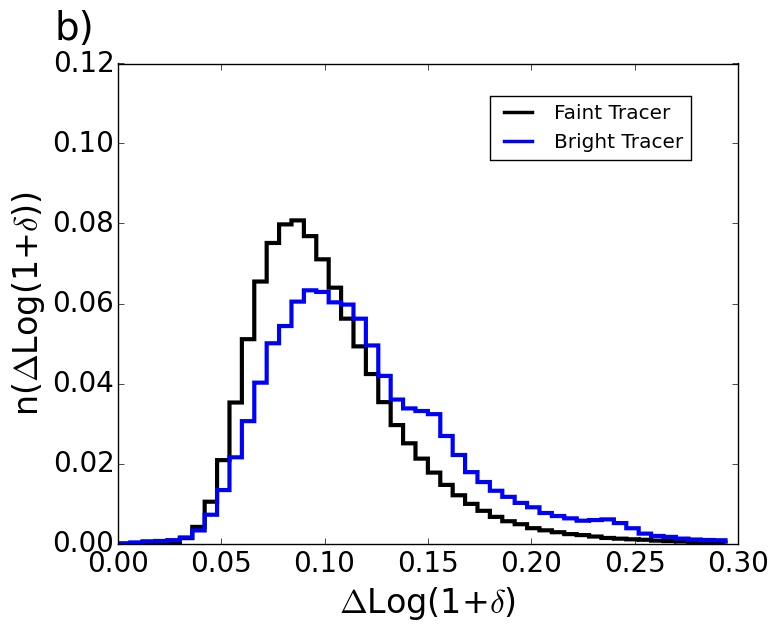}
\end{minipage}
\begin{minipage}[t]{0.49\linewidth}
  \centering  
  \includegraphics[width=.99\linewidth]{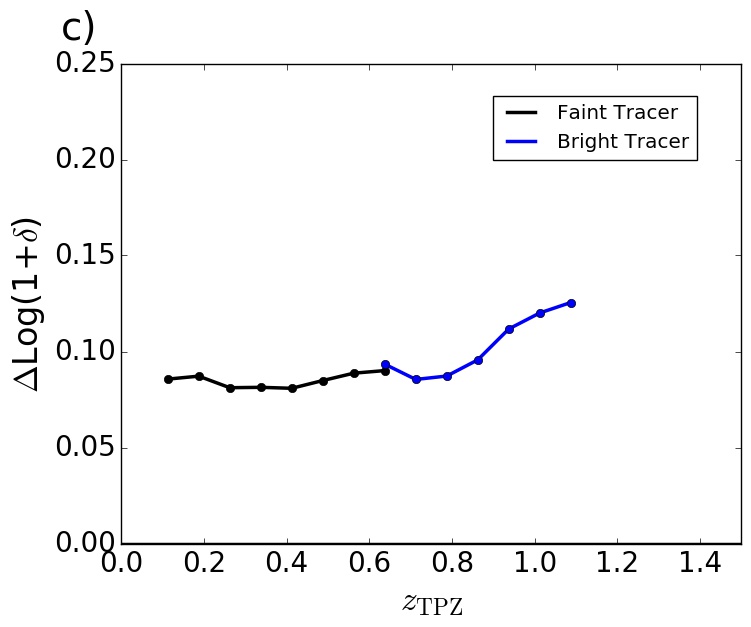}
\end{minipage} 
\begin{minipage}[t]{0.49\linewidth}
   \centering 
   \includegraphics[width=.99\linewidth]{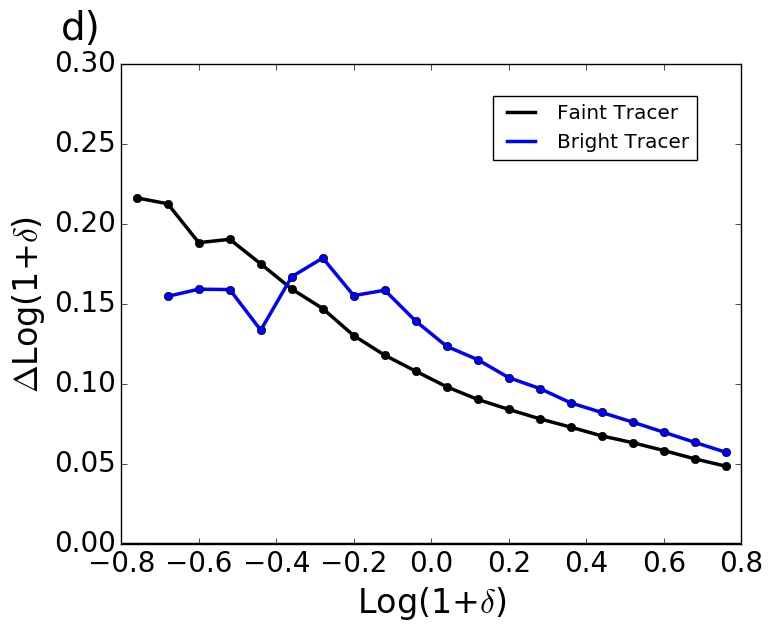}
\end{minipage}
\caption{a) environment distribution, b) distribution of environment error, c) TPZ redshift dependence of the environment error and d) environment error as a function of environment for the faint (black) and bright (blue) tracers.}  
\label{fig:envs_character}
\end{figure*}

\subsection{Environments as a function of the aperture parameters} \label{subsec:environment_vs_parameters}

To investigate the impact of the aperture parameters on the environment measurements we tested apertures with radii of $0.1-3.0\;$Mpc at $0.1\;$Mpc increments and half-depths of $0.08-0.3$ at $0.02$ increments in redshift. 

Fig \ref{fig:envs_ratio_by_param} shows the range of the environment distribution of the galaxies in the density defining population as a function of the aperture parameters in the left column; the median error of the environment measurements as a function of the aperture parameters in the middle column and the ratio of the range to the environment error as a function of the aperture parameters in the right column. The top row is for the faint, low-z tracer. The bottom row is for the bright, high-z tracer. We define the range of the environment distribution as the difference between the $84^{\textrm{th}}$ and $16^{\textrm{th}}$ percentiles of the distribution.

The range of the environment distribution of the density defining population depends strongly on aperture radius and weakly on the aperture half-depth for both the faint and bright tracers. For both tracers the environment range becomes larger as the radius decreases from $3.0\;$Mpc. This trend continues for the faint tracers until a radius of $0.2\;$Mpc and for the bright tracer until a radius of $0.6\;$Mpc. Apertures with radii smaller than these values poorly sample the tracing population, because many apertures are devoid of galaxies. A larger aperture is required for the bright tracer because this tracer samples the density field more sparsely. 

The middle column shows that the median environment error is a smooth function of both the aperture radius and half depth. The general trend is that the median environment error decreases as the volume of the aperture increases. Probing environments on large scales homogenises the measurements as local contrasts are smoothed out. The median error is also small for radii less than $0.3\;$Mpc for the faint tracer and less than $0.6\;$Mpc for the bright tracer. These scales are approaching or beneath the average sampling scale of the tracing populations. The small environment errors at these scales are an artificial effect and these scales should not be employed for scientific analysis. 
 
The ideal scenario is to have a large environment range and small measurement errors. However the trends of increasing range and decreasing error depend on the aperture parameters, especially the radius in a counteracting fashion. 

The right column shows the ratio of the tracing population environment range to the median environment error as a function of the aperture parameters. The black lines over-plotted mark a constant ratio of $3.8$. The black crosses mark the aperture parameters that we select to employ later. The plots show that the ratio increases with increasing depth and radius for both the faint and bright tracers. However the trend is stronger for the faint tracer, illustrated with the stronger colour gradient. On the one hand it is desirable for this ratio to be as large as possible to minimize contamination between environment bins but on the other hand it is necessary to probe signal from the scales where environmental processes have a role.  

Previous studies have reported that environmental processes occur most readily on scales of {\raise.17ex\hbox{$\scriptstyle\sim$}}$1\;$Mpc or less \citep{Blanton2007,Wilman2010}. For scientific analysis employing the faint tracer we therefore opt for a radius of $1\;$Mpc. The choice of depth for the faint tracer is a trade off between maximising the ratio (between the range of the environment distribution and the median environment error) and the goal of measuring \textit{local} environment. We opt for a half-depth of $0.1$ (in redshift). The ratio with these aperture parameter values is {\raise.17ex\hbox{$\scriptstyle\sim$}}$3.8$. The choice for the bright tracer is more constrained. For the purposes of comparison between the faint and bright tracers we choose aperture parameter values for the bright that lead to a similar ratio. For the bright tracer we therefore choose a radius of $1.4\;$Mpc and a half-depth of $0.2$ (in redshift) which also gives a ratio of {\raise.17ex\hbox{$\scriptstyle\sim$}}$3.8$. We note that with this choice of parameters the half-depths of the aperture are at least twice as large as the 1-sigma photometric redshift errors (see Fig. \ref{fig:propagation_errors_properties}) across the whole range of redshifts we study and this ensures that the environment measurements are not severely affected by signal to noise issues. The number of apertures that are devoid of density defining population galaxies is less than $0.2$ and $4.0$ percent for the faint and bright tracers respectively.

The ratio between the range of the distribution and the average error for the other galaxy properties (redshift, mass and i-band absolute magnitude) was {\raise.17ex\hbox{$\scriptstyle\sim$}}$10$-$12$ as shown in Section \ref{subsubsec:property_errors}. The environment measurements therefore have a distinguishing power that is only about $3$ times less than the other parameters.

\subsection{Environment characterisation} \label{subsec:environment_characterization}

We now characterise in more detail the environment measurements obtained with the aperture parameters chosen in Section \ref{subsec:environment_vs_parameters}. The key properties, such as the redshift ranges, aperture parameters and environment properties are listed in Table \ref{table:tracer_properties}. 

Fig \ref{fig:envs_character} shows the environment distribution in the top left; the median environment error distribution in the top right; the median environment error as a function of the TPZ redshift in the bottom left and the median environment error as a function of the median environment in the bottom right for the faint (black) and bright (blue) tracers. 

The environment distributions for both the faint and the bright tracers are approximately Gaussian and have widths of $0.36$ and $0.42$ respectively. The apertures are sufficiently large that a negligible number are devoid of galaxies and the distributions are roughly symmetrical. The shape and range of the faint and bright environment distributions are similar despite the fact that the apertures probe different volumes and different tracing populations. The environment distribution for the faint tracer is slightly narrower and more peaked than the distribution for the bright tracer. The error distributions shown in plot b) have extended tails at the high error end. The median environment error for the faint and brighter tracer are $0.096$ and $0.11$. The faint tracer has a slightly smaller median error than the bright tracer. The ratio of the distribution width to the median environment error is approximately $3.8$ for both tracers by construction. This value is sufficient to study trends with environment. The similarities in the overall properties of the environments for the faint and bright tracer are due to the constraint on this ratio. Plot c) shows that the error on the environment measurements is relatively constant as a function of the TPZ redshift. The error increases slightly for each tracer as the TPZ redshift increases.

The errors for the faint and bright tracers are approximately the same in the redshift region ($z=0.65$) where the tracers overlap. Plot d) shows that the median error on the environment measurements decreases with increasing environment for both tracers from $0.2$ for sparse environments to $0.05$ for the most dense environments. The main reason for this is that high density environments by definition contain many galaxies. Perturbing the number of galaxies in high density regions due to the imprecise redshift measurements therefore has a much smaller effect than perturbing the number of galaxies in a low density region because of the logarithmic definition of environment that is adopted in this work.

\begin{figure}
\begin{minipage}[t]{0.99\linewidth}
   \centering 
   \includegraphics[width=.99\linewidth]{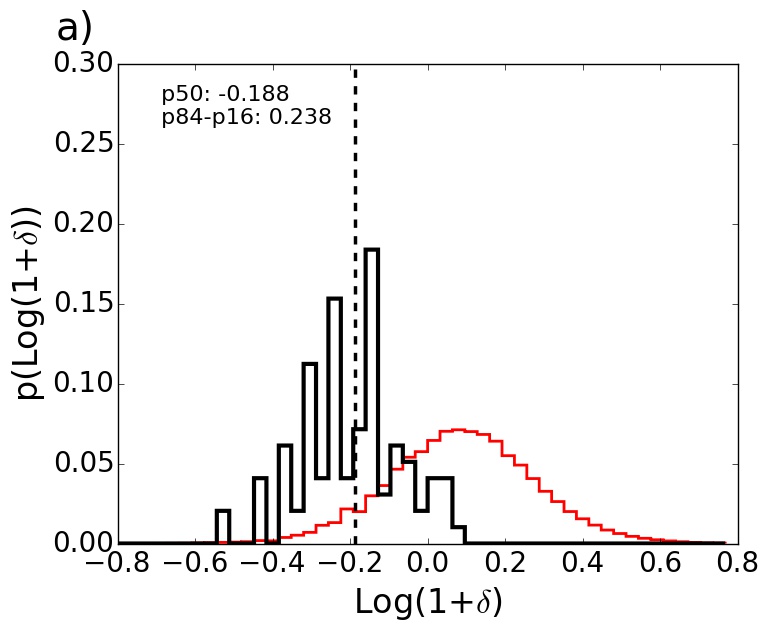}
\end{minipage}
\begin{minipage}[t]{0.99\linewidth}
  \centering  
  \includegraphics[width=.99\linewidth]{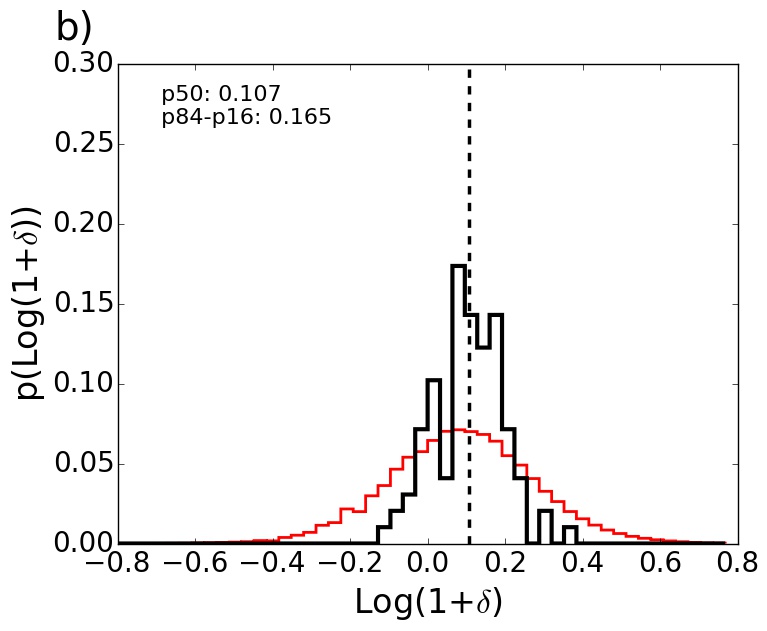}
\end{minipage}
\begin{minipage}[t]{0.99\linewidth}
  \centering  
  \includegraphics[width=.99\linewidth]{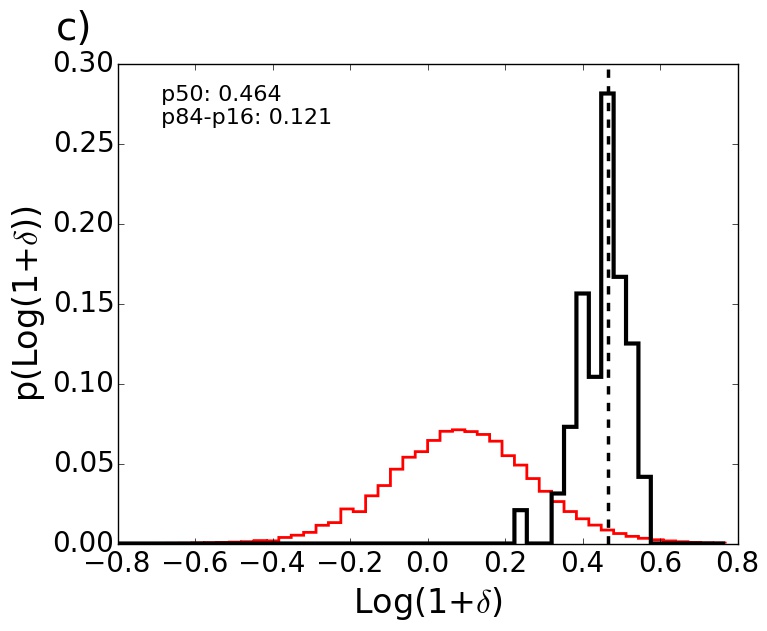}
\end{minipage}
\caption{Three examples of environment PDFs: a) low density (large error), b) intermediate density (medium error) and c) high density (small error) are shown in black. The environment distribution for the whole population of galaxies is shown in red in each plot. The vertical dashed lines marks the median of the environment PDFs. The difference between the $84^{\textrm{th}}$ and $16^{\textrm{th}}$ percentiles for the environment PDFs are quoted to quantify the width of the PDFs.}  
\label{fig:env_pdfs}
\end{figure}

\subsection{Environment PDFs} \label{subsec:environment_pdfs}

Using the results from the $100$ Monte-Carlo realizations the environment measurements for each galaxy can be presented as PDFs. This is achieved for a galaxy by constructing a histogram of the relative frequencies of each environment from the $100$ measurements. Fig \ref{fig:env_pdfs} shows three examples of environment PDFs based on the faint tracer. The top plot shows a galaxy in a low density environment, the middle plot shows a galaxy in an intermediate density environment and the bottom plot shows a galaxy in a high density environment. The red histogram shows the distribution of median environments for the whole population of galaxies (i.e. it is the same as plot a) in Fig \ref{fig:envs_character}). The vertical dashed line marks the median values of the environment PDFs. The distributions for these galaxies are peaked and their widths are clearly smaller than the distribution of environments for the entire population. With such environment PDFs it is possible to split the galaxies into bins of environment, albeit some unavoidable contamination. The impact of any contamination, however, is negligible given the large statistical sample in this study.

These three galaxies also illustrate the trend that as the environmental density increases the error on the environment measurement decreases. As the environmental density increases the environment PDFs become more peaked and narrower.

The environment PDFs presented here enable more sophisticated statistical studies of galaxy environment in photometric surveys. The median environment measurement for each galaxy can be employed with an associated error or the complete environment PDFs can be folded into analyses. We demonstrate such an analysis in Section \ref{sec:mass_function} by studying the environmental components of the galaxy stellar mass function.

\section{Mass Function analysis} \label{sec:mass_function}

Analysis of the total galaxy stellar mass function for the DES SV data together with a detailed comparison with the literature is presented in Capozzi et al. (in prep). In this Section we present an analysis of the environmental components of the galaxy stellar mass function using the mass and environment measurements we have described in the previous Sections. We adopt a similar approach to \cite{Bundy2006}, \cite{Bolzonella2010} and \cite{Davidzon2016}. In Section \ref{subsec:mf_method} we describe the method we use to compute the mass functions. In Section \ref{subsec:mf_results} we present the results which are split into four parts: (i) the local universe in Section \ref{subsubsec:local_universe}, (ii) environmental components of the mass function for complete mass ranges in Section \ref{subsubsec:components_for_complete_range}, (iii) the redshift evolution of the environmental components of the mass function for common mass ranges in Section \ref{subsubsec:evolution_for_common_range} and (iv) the evolution of the environmental ratio of effective number of galaxies per unit volume in Section \ref{subsubsec:evolution_of_the_environmental_ratio}.

\subsection{Method} \label{subsec:mf_method}

We adopt the standard Schmidt-Eales (1/V$_{\rm max}$, \citealp{Schmidt1968}) method to calculate the galaxy stellar mass function. The number of galaxies per comoving volume $\phi(M)$ for the mass interval $\Delta M$ is given by the sum over the N galaxies observed within this interval:

\begin{equation} \label{eq:mass_function}
\phi (M) = \frac{1}{\Delta M} \sum_{i=1}^{N} \frac{1}{V_{\mathrm{max},i} \cdot C_{i}}
\end{equation}

In this equation V$_{\rm max, i}$ is the maximum volume accessible by the $i^{\rm th}$ galaxy. It is calculated by determining the minimum and maximum redshifts ($z_{\rm max, i}$ and $z_{\rm min, i}$) at which the galaxy could be detected in the survey, given the flux detection limits. These minimum and maximum redshifts are dependent on the galaxy SED and in particular on k-correction. $C_{\rm i}$ is the completeness factor of the $i^{\rm th}$ galaxy. It depends on the galaxy's surface brightness and apparent magnitude and takes a value between 0 and 1. The quantities used for determining V$_{\rm max, i}$ (i.e. $z_{\rm max, i}$ and $z_{\rm min, i}$) and the completeness factors $C_{\rm i}$ were provided by the COMMODORE catalogue as described in Section \ref{subsec:stellar_masses_fitting} and in more detail in Capozzi et al. (in prep).

\begin{table*} 
\caption{$16.67^{\mathrm{th}}$, $33.33^{\mathrm{rd}}$, $50.0^{\mathrm{th}}$, $66.67^{\mathrm{th}}$ and $83.33^{\mathrm{rd}}$ percentile boundaries for the environment bins for the faint and bright tracers employing the complete mass range for each redshift bin. The numbers in brackets are the 1-sigma errors on the bin boundaries.} 
\begin{center}
    \begin{tabular}{@{}llllllll}
    \hline
  Tracer & Redshift range & Mass range & $16.67^{\mathrm{th}}$ & $33.33^{\mathrm{rd}}$ & $50.0^{\mathrm{th}}$ & $66.67^{\mathrm{th}}$ & $83.33^{\mathrm{rd}}$  \\
     \hline 
Faint & $0.15, 0.225$ & $8.11, 11.35$ & -0.204 (0.0015) & -0.0776 (0.00096) & 0.0189 (0.0010) & 0.110 (0.0014) & 0.224 (0.0012) \\ 
 & $0.225, 0.3$ & $8.61, 11.70$ & -0.173 (0.00023) & -0.0669 (0.0016) & 0.0365 (0.00018) & 0.126 (0.00038) & 0.229 (0.00079) \\ 
 & $0.3, 0.375$ & $8.88, 12.18$ & -0.163 (0.00093) & -0.0529 (0.00039) & 0.0328 (0.00062) & 0.127 (0.00043) & 0.243 (0.0012) \\ 
 & $0.375, 0.45$ & $9.33, 12.37$ & -0.165 (0.00086) & -0.0332 (0.00077) & 0.0631 (0.0011) & 0.154 (0.0013) & 0.266 (0.00087) \\ 
 & $0.45, 0.525$ & $9.45, 12.61$ & -0.144 (0.00070) & -0.0265 (0.00055) & 0.0640 (0.00061) & 0.151 (0.0013) & 0.258 (0.00065) \\ 
 & $0.525, 0.6$ & $9.52, 12.68$ & -0.123 (0.00087) & -0.0130 (0.00055) & 0.0687 (0.0013) & 0.150 (0.00088) & 0.251 (0.0014) \\ 
 & $0.6, 0.675$ & $9.72, 12.68$ & -0.126 (0.00081) & -0.00392 (0.00076) & 0.0836 (0.00068) & 0.167 (0.00080) & 0.267 (0.0010) \\ 
 & $0.675, 0.75$ & $9.87, 12.51$ & -0.113 (0.0015) & 0.0158 (0.0013) & 0.111 (0.0010) & 0.202 (0.00098) & 0.310 (0.00096) \\ 
Bright & $0.75, 0.825$ & $9.99, 12.54$ & -0.204 (0.00024) & -0.0695 (0.00085) & 0.0399 (0.00093) & 0.164 (0.00037) & 0.275 (0.00049) \\ 
 & $0.825, 0.9$ & $10.2, 12.55$ & -0.194 (0.00048) & -0.0465 (0.0020) & 0.0756 (0.0021) & 0.175 (0.00055) & 0.289 (0.0013) \\ 
 & $0.9, 0.975$ & $10.5, 12.65$ & -0.173 (0.0027) & 0.0127 (0.0027) & 0.130 (0.0014) & 0.228 (0.0017) & 0.342 (0.0015) \\ 
 & $0.975, 1.05$ & $10.7, 12.83$ & -0.0985 (0.0028) & 0.0501 (0.0033) & 0.170 (0.0016) & 0.276 (0.0020) & 0.392 (0.0028) \\ 
    \hline
    \end{tabular}
\end{center}
\label{table:env_boundaries}
\end{table*}

\begin{table*} 
\caption{$25.0^{\mathrm{th}}$, $50.0^{\mathrm{rd}}$ and $75.0^{\mathrm{th}}$ percentile boundaries for the environment bins for the faint and bright tracers employing a common mass range for each tracer. The numbers in brackets are the 1-sigma errors on the bin boundaries. This binning scheme is used to study the redshift evolution of the components of the mass function.} 
\begin{center}
    \begin{tabular}{@{}lllllll}
    \hline
  Tracer & Redshift range & Mass range & $25.0^{\mathrm{th}}$ & $50.0^{\mathrm{rd}}$ & $75.0^{\mathrm{th}}$  \\
     \hline 
Faint & $0.3, 0.45$ & $10.00, 12.00$ & -0.0571 (0.00061) & 0.0843 (0.00084) & 0.223 (0.00077) &  \\ 
 & $0.45, 0.55$ & $10.00, 12.00$ & -0.0550 (0.0022) & 0.0845 (0.0012) & 0.218 (0.00096) &  \\ 
 & $0.55, 0.65$ & $10.00, 12.00$ & -0.0471 (0.00050) & 0.0919 (0.00059) & 0.219 (0.00055) &  \\ 
 & $0.65, 0.75$ & $10.00, 12.00$ & -0.0391 (0.00089) & 0.113 (0.00085) & 0.251 (0.00085) &  \\ 
Bright & $0.65, 0.75$ & $10.80, 12.40$ & -0.0373 (0.0010) & 0.133 (0.0011) & 0.278 (0.0024) &  \\ 
 & $0.75, 0.85$ & $10.80, 12.40$ & -0.0645 (0.00078) & 0.110 (0.00068) & 0.272 (0.00078) &  \\ 
 & $0.85, 0.95$ & $10.80, 12.40$ & -0.0409 (0.0017) & 0.127 (0.0020) & 0.277 (0.0018) &  \\ 
 & $0.95, 1.05$ & $10.80, 12.40$ & -0.0199 (0.0046) & 0.161 (0.0030) & 0.318 (0.0023) &  \\ 
    \hline
    \end{tabular}
\end{center}
\label{table:env_boundaries_for_evolution}
\end{table*}

\begin{figure}
   \centering 
   \includegraphics[width=.99\linewidth]{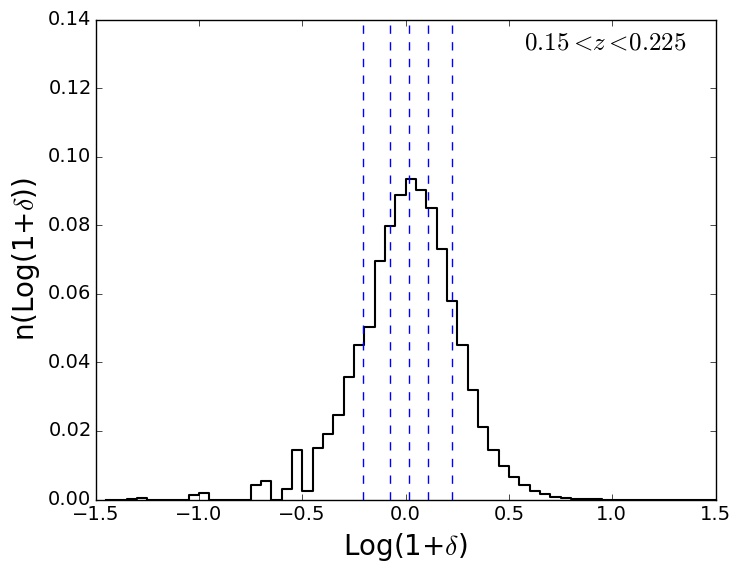}
\caption{Environment distribution and dividing values for environment bins for the redshift range $0.15<z<0.225$. Environments are based on the faint tracer.}  
\label{fig:env_bins}
\end{figure}

\begin{figure*}
\begin{minipage}[t]{0.33\linewidth}
   \centering 
   \includegraphics[width=.99\linewidth]{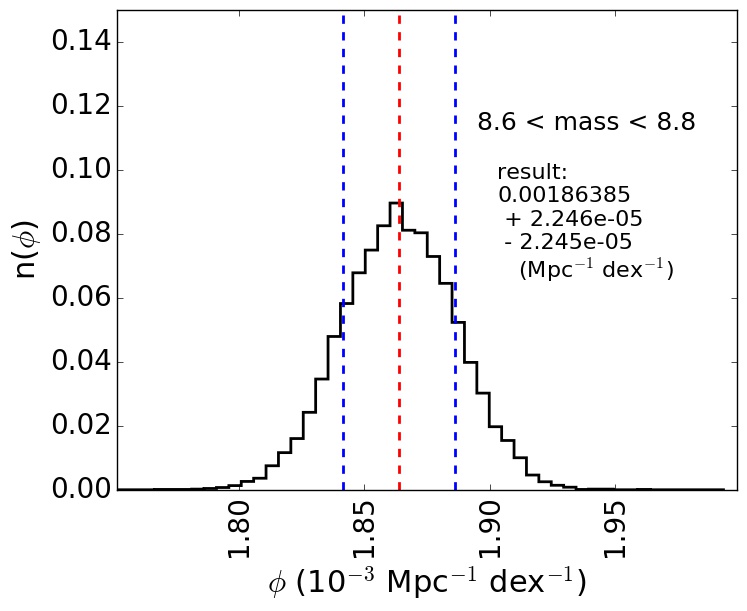}
\end{minipage}
\begin{minipage}[t]{0.33\linewidth}
  \centering  
  \includegraphics[width=.99\linewidth]{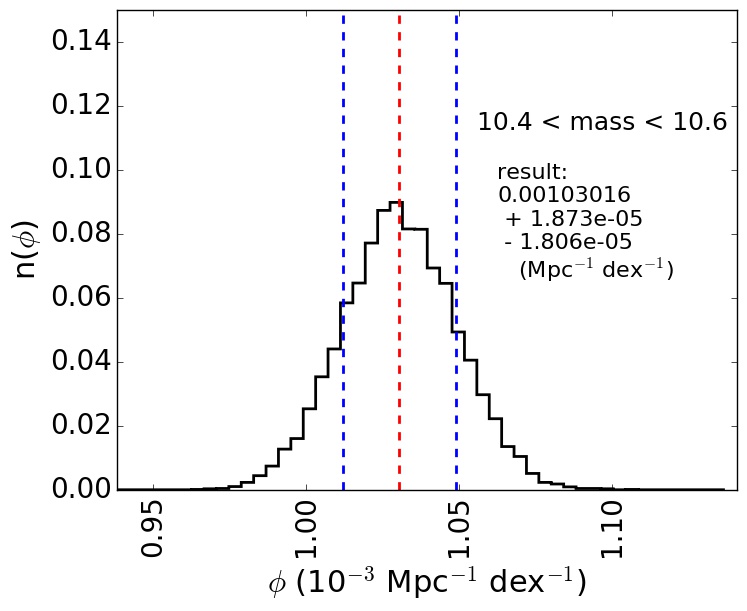}
\end{minipage}
\begin{minipage}[t]{0.33\linewidth}
  \centering  
  \includegraphics[width=.99\linewidth]{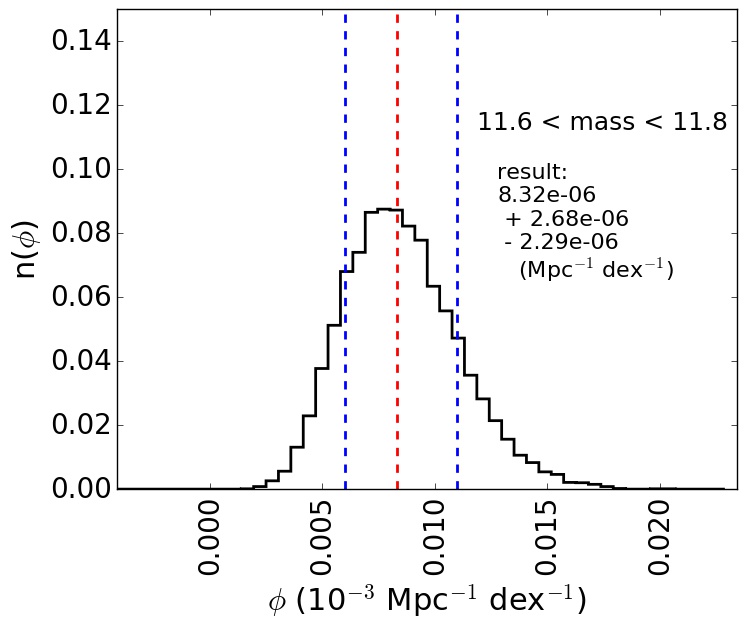}
\end{minipage}
\caption{Three examples of the error distribution of the effective number density of galaxies in the redshift range: $0.16<z<0.25$ for a low mass bin (left), intermediate mass bin (middle) and a high mass bin (right).}  
\label{fig:mf_errors_examples}
\end{figure*}

\begin{figure*}
\begin{minipage}[t]{0.49\linewidth}
  \centering 
  \includegraphics[width=.99\linewidth]{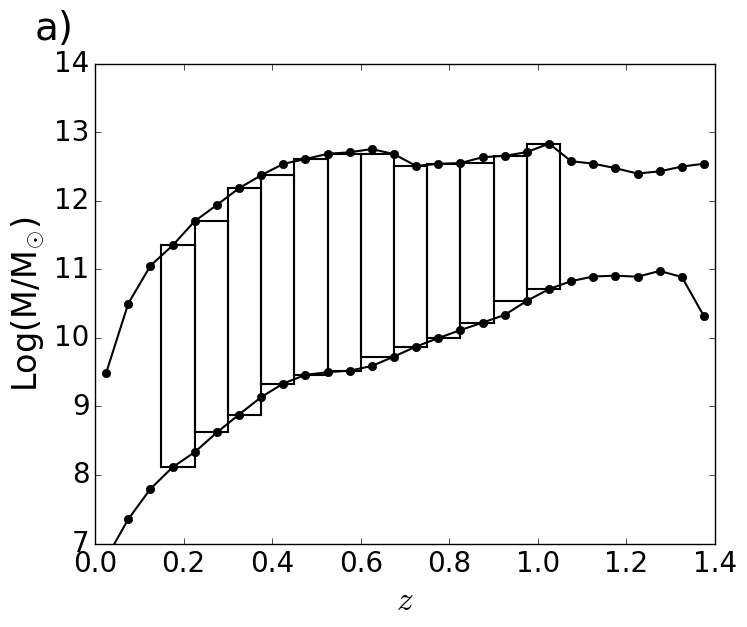}
\end{minipage}
\begin{minipage}[t]{0.49\linewidth}
  \centering      
  \includegraphics[width=.99\linewidth]{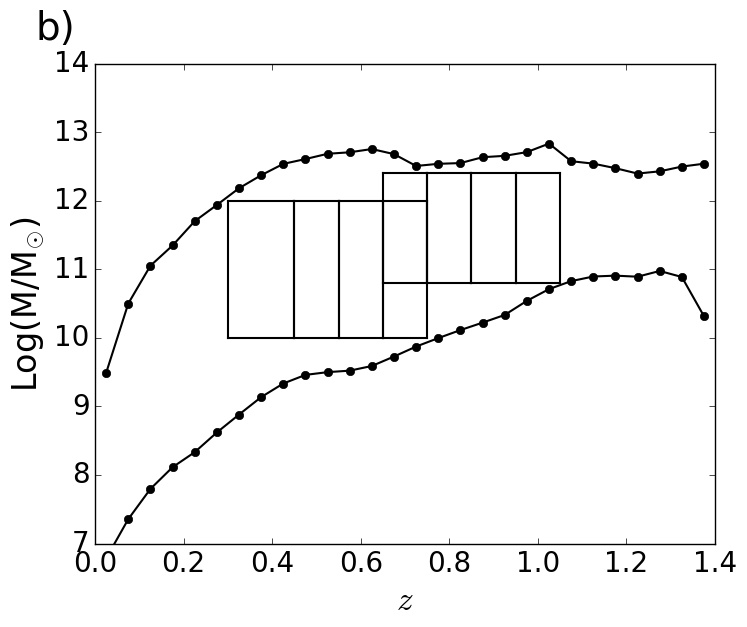}
\end{minipage}
\caption{Curves show the mass completeness limits. The rectangles shows the complete mass range for each redshift bin (left) and common mass ranges for the faint ($z<0.75$) and bright ($z>0.65$) tracers for the redshift evolution analysis (right).}
\label{fig:mass_completeness}
\end{figure*}

\begin{figure*}
\begin{minipage}[t]{0.32\linewidth}
\centering 
\includegraphics[width=.99\linewidth]{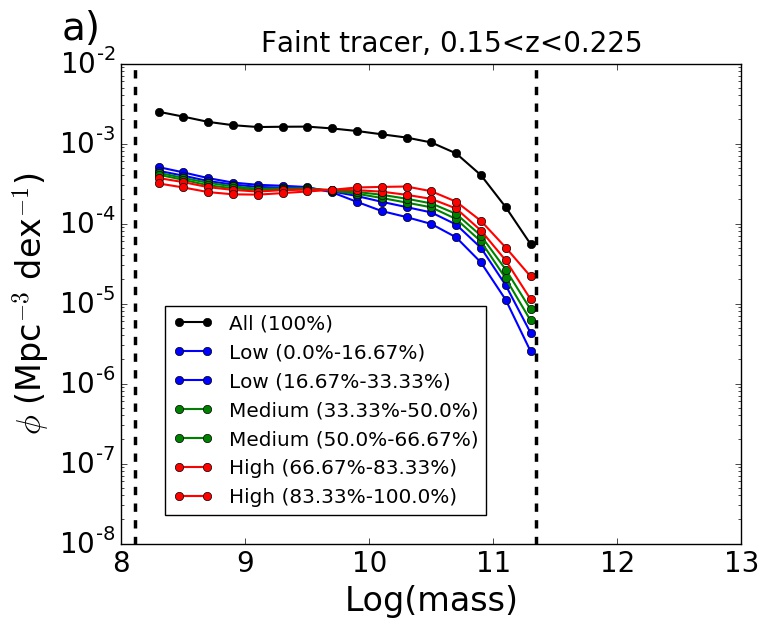}
\end{minipage}
\begin{minipage}[t]{0.32\linewidth}
\centering 
\includegraphics[width=.99\linewidth]{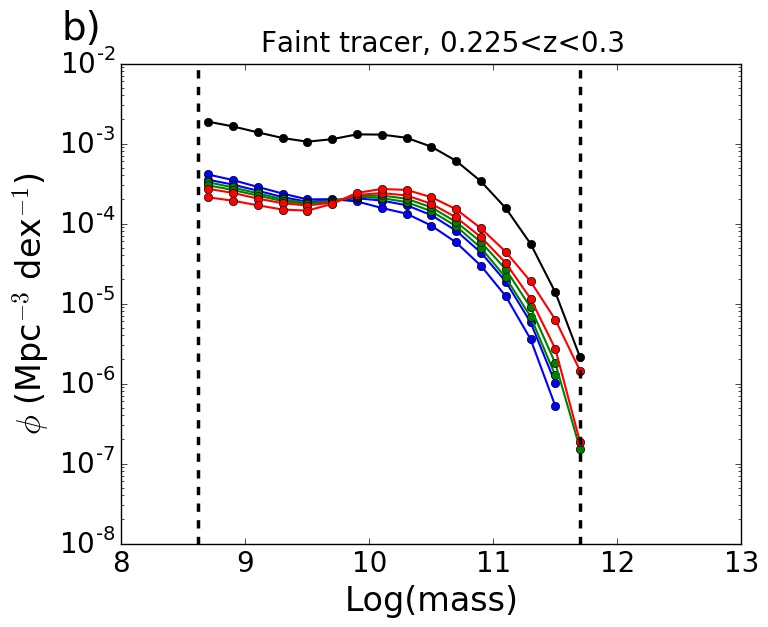}
\end{minipage}
\begin{minipage}[t]{0.32\linewidth}
\centering 
\includegraphics[width=.99\linewidth]{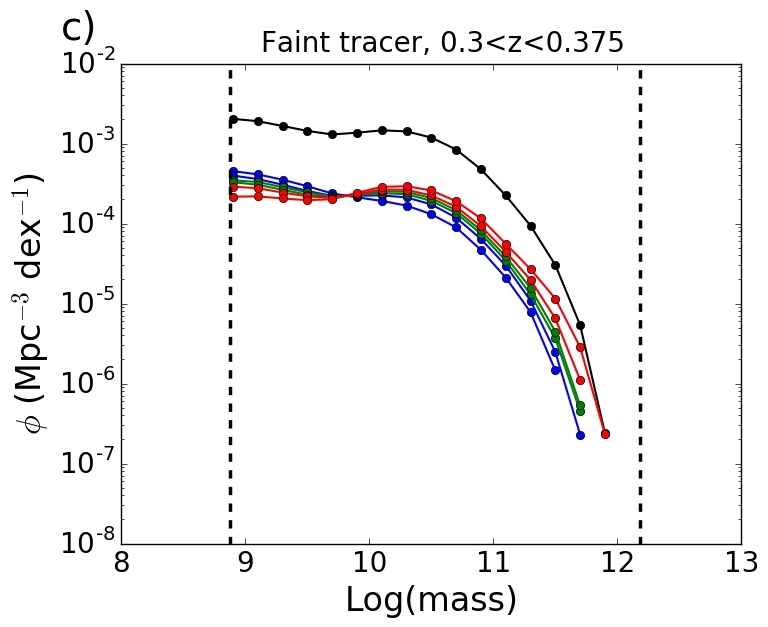}
\end{minipage}
\begin{minipage}[t]{0.32\linewidth}
\centering 
\includegraphics[width=.99\linewidth]{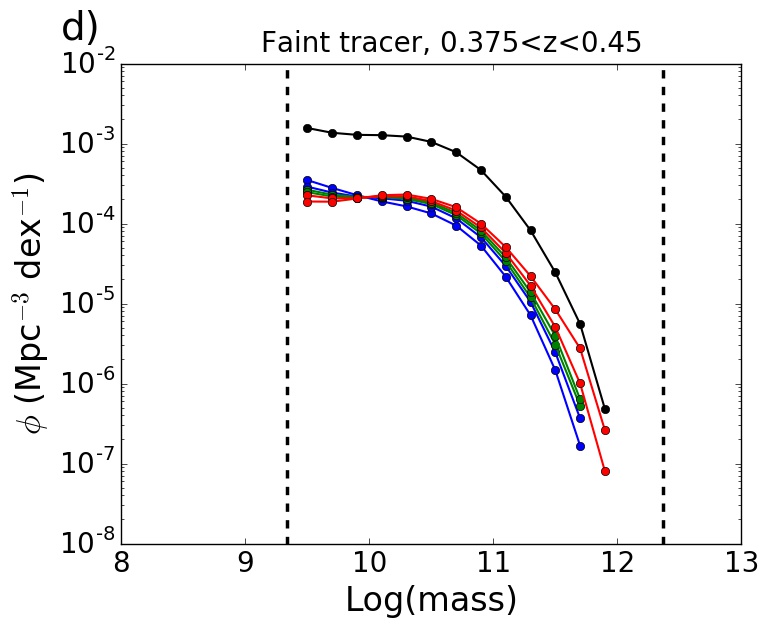}
\end{minipage}
\begin{minipage}[t]{0.32\linewidth}
\centering 
\includegraphics[width=.99\linewidth]{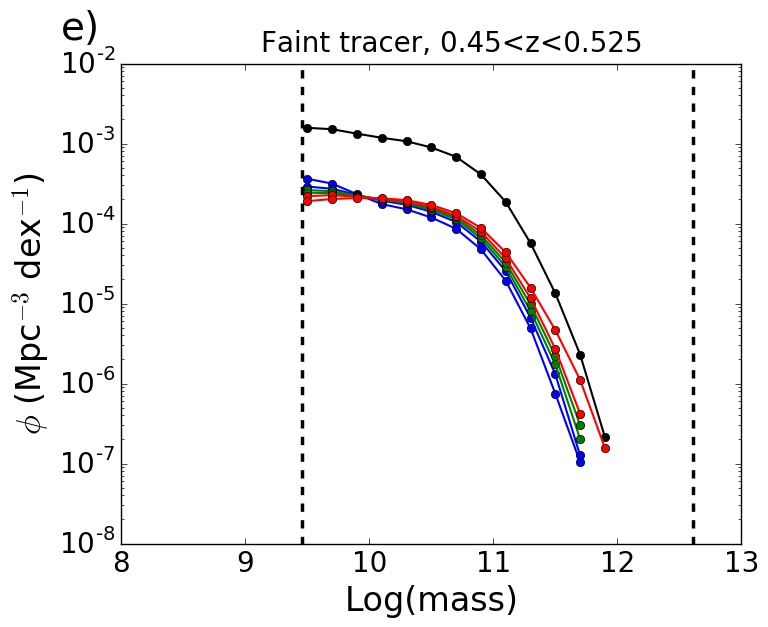}
\end{minipage}
\begin{minipage}[t]{0.32\linewidth}
\centering 
\includegraphics[width=.99\linewidth]{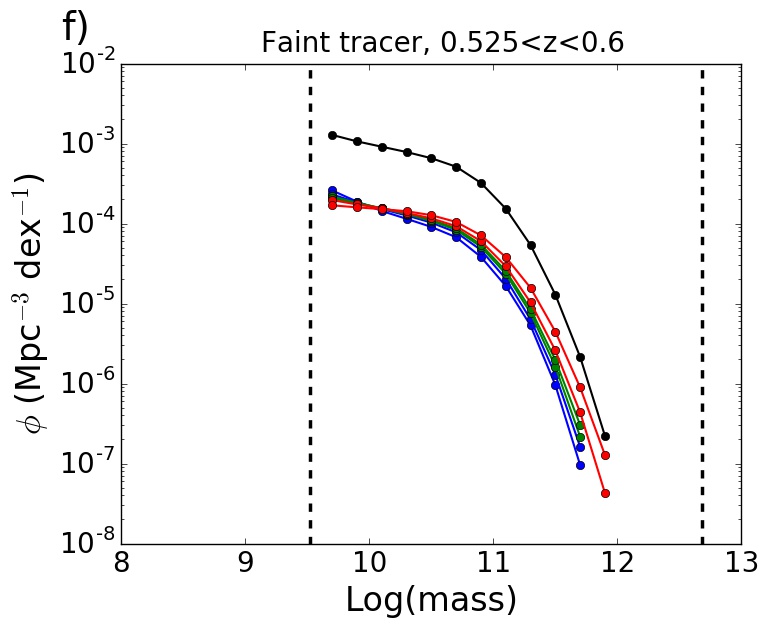}
\end{minipage}
\begin{minipage}[t]{0.32\linewidth}
\centering 
\includegraphics[width=.99\linewidth]{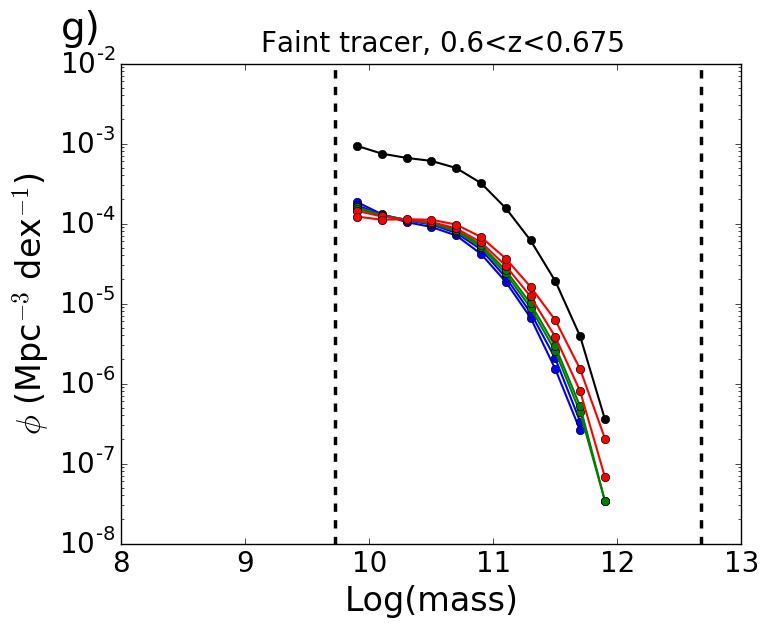}
\end{minipage}
\begin{minipage}[t]{0.32\linewidth}
\centering 
\includegraphics[width=.99\linewidth]{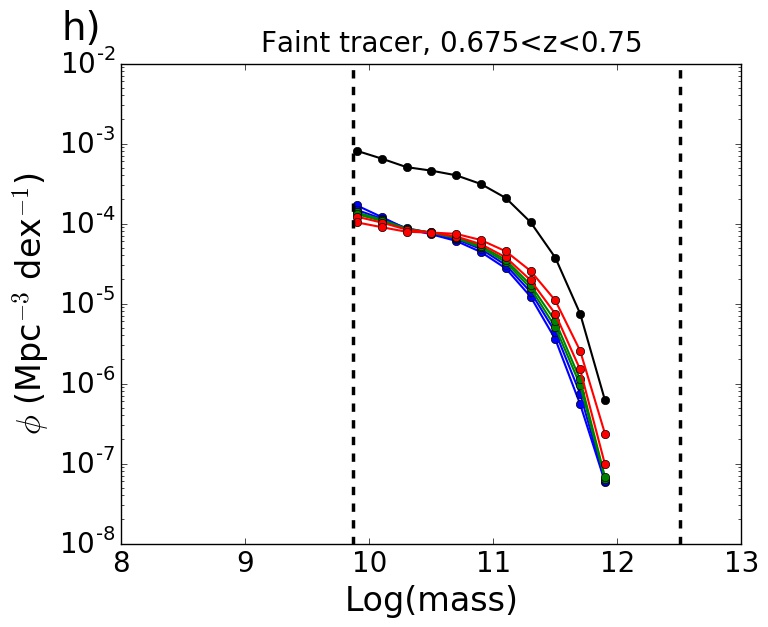}
\end{minipage}
\begin{minipage}[t]{0.32\linewidth}
\centering 
\includegraphics[width=.99\linewidth]{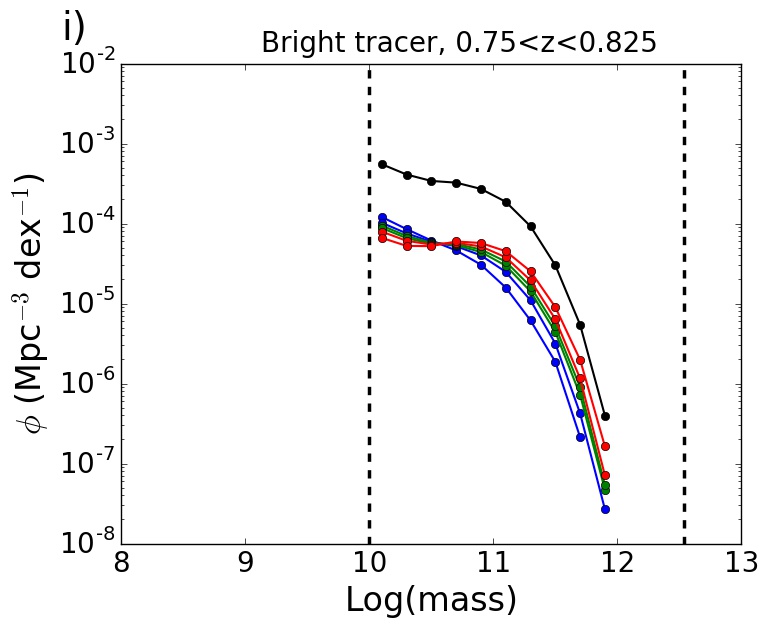}
\end{minipage}
\begin{minipage}[t]{0.32\linewidth}
\centering 
\includegraphics[width=.99\linewidth]{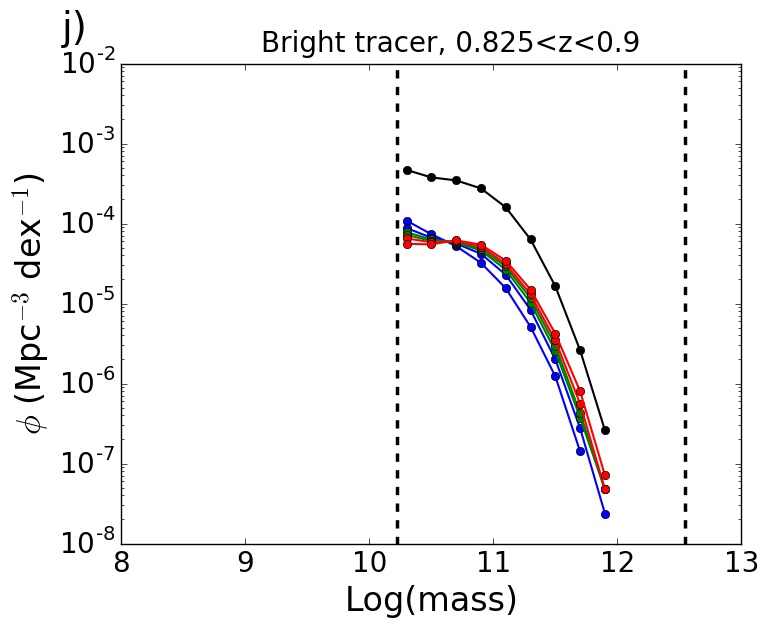}
\end{minipage}
\begin{minipage}[t]{0.32\linewidth}
\centering 
\includegraphics[width=.99\linewidth]{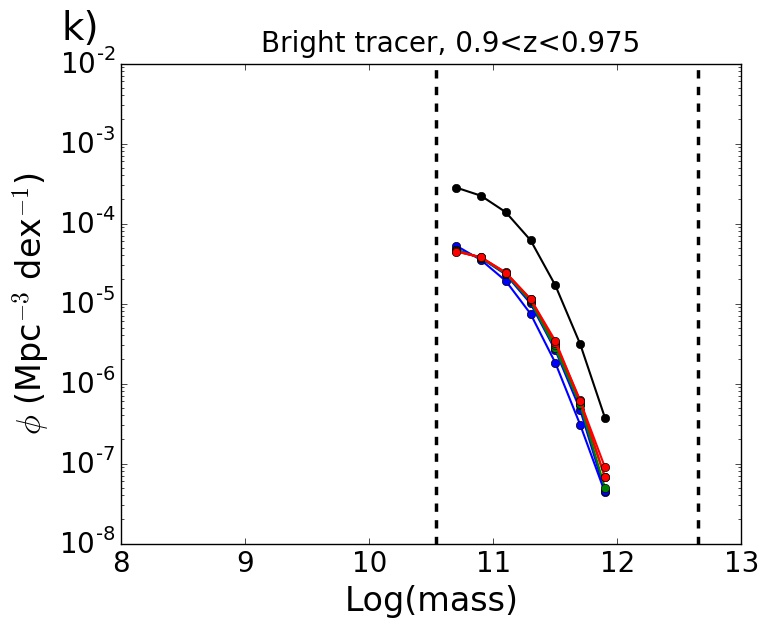}
\end{minipage}
\begin{minipage}[t]{0.32\linewidth}
\centering 
\includegraphics[width=.99\linewidth]{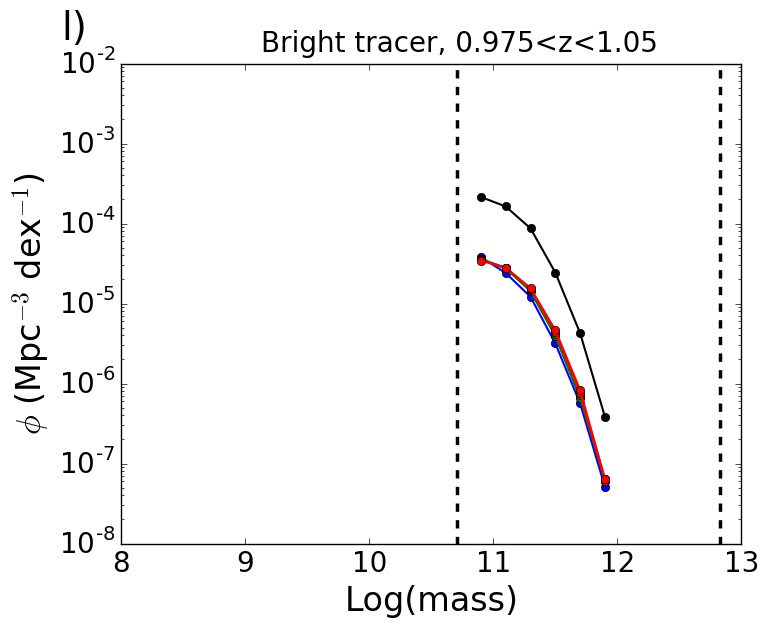}
\end{minipage}
\caption{Galaxy stellar mass functions for the lowest (blue), intermediate (green), highest (red) and all (black) environment bins for $12$ redshift bins. The environments for plots a-h are based on the faint tracer and the environments for plots i-l are based on the bright tracer. The vertical dashed lines indicate the mass completeness limits of each redshift bin. For clarity error bars are not shown in this figure.}  
\label{fig:mass_function_by_env_bin_by_bin}
\end{figure*}

\begin{figure*}
\begin{minipage}[t]{0.49\linewidth}
\centering 
\includegraphics[width=.80\linewidth]{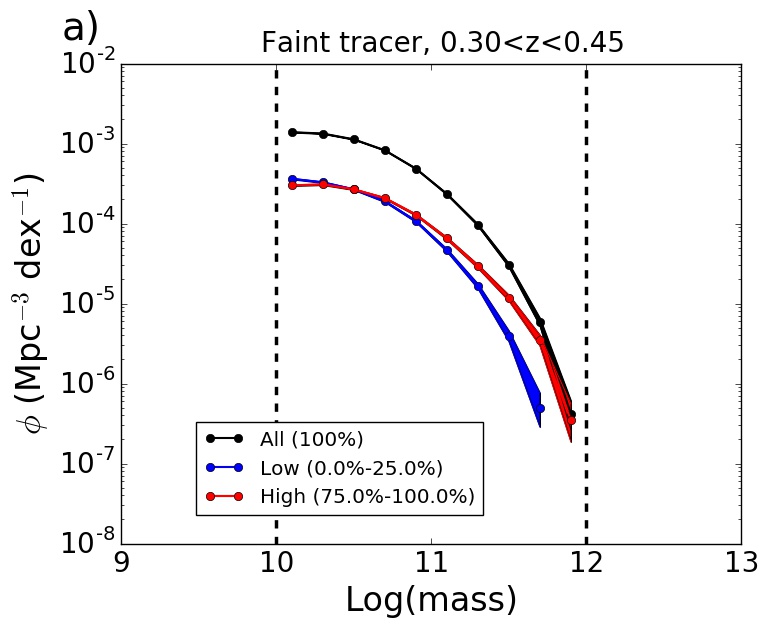}
\end{minipage}
\begin{minipage}[t]{0.49\linewidth}
\centering 
\includegraphics[width=.80\linewidth]{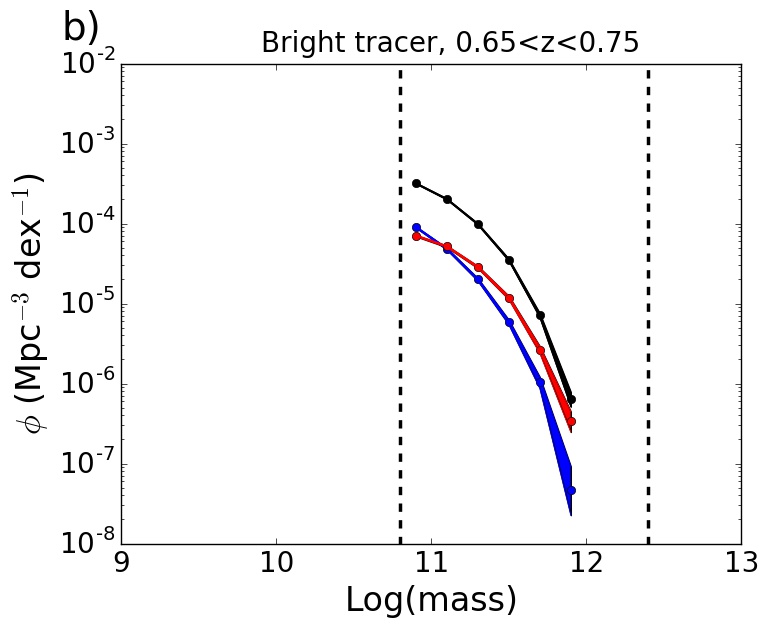}
\end{minipage}
\begin{minipage}[t]{0.49\linewidth}
\centering 
\includegraphics[width=.80\linewidth]{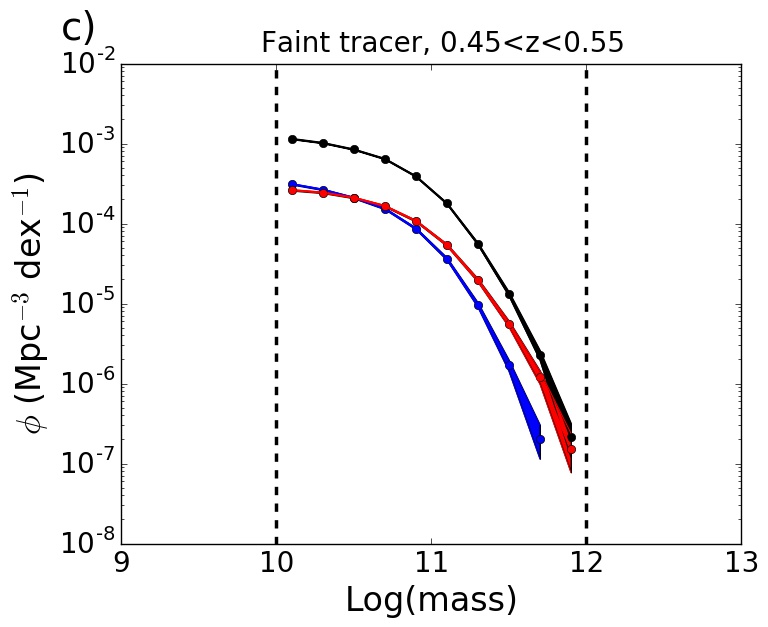}
\end{minipage}
\begin{minipage}[t]{0.49\linewidth}
\centering 
\includegraphics[width=.80\linewidth]{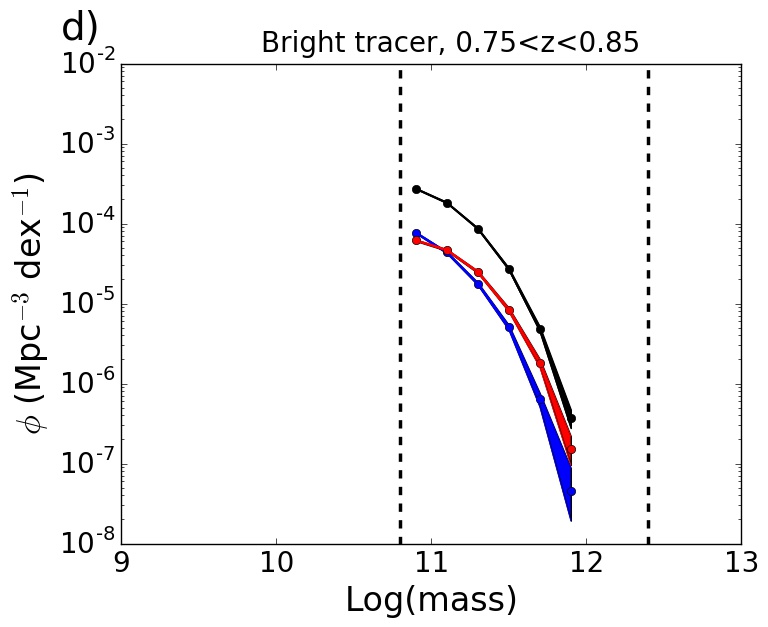}
\end{minipage}
\begin{minipage}[t]{0.49\linewidth}
\centering 
\includegraphics[width=.80\linewidth]{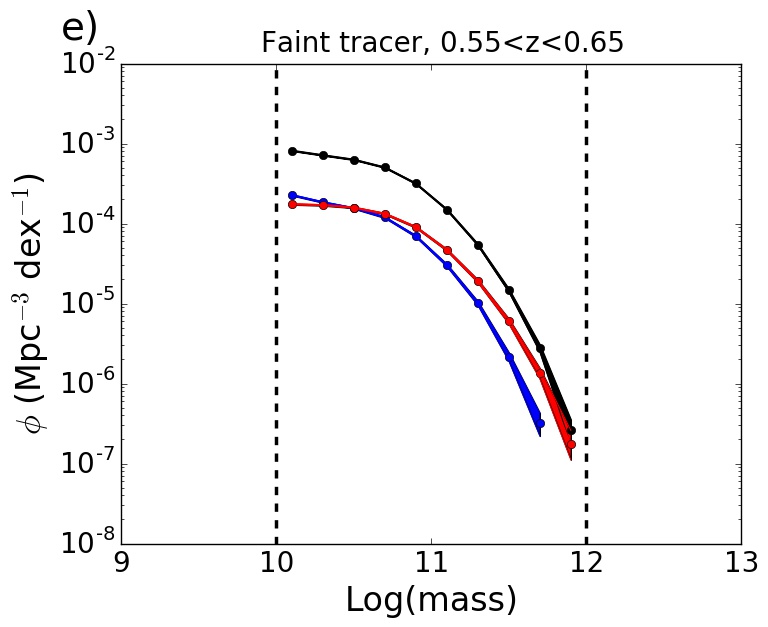}
\end{minipage}
\begin{minipage}[t]{0.49\linewidth}
\centering 
\includegraphics[width=.80\linewidth]{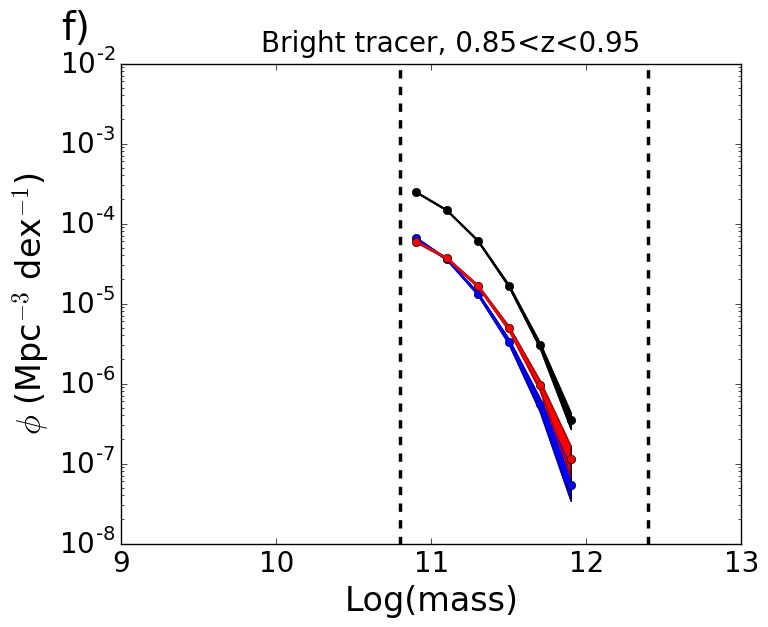}
\end{minipage}
\begin{minipage}[t]{0.49\linewidth}
\centering 
\includegraphics[width=.80\linewidth]{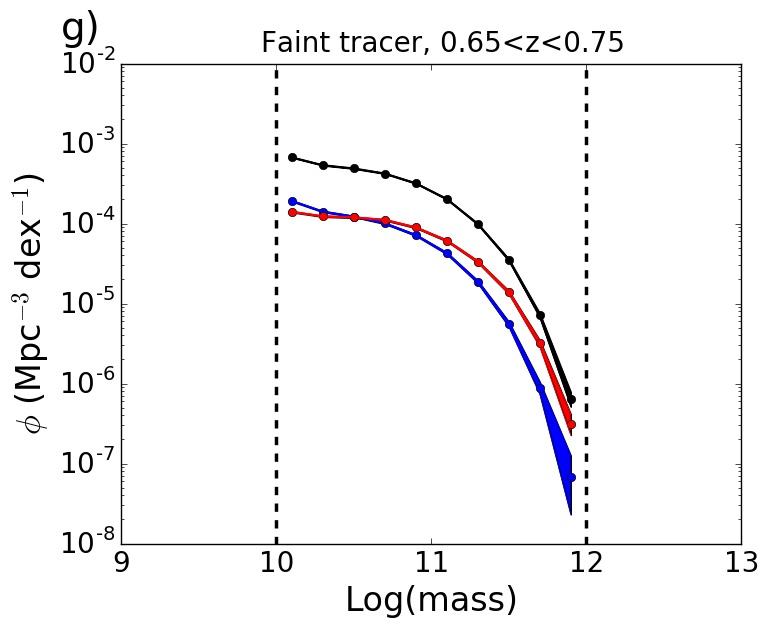}
\end{minipage}
\begin{minipage}[t]{0.49\linewidth}
\centering 
\includegraphics[width=.80\linewidth]{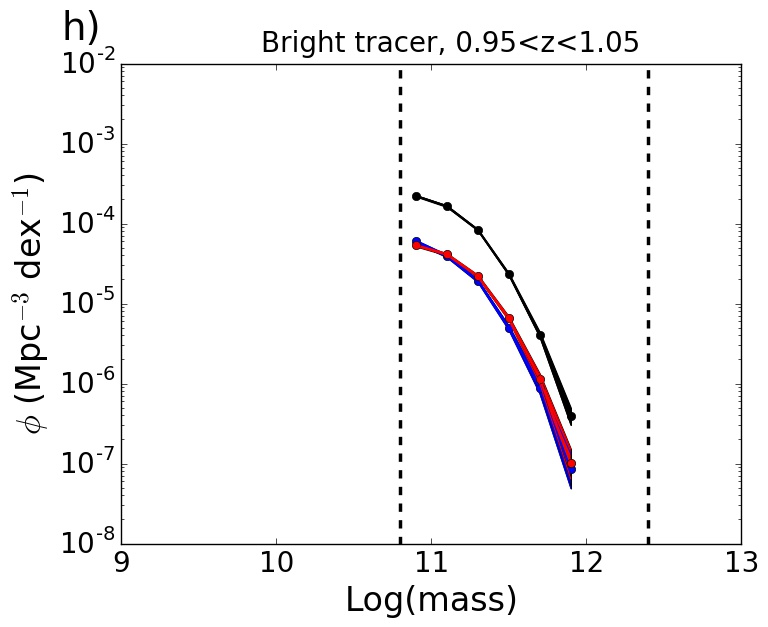}
\end{minipage}
\caption{Galaxy stellar mass functions for the lowest (blue), highest (red) and all (black) environment bins for $7$ redshift bins. The environments for the plots on the left are based on the faint tracer and the environments for the plots on the right are based on the bright tracer. The vertical dashed lines mark the bounds of the common mass range. The shaded regions show the 1-sigma errors on the low and high density environment mass functions.}  
\label{fig:mass_function_for_evolution}
\end{figure*}

\begin{figure*}
   \centering 
   \includegraphics[width=.49\linewidth]{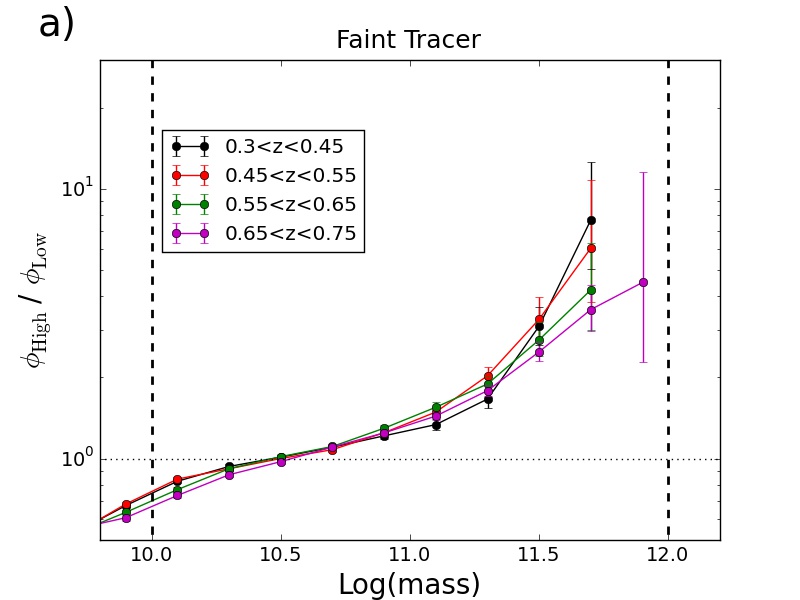}
   \includegraphics[width=.49\linewidth]{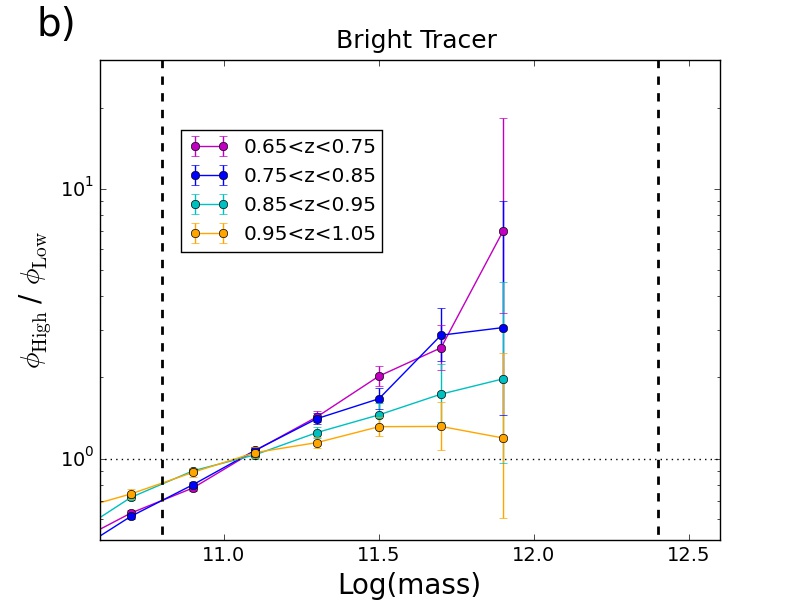}
\caption{Ratio of the effective number of galaxies in high and low environment bins as a function of mass for the faint tracer (left) and the bright tracer (right) for different redshift bins. The error bars show the 1-sigma errors on the ratios. The vertical dashed lines mark common mass ranges within the completeness limits for the range of redshifts for the faint and bright tracers. The horizontal dotted line marks a ratio of unity.}  
\label{fig:neffective_ratio}
\end{figure*}

\begin{figure}
\begin{minipage}[t]{0.99\linewidth}
\centering 
\includegraphics[width=.99\linewidth]{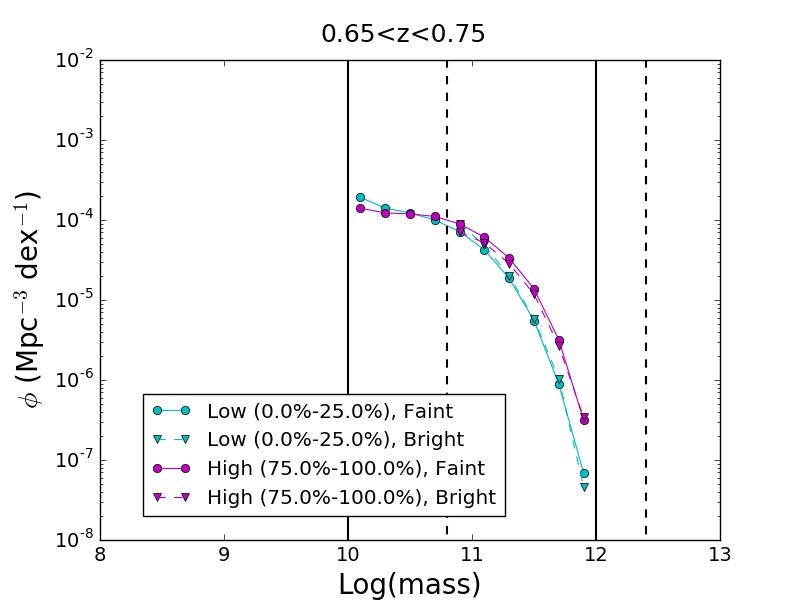}
\end{minipage}
\caption{Comparison of the low (cyan) and high (magenta) density mass functions components for the faint (solid) and bright (dashed) tracers for the redshift range $0.65<z<0.75$ where the tracers overlap. The black vertical lines show the mass limits for the faint (solid) and bright (dashed) tracers.}
\label{fig:mf_tracer_comparison}
\end{figure}

\subsubsection{Evaluation of errors} \label{subsubsec:mf_error_evaluation}

In this analysis we consider two main sources of errors: statistical errors and the propagated errors due to the imprecise photometric redshift measurements. 

The statistical errors depend upon the number of galaxies in the sample in each bin of mass, redshift and environment. Analytical expressions for the statistical errors (Poisson statistics) are available but these usually assume the errors follow a Gaussian distribution. This is untrue particularly at the high mass end. A further difficulty of adopting an analytical form for the statistical errors is incorporating the photometric redshift errors.

Therefore we employ a bootstrap resampling scheme to evaluate the combined errors (statistical and redshift). This ensures that the redshift and environment PDFs of each galaxy are incorporated into the analysis. The $100$ catalogue realizations form the basis of this scheme. We drew galaxies at random (with replacement) from the 100 catalogues to create $10,000$ new catalogues. We divided the redshift range for each tracer into a number of bins. Each galaxy was weighted appropriately for the volume and surface brightness corrections and the environment distributions for each redshift bin in each resampled catalogue using only those galaxies within a particular mass range. We then split the environment distributions into a number (4 or 6) of equi-percentage environment bins (for each redshift bin). Each environment bin therefore contained the same effective number of galaxies. To do this for each resampled catalogue we determine the environments at the $25.0\textrm{th}$, $50.0\textrm{th}$ and $75.0\textrm{th}$ percentiles for 4 environment bins or the $16.7\textrm{th}$, $33.3\textrm{rd}$, $50.0\textrm{th}$, $66.7\textrm{th}$ and $83.3\textrm{th}$ percentiles for 6 environment bins. We computed the mean and standard deviation for each of these percentiles from the $10,000$ catalogues to obtain robust dividing values between the environment bins. Fig \ref{fig:env_bins} shows the environment distribution for the redshift range $0.15<z<0.225$ divided into $6$ environment bins. The dividing values between the bins are shown with the vertical dashed lines. Tables \ref{table:env_boundaries} and \ref{table:env_boundaries_for_evolution} lists the limits of redshift, mass and environment bins for the analyses presented in Section \ref{subsec:mf_results}. We note here that the percentile environment binning used in this study is not an evolving (in redshift) density cut as the large scale density contrast of the Universe evolves with time. Binning in this way allows us to study the relative shapes of the mass function components at different redshifts but not their absolute normalizations.

To calculate the number density error distributions we identified the galaxies in each redshift, mass and environment bin in each of the resampled catalogues. We calculate the effective number of galaxies in each bin using the volume and surface brightness corrections and divide this by the survey volume for the associated redshift range. The variability in each bin between the $10,000$ resampled catalogues are the number density error distributions. Fig \ref{fig:mf_errors_examples} shows three examples of the number density error distributions for the total mass function for the redshift range $0.15<z<0.225$. The left hand plot is for a low mass bin, the middle plot is for a intermediate mass bin and the right plot is for a high mass bin. The error as a fraction of the number density increases with mass as expected as the most massive galaxies in the Universe are the most rare. Nevertheless the large number of galaxies within this sample lead to exquisite statistical errors. The error distributions are essentially Gaussian for the low and intermediate mass bins. The error distribution for the high mass bin is not symmetrical and is skewed to larger values.

\subsection{Results} \label{subsec:mf_results}

We present two analyses of the environmental components of the galaxy stellar mass function. In the first analysis we look at each redshift bin in turn and use the largest possible complete mass range for each redshift bin. We examine 12 redshift bins the first starting at $z=0.15$ and the last ending at $z=1.05$. We employ the faint tracer for the 8 lowest redshift bins and the bright tracer for the 4 highest redshift bins. In this analysis all of the galaxies within the complete mass range are used to determine 6 equi-percentage environment bins for each redshift bin. The redshift bins and the mass limits for this analysis are shown on the left hand side of Fig \ref{fig:mass_completeness} and listed together with the environment boundaries in Table \ref{table:env_boundaries}. This analysis enables a detailed examination of the mass function components for each redshift bin, but because different mass ranges are employed this analysis cannot be used to study the redshift evolution of the components of the mass function. 

The second analysis employs common mass ranges for the redshift ranges traced by the faint and bright density defining populations. In this analysis we split the environments into 4 equi-percentage environment bins and investigate the redshift evolution of the lowest and highest environment components. The redshift bins and the common mass ranges for this analysis are shown on the right hand side of Fig \ref{fig:mass_completeness} and listed together with the environment boundaries in Table \ref{table:env_boundaries_for_evolution}.

\subsubsection{Local Universe} \label{subsubsec:local_universe}

Fig \ref{fig:mass_function_by_env_bin_by_bin} shows the environmental components of the galaxy stellar mass function for 12 redshift bins covering the range $0.15<z<1.05$. Each plot shows a different redshift bin. The first 8 bins employ the faint tracer and the last 4 bins employ the bright tracer. These tracing populations were used to compute $6$ equi-percentage environment bins. The redshift, mass and environment bins used are listed in Table \ref{table:env_boundaries} and illustrated in the left hand plot in Fig \ref{fig:mass_completeness}. The total mass function is shown in black. The lowest environment components are shown in blue, the intermediate environment components are shown in green and the densest environment components in red. The vertical dashed black lines mark the mass completeness limits. These limits change with redshift. The range of complete masses generally decreases with redshift. This is mainly due to the increase in the low mass limit with redshift. At higher redshifts galaxies must be brighter (more massive) to be detected. The upper mass limit also increases with redshift, particularly for the first few redshift bins. This is due to the increase in detectable volume as the redshift increases; enabling rarer species of galaxies to be found (e.g. the most massive galaxies).

In this Section we focus on the four lowest redshift bins (i.e. $0.15<z<0.45$) which are shown in plots a) to d). Comparisons with other studies are tricky because of different definitions of environment, measurement methods (photometric vs spectroscopic) and mass completeness limits. Nevertheless it is interesting to compare the environment components of the mass function with those obtained by \cite{Bolzonella2010} from zCOSMOS data shown in their Fig 3. The zCOSMOS study is based on spectroscopic measurements for an area of {\raise.17ex\hbox{$\scriptstyle\sim$}}1.5 deg$^{2}$ and the $5^{\textrm{th}}$ nearest neighbour method is employed to quantify environment. Conversely in this study we have photometric measurements only, but for a much larger area {\raise.17ex\hbox{$\scriptstyle\sim$}}78 deg$^{2}$ and we employ a fixed aperture method to quantify environment. The mass function components in \cite{Bolzonella2010} are anchored at high mass, whereas the environment components in this study separate at high mass. The anchoring seen in \cite{Bolzonella2010} is by construction because the environments of only the massive galaxies ($\textrm{Log(M)}>10.51$) are used to determine the boundaries of the environment bins. As noted by the authors the effective (i.e. weighted by the volume corrections etc.) number of galaxies in each of their environment bins for the complete mass range is therefore not equal. In this study we used Monte-Carlo simulations to derive statistically robust equi-percentage environment bins for the mass range between the completeness limits. This difference accounts for the larger (smaller) separation we see at high (low) mass compared with \cite{Bolzonella2010}. Strikingly the shapes of the components of the mass functions are very similar. The difference is the relative normalizations of the low and high environment curves. \cite{Bolzonella2010} finds an upturn in the highest density component at low mass (Log(M)=9.5). There is evidence of this upturn in our data too (particularly in plot b)), but the upturn appears slightly less pronounced. Importantly, consistent with \citep{Bolzonella2010} and also SDSS \citep{Baldry2006} and GAMA \citep{McNaught-Roberts2014} studies, for the low redshift regime, we find that the fraction of massive galaxies is larger in high density environments than low density environments and the converse for the fraction of less massive galaxies (e.g. Log(M)=9.0). The normalized mass function components for the intermediate environments uniformly (and in order) populate the range in between the lowest and highest components. 

Despite the cruder redshift and environment measurements for individual galaxies in this study, because of the large sample we are able to distinguish between the lowest and highest environment components of the galaxy stellar mass function in the local Universe.

\subsubsection{Environmental components for complete mass ranges} \label{subsubsec:components_for_complete_range}

We now move on to examine the environment components of the mass function at higher redshifts by continuing the discussion of Fig \ref{fig:mass_function_by_env_bin_by_bin}. The figure shows that the environmental trends found at low redshift are maintained to high redshifts and the shapes of the environment components are distinguishable up to $z${\raise.17ex\hbox{$\scriptstyle\sim$}}0.8. However the narrowing of the complete mass range with redshift tends to increasingly anchor the mass function components, in a similar way to that discussed in the previous Section. Since the environment components are constructed to contain the same effective number of galaxies this has the effect of driving the mass function components together. The final two redshift bins hint that the shapes of the environments components of the mass function increasingly converge with redshift. However, from these plots it is difficult to disentangle whether this is a real effect or due to the narrowing of the complete mass range. In the next Section we investigate this further by adopting common mass ranges over redshift bins. 

\subsubsection{Redshift evolution for common mass ranges} \label{subsubsec:evolution_for_common_range}

In this Section we repeat the analysis but use two common mass ranges, one for the faint tracer: $10.0<\textrm{Log(M)}<12.0$ and one for the bright tracer: $10.8<\textrm{Log(M)}<12.4$ for the redshift ranges $0.3<z<0.75$ and $0.65<z<1.05$ respectively. These common mass ranges are within the completeness limits for the corresponding redshift ranges. We split each of the redshift ranges into 4 redshift bins as shown in the right hand plot of Fig \ref{fig:mass_completeness}. Using these mass and redshift bins we compute the environment boundaries to give 4 equi-percentage environment bins. The details of the redshift, mass and environment bins are listed in Table \ref{table:env_boundaries_for_evolution}. 

Fig \ref{fig:mass_function_for_evolution} shows the lowest (blue) and highest (red) density environment components of the galaxy stellar mass function for the faint tracer (left) and the bright tracer (right). The lowest bin consists of the bottom 25 percent of the environment distribution whereas the highest bin consists of the top 25 percent of the environment distribution. The red and blue shaded regions show the 1-sigma errors on the effective number density of galaxies in each bin. The vertical dashed lines mark the limits of the common mass ranges. The low and high density environment components behave consistently with the trends shown in Figure \ref{fig:mass_function_by_env_bin_by_bin}. It is clear, especially for the bright tracer, that the shape of the environmental components converge with increasing redshift. For $\textrm{Log(M)}>11.0$ and within the limits of the 1-sigma errors at a redshift of $z${\raise.17ex\hbox{$\scriptstyle\sim$}}1.0 the environmental components are indistinguishable. This result is consistent with recent work from \cite{Davidzon2016} which examines the redshift range: $0.51<z<0.9$ using data from VIPERS \citep{Garilli2014} which contains $57,204$ spectra and covers {\raise.17ex\hbox{$\scriptstyle\sim$}}10 deg$^{2}$ of the sky. 

To illustrate this further we examine the ratio between the effective number density of galaxies in the high and low environment components. Fig \ref{fig:neffective_ratio} shows this ratio as a function of mass for different redshift bins for the faint tracer (left) and the bright tracer (right). The vertical dashed lines mark the limits of the common mass ranges. The error bars show the 1-sigma errors on the ratios. For low masses the ratio at all redshifts and both tracers is $<1.0$. In this regime per unit volume of space there is a larger number of galaxies in low density environments than in high density environments. As the mass increases the ratio becomes $>1.0$ and here the opposite is true. Per unit volume of space there is a larger number of galaxies in high density environments than low density environments. For $\textrm{Log(M)}<11.2$ the ratio varies little with redshift. This changes for $\textrm{Log(M)}>11.2$. Although the errors are large, in this mass range the ratio between the effective number density of galaxies in the high and low environment components decreases with redshift for both tracers, falling to nearly unity for the highest (mass and) redshift bin.

In an effort to connect the results based on the faint tracer to those on the bright tracer we now compare the mass functions for the two tracers in the overlapping redshift range: $0.65<z<0.75$. Fig \ref{fig:mf_tracer_comparison} shows the low (cyan) and high (magenta) density mass function components for the faint (solid) and bright (dashed) tracers. The vertical black lines show the common mass ranges associated with the faint (solid) and bright (dashed) tracers. Despite using different sized apertures to quantify galaxy environment and employing different common mass ranges the low and high density environmental components of the mass function for $\textrm{Log(M)}>11.2$ are strikingly similar for the faint and bright tracers. The difference in number density between the faint and bright tracers is smaller than the 1-sigma errors for the massive galaxies. 

\subsubsection{Evolution of the environmental ratio of effective number of galaxies per unit volume} \label{subsubsec:evolution_of_the_environmental_ratio}

\begin{figure}
   \centering 
   \includegraphics[width=.99\linewidth]{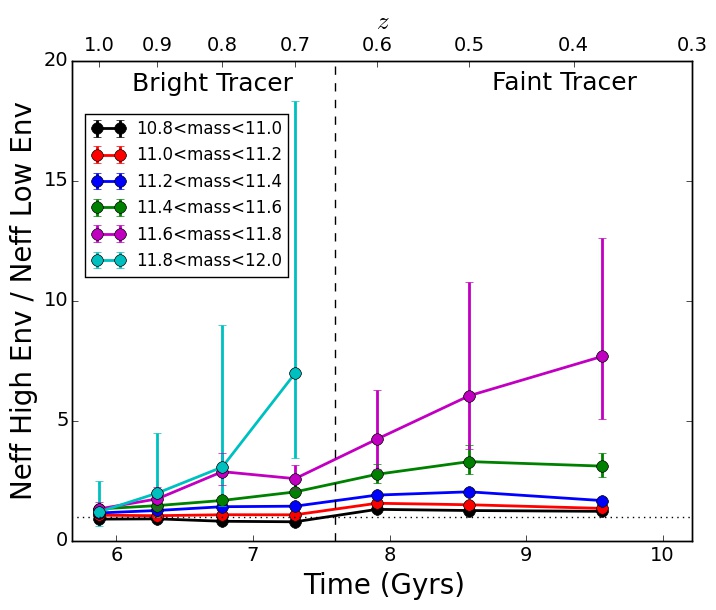}
\caption{Ratio of the effective number of galaxies in high and low environment as a function of cosmic time for different mass bins. The bright tracer was used for the points on the left and the faint tracer was used for the points on the right of the vertical dashed line. The upper horizontal axis shows the redshift. The error bars show the 1-sigma errors on the ratios. The vertical dashed line marks the redshift that separates the measurements made with the faint and bright tracers.}  
\label{fig:neffective_ratio_evolution}
\end{figure}

We are now in the position to investigate the evolution of the ratio of the effective number of galaxies per unit volume in the high and low density environment components. Fig \ref{fig:neffective_ratio_evolution} shows the ratio of the effective number of galaxies per unit volume in the high and low density environment components as a function of cosmic time for a range of different masses. The redshift is shown on the upper horizontal axis. Exploiting the good agreement between the mass function components for the faint and bright tracers (for $\textrm{Log(M)}>11.2$) shown in Fig \ref{fig:mf_tracer_comparison} in this figure we connect the results from the two tracers. The results on the left of the vertical dashed line ($z=0.65$) are based on the bright tracer and the those on the right are based on the faint tracer. The ratio of the number of galaxies per unit volume in high density environments to the number in low density environments does not evolve with cosmic time for galaxies with $\textrm{Log(M)}<11.2$. Conversely this ratio evolves considerably for more massive galaxies and increases with cosmic time. For example for galaxies with masses in range: $11.6<\textrm{Log(M)}<11.8$ (purple) the ratio increases from {\raise.17ex\hbox{$\scriptstyle\sim$}}1 to {\raise.17ex\hbox{$\scriptstyle\sim$}}8 between z=1.0 (6 Gyrs) and z=0.375 (9.5 Gyrs). At $z${\raise.17ex\hbox{$\scriptstyle\sim$}}1 the lines for the different mass bins converge to a ratio of {\raise.17ex\hbox{$\scriptstyle\sim$}}1.0. At this redshift the number density of galaxies in the low and high environment components becomes equal. Stated another way at $z${\raise.17ex\hbox{$\scriptstyle\sim$}}1.0 the probability of finding a massive galaxy in the highest density quartile is the same as finding it in the lowest density quartile, whilst at low redshift massive galaxies preferentially reside in the high density quartile. Assuming that most of the massive galaxies ($\textrm{Log(M)}>11.8$) have formed at $z>1.0$ \citep[i.e. downsizing - ][]{Thomas2010,Pozzetti2010} this figure suggests that as cosmic time proceeds high density structures form around the massive galaxies, such that at $z=0.375$ the fraction of massive galaxies is {\raise.17ex\hbox{$\scriptstyle\sim$}}$8$ times larger in high density environments than in low density environments. The convergence point at $z${\raise.17ex\hbox{$\scriptstyle\sim$}}1.0 is important because it marks the transition between an earlier epoch where the mass distribution of galaxies is independent of galaxy environment \citep{Mortlock2015} and the later epoch where the mass distribution of galaxies does depend on environment. 

\section{Discussion} \label{sec:discussion}

We have tried to diligently deal with the errors on the galaxy mass, redshift and especially environment by using bins in these quantities which have a similar scale or are larger than the errors. Nevertheless it is important to consider the possibility that the results for the highest redshift bin (i.e. Fig \ref{fig:mass_function_for_evolution}h and the convergence point in Fig \ref{fig:neffective_ratio_evolution} at $z${\raise.17ex\hbox{$\scriptstyle\sim$}}1.0) could be partially driven by contamination due to the scattering of galaxies from adjacent bins. The highest redshift bin is potentially the most vulnerable to this effect because the photometric redshift error increases with redshift. For example Fig \ref{fig:propagation_errors_properties}g shows that $\Delta z${\raise.17ex\hbox{$\scriptstyle\sim$}}0.08 at $z${\raise.17ex\hbox{$\scriptstyle\sim$}}1.0. There is a slight upturn in the error on the mass estimates with $\Delta$log(mass){\raise.17ex\hbox{$\scriptstyle\sim$}}0.2 at $z${\raise.17ex\hbox{$\scriptstyle\sim$}}1.0 as shown in Fig \ref{fig:propagation_errors_properties}i and the environment error also increases with redshift for the bright tracer with $\Delta$log(1+ $\delta$){\raise.17ex\hbox{$\scriptstyle\sim$}}0.12 at $z${\raise.17ex\hbox{$\scriptstyle\sim$}}1.0 as shown in Fig \ref{fig:envs_character}c.

We think our result is robust at $z${\raise.17ex\hbox{$\scriptstyle\sim$}}1.0 because we have employed an aperture with a half depth, $\delta z=0.2$, mass bins of size $0.2$ dex and split the environments, low from high, using the first and fourth quartiles. These choices ensure we capture the vast majority of galaxies at $z${\raise.17ex\hbox{$\scriptstyle\sim$}}1.0 and even those with the largest photometric errors.

Since we have not considered the errors in the mass estimates attributed to different SED modelling efforts in this analysis it is conceivable that there is some scattering between adjacent mass bins. This would lead to a shallowing (i.e. the Eddington bias) of both the low and high environmental components of the mass function. We would not expect there to be a differential effect between the components adversely effecting our result.

Table \ref{table:env_boundaries_for_evolution} shows that the difference between the $25^\textrm{th}$ and $75^\textrm{th}$ environmental percentiles at $z${\raise.17ex\hbox{$\scriptstyle\sim$}}1.0 is {\raise.17ex\hbox{$\scriptstyle\sim$}}0.3. This is $2.5$ times larger than the median environmental error of $0.12$ at $z${\raise.17ex\hbox{$\scriptstyle\sim$}}1.0. This means that there should be minimal contamination between the low and high environmental bins. We therefore believe that our results are robust and that the low and high components of the mass function converge with increasing redshift. 

We find that the convergence point is $z${\raise.17ex\hbox{$\scriptstyle\sim$}}1.0. It is possible that contamination between environments, mass and redshift bins partially contributes to the convergence meaning that the true transition is at a slightly higher redshift. This may explain the difference between the converge redshift found in this paper compared to \cite{Mortlock2015} that find $z${\raise.17ex\hbox{$\scriptstyle\sim$}}1.5.

\section{Conclusions} \label{sec:conclusions}

The main objective of this paper is to analyse the environmental components of the galaxy stellar mass function using the DES science verification dataset. The DES is wide area photometric survey imaging galaxies to a redshift of about $1.4$.

Specifically we studied the SPT-E field which is the largest contiguous field in the dataset and has an area of approximately 130 squared degrees. The SPT-E field contains approximately 3.2 million galaxies. 

We adopted a Monte-Carlo approach and used the photometric redshift PDFs to propagate the errors into the derived galaxy properties: the stellar masses and i-band absolute magnitudes. We found that the ratio between the range of the property distributions and the median errors was approximately $10-12$ and this was sufficient for further analyses.

We constructed two density defining populations: one for the redshift range $0.15<z<0.75$ which we call the faint tracer, and one for $0.6<z<1.05$ which we call the bright tracer. We used the tracing populations and a fixed aperture method to compute galaxy environments for a range of aperture parameters. We used Monte-Carlo realizations to quantified the errors on the environment measurements. This enabled the selection of a set of aperture parameters for the faint and bright tracers that resulted in similar environmental properties. The ratio between the range of environments and the median error on the environments was $3.8$, only 3 times smaller than the ratios for the absolute magnitudes and stellar masses. We showed that environment PDFs could be constructed from the Monte-Carlo realizations. We found that the error on the environment measurements increased as a function of environment but was relatively constant as a function of redshift. 

We calculated volume and surface brightness corrections for each galaxy and used them to construct weighted environment distributions. We used Monte-Carlo realizations to derive statistically robust equi-percentage environment bins for a set of redshift bins. We carefully controlled the mass ranges (for each redshift bin) and used the environmental components to conduct two analyses of the galaxy stellar mass function. In the first we studied the environmental components using the largest possible complete mass range for each redshift bin. In the second we employed common mass ranges across redshift bins to study the redshift evolution of the environmental components. We computed the environment components of the galaxy stellar mass function for the redshift range $0.15<z<1.05$. We found a clear separation between the shapes of the environmental components of the stellar mass function at low and intermediate redshift. For $z<0.75$ we found that the fraction of massive galaxies is larger in high density environments than low density environments and the converse for the fraction of less massive galaxies (${\textrm{Log(M)}<9.0}$). The low and high density components converge with increasing redshift up to $z${\raise.17ex\hbox{$\scriptstyle\sim$}}1.0 where their shapes are indistinguishable. This redshift is important because it marks the transition between an earlier epoch where the mass distribution of galaxies is independent of environment and a later epoch where the mass distribution does depend on galaxy environment. We studied the ratio between the high and low environment components of the stellar mass function as a function of cosmic time and showed the build up of high density structures around the most massive galaxies.

The science verification data is the first dataset from the DES. We have demonstrated with approximately $2$ percent of the total area of the full survey an analysis of the evolving population of galaxies and their environments. Future datasets from the DES will provide the opportunity to study different components of the galaxy stellar mass function including colour, star formation rate and morphology, unlocking more clues.  

\section*{Acknowledgements}

Funding for the DES Projects has been provided by the U.S. Department of Energy, the U.S. National Science Foundation, the Ministry of Science and Education of Spain, the Science and Technology Facilities Council of the United Kingdom, the Higher Education Funding Council for England, the National Center for Supercomputing Applications at the University of Illinois at Urbana-Champaign, the Kavli Institute of Cosmological Physics at the University of Chicago, the Center for Cosmology and Astro-Particle Physics at the Ohio State University, the Mitchell Institute for Fundamental Physics and Astronomy at Texas A\&M University, Financiadora de Estudos e Projetos, Funda{\c c}{\~a}o Carlos Chagas Filho de Amparo {\`a} Pesquisa do Estado do Rio de Janeiro, Conselho Nacional de Desenvolvimento Cient{\'i}fico e Tecnol{\'o}gico and the Minist{\'e}rio da Ci{\^e}ncia, Tecnologia e Inova{\c c}{\~a}o, the Deutsche Forschungsgemeinschaft and the Collaborating Institutions in the Dark Energy Survey. 

The Collaborating Institutions are Argonne National Laboratory, the University of California at Santa Cruz, the University of Cambridge, Centro de Investigaciones Energ{\'e}ticas, Medioambientales y Tecnol{\'o}gicas-Madrid, the University of Chicago, University College London, the DES-Brazil Consortium, the University of Edinburgh, the Eidgen{\"o}ssische Technische Hochschule (ETH) Z{\"u}rich, Fermi National Accelerator Laboratory, the University of Illinois at Urbana-Champaign, the Institut de Ci{\`e}ncies de l'Espai (IEEC/CSIC), the Institut de F{\'i}sica d'Altes Energies, Lawrence Berkeley National Laboratory, the Ludwig-Maximilians Universit{\"a}t M{\"u}nchen and the associated Excellence Cluster Universe, the University of Michigan, the National Optical Astronomy Observatory, the University of Nottingham, The Ohio State University, the University of Pennsylvania, the University of Portsmouth, SLAC National Accelerator Laboratory, Stanford University, the University of Sussex, and Texas A\&M University.

The DES data management system is supported by the National Science Foundation under Grant Number AST-1138766. The DES participants from Spanish institutions are partially supported by MINECO under grants AYA2012-39559, ESP2013-48274, FPA2013-47986, and Centro de Excelencia Severo Ochoa SEV-2012-0234. Research leading to these results has received funding from the European Research Council under the European Union’s Seventh Framework Programme (FP7/2007-2013) including ERC grant agreements 240672, 291329, and 306478.

We are grateful for the extraordinary contributions of our CTIO colleagues and the DECam Construction, Commissioning and Science Verification teams in achieving the excellent instrument and telescope conditions that have made this work possible.  The success of this project also relies critically on the expertise and dedication of the DES Data Management group.

\bibliography{paperbib}{}

\begin{thebibliography}{92}
\expandafter\ifx\csname natexlab\endcsname\relax\def\natexlab#1{#1}\fi

\bibitem[{{Ahn} {et~al}\mbox{.}(2012){Ahn}, {Alexandroff}, {Allende Prieto},
  {Anderson}, {Anderton}, {Andrews}, {Aubourg}, {Bailey}, {Balbinot}, {Barnes},
  \& et~al.}]{Ahn2012}
{Ahn} C.~P. {et~al.}, 2012, ApJS, 203, 21

\bibitem[{Baldry \& Balogh(2006)}]{Baldry2006}
Baldry I., Balogh M., 2006, MNRAS, 373, 19

\bibitem[{{Baldry} {et~al}\mbox{.}(2012){Baldry}, {Driver}, {Loveday},
  {Taylor}, {Kelvin}, {Liske}, {Norberg}, {Robotham}, {Brough}, {Hopkins},
  {Bamford}, {Peacock}, {Bland-Hawthorn}, {Conselice}, {Croom}, {Jones},
  {Parkinson}, {Popescu}, {Prescott}, {Sharp}, \& {Tuffs}}]{Baldry2012}
{Baldry} I.~K. {et~al.}, 2012, MNRAS, 421, 621

\bibitem[{{Baldry} {et~al}\mbox{.}(2008){Baldry}, {Glazebrook}, \&
  {Driver}}]{Baldry2008}
{Baldry} I.~K., {Glazebrook} K., {Driver} S.~P., 2008, MNRAS, 388, 945

\bibitem[{{Balogh} {et~al}\mbox{.}(2001){Balogh}, {Christlein}, {Zabludoff}, \&
  {Zaritsky}}]{Balogh2001}
{Balogh} M.~L., {Christlein} D., {Zabludoff} A.~I., {Zaritsky} D., 2001, ApJ,
  557, 117

\bibitem[{{Bell} {et~al}\mbox{.}(2003){Bell}, {McIntosh}, {Katz}, \&
  {Weinberg}}]{Bell2003}
{Bell} E.~F., {McIntosh} D.~H., {Katz} N., {Weinberg} M.~D., 2003, ApJS, 149,
  289

\bibitem[{{Blanton} \& {Berlind}(2007)}]{Blanton2007}
{Blanton} M.~R., {Berlind} A.~A., 2007, ApJ, 664, 791

\bibitem[{{Blanton} \& {Moustakas}(2009)}]{Blanton2009}
{Blanton} M.~R., {Moustakas} J., 2009, ARA\&A, 47, 159

\bibitem[{{Bolzonella} {et~al}\mbox{.}(2010){Bolzonella}, {Kova{\v c}},
  {Pozzetti}, {Zucca}, {Cucciati}, {Lilly}, {Peng}, {Iovino}, {Zamorani},
  {Vergani}, {Tasca}, {Lamareille}, {Oesch}, {Caputi}, {Kampczyk}, {Bardelli},
  {Maier}, {Abbas}, {Knobel}, {Scodeggio}, {Carollo}, {Contini}, {Kneib}, {Le
  F{\`e}vre}, {Mainieri}, {Renzini}, {Bongiorno}, {Coppa}, {de la Torre}, {de
  Ravel}, {Franzetti}, {Garilli}, {Le Borgne}, {Le Brun}, {Mignoli},
  {Pell{\'o}}, {Perez-Montero}, {Ricciardelli}, {Silverman}, {Tanaka},
  {Tresse}, {Bottini}, {Cappi}, {Cassata}, {Cimatti}, {Guzzo}, {Koekemoer},
  {Leauthaud}, {Maccagni}, {Marinoni}, {McCracken}, {Memeo}, {Meneux},
  {Porciani}, {Scaramella}, {Aussel}, {Capak}, {Halliday}, {Ilbert},
  {Kartaltepe}, {Salvato}, {Sanders}, {Scarlata}, {Scoville}, {Taniguchi}, \&
  {Thompson}}]{Bolzonella2010}
{Bolzonella} M. {et~al.}, 2010, A\&A, 524, A76

\bibitem[{Bolzonella {et~al}\mbox{.}(2000)Bolzonella, Miralles, \&
  Pell\'{o}}]{Bolzonella2000}
Bolzonella M., Miralles J., Pell\'{o} R., 2000, A\&A, 492, 476

\bibitem[{{Boselli} \& {Gavazzi}(2014)}]{Boselli2014}
{Boselli} A., {Gavazzi} G., 2014, A\&A Rev., 22, 74

\bibitem[{{Boylan-Kolchin} {et~al}\mbox{.}(2009){Boylan-Kolchin}, {Springel},
  {White}, {Jenkins}, \& {Lemson}}]{Boylan-Kolchin2009}
{Boylan-Kolchin} M., {Springel} V., {White} S.~D.~M., {Jenkins} A., {Lemson}
  G., 2009, MNRAS, 398, 1150

\bibitem[{{Bundy} {et~al}\mbox{.}(2006){Bundy}, {Ellis}, {Conselice}, {Taylor},
  {Cooper}, {Willmer}, {Weiner}, {Coil}, {Noeske}, \& {Eisenhardt}}]{Bundy2006}
{Bundy} K. {et~al.}, 2006, ApJ, 651, 120

\bibitem[{{Capozzi} {et~al}\mbox{.}(2016){Capozzi}, {Maraston}, {Daddi},
  {Renzini}, {Strazzullo}, \& {Gobat}}]{Capozzi2016}
{Capozzi} D., {Maraston} C., {Daddi} E., {Renzini} A., {Strazzullo} V., {Gobat}
  R., 2016, MNRAS, 456, 790

\bibitem[{{Carollo} {et~al}\mbox{.}(2013){Carollo}, {Cibinel}, {Lilly},
  {Miniati}, {Norberg}, {Silverman}, {van Gorkom}, {Cameron}, {Finoguenov},
  {Peng}, {Pipino}, \& {Rudick}}]{Carollo2013}
{Carollo} C.~M. {et~al.}, 2013, ApJ, 776, 71

\bibitem[{{Carrasco Kind} \& {Brunner}(2013)}]{Carrasco-Kind2013}
{Carrasco Kind} M., {Brunner} R.~J., 2013, MNRAS, 432, 1483

\bibitem[{{Cibinel} {et~al}\mbox{.}(2013){Cibinel}, {Carollo}, {Lilly},
  {Miniati}, {Silverman}, {van Gorkom}, {Cameron}, {Finoguenov}, {Norberg},
  {Peng}, {Pipino}, \& {Rudick}}]{Cibinel2013}
{Cibinel} A. {et~al.}, 2013, ApJ, 776, 72

\bibitem[{{Cole} {et~al}\mbox{.}(2001){Cole}, {Norberg}, {Baugh}, {Frenk},
  {Bland-Hawthorn}, {Bridges}, {Cannon}, {Colless}, {Collins}, {Couch},
  {Cross}, {Dalton}, {De Propris}, {Driver}, {Efstathiou}, {Ellis},
  {Glazebrook}, {Jackson}, {Lahav}, {Lewis}, {Lumsden}, {Maddox}, {Madgwick},
  {Peacock}, {Peterson}, {Sutherland}, \& {Taylor}}]{Cole2001}
{Cole} S. {et~al.}, 2001, MNRAS, 326, 255

\bibitem[{{Colless} {et~al}\mbox{.}(2001){Colless}, {Dalton}, {Maddox},
  {Sutherland}, {Norberg}, {Cole}, {Bland-Hawthorn}, {Bridges}, {Cannon},
  {Collins}, {Couch}, {Cross}, {Deeley}, {De Propris}, {Driver}, {Efstathiou},
  {Ellis}, {Frenk}, {Glazebrook}, {Jackson}, {Lahav}, {Lewis}, {Lumsden},
  {Madgwick}, {Peacock}, {Peterson}, {Price}, {Seaborne}, \&
  {Taylor}}]{Colless2001}
{Colless} M. {et~al.}, 2001, MNRAS, 328, 1039

\bibitem[{{Collister} \& {Lahav}(2004)}]{Collister2004}
{Collister} A.~A., {Lahav} O., 2004, PASP, 116, 345

\bibitem[{{Conroy} \& {Wechsler}(2009)}]{Conroy2009}
{Conroy} C., {Wechsler} R.~H., 2009, ApJ, 696, 620

\bibitem[{{Conselice} {et~al}\mbox{.}(2007){Conselice}, {Bundy}, {Trujillo},
  {Coil}, {Eisenhardt}, {Ellis}, {Georgakakis}, {Huang}, {Lotz}, {Nandra},
  {Newman}, {Papovich}, {Weiner}, \& {Willmer}}]{Conselice2007}
{Conselice} C.~J. {et~al.}, 2007, MNRAS, 381, 962

\bibitem[{{Cooper} {et~al}\mbox{.}(2006){Cooper}, {Newman}, {Croton}, {Weiner},
  {Willmer}, {Gerke}, {Madgwick}, {Faber}, {Davis}, {Coil}, {Finkbeiner},
  {Guhathakurta}, \& {Koo}}]{Cooper2006}
{Cooper} M.~C. {et~al.}, 2006, MNRAS, 370, 198

\bibitem[{Cooper {et~al}\mbox{.}(2005)Cooper, Newman, Madgwick, Gerke, Yan,
  Davis, \& Al}]{Cooper2005}
Cooper M.~C., Newman J.~A., Madgwick D.~S., Gerke B.~F., Yan R., Davis M., Al
  C. E.~T., 2005, AJ, 634, 833

\bibitem[{{Cooper} {et~al}\mbox{.}(2012){Cooper}, {Yan}, {Dickinson}, {Juneau},
  {Lotz}, {Newman}, {Papovich}, {Salim}, {Walth}, {Weiner}, \&
  {Willmer}}]{Cooper2012}
{Cooper} M.~C. {et~al.}, 2012, MNRAS, 425, 2116

\bibitem[{{Cowie} {et~al}\mbox{.}(1996){Cowie}, {Songaila}, {Hu}, \&
  {Cohen}}]{Cowie1996}
{Cowie} L.~L., {Songaila} A., {Hu} E.~M., {Cohen} J.~G., 1996, AJ, 112, 839

\bibitem[{{Cristiani} {et~al}\mbox{.}(2004){Cristiani}, {Alexander}, {Bauer},
  {Brandt}, {Chatzichristou}, {Fontanot}, {Grazian}, {Koekemoer}, {Lucas},
  {Monaco}, {Nonino}, {Padovani}, {Stern}, {Tozzi}, {Treister}, {Urry}, \&
  {Vanzella}}]{Cristiani2004}
{Cristiani} S. {et~al.}, 2004, ApJ, 600, L119

\bibitem[{{Crocce} {et~al}\mbox{.}(2016){Crocce}, {Carretero}, {Bauer}, {Ross},
  {Sevilla-Noarbe}, {Giannantonio}, {Sobreira}, {Sanchez}, {Gaztanaga}, {Kind},
  {S{\'a}nchez}, {Bonnett}, {Benoit-L{\'e}vy}, {Brunner}, {Rosell}, {Cawthon},
  {Fosalba}, {Hartley}, {Kim}, {Leistedt}, {Miquel}, {Peiris}, {Percival},
  {Rosenfeld}, {Rykoff}, {S{\'a}nchez}, {Abbott}, {Abdalla}, {Allam},
  {Banerji}, {Bernstein}, {Bertin}, {Brooks}, {Buckley-Geer}, {Burke},
  {Capozzi}, {Castander}, {Cunha}, {D'Andrea}, {da Costa}, {Desai}, {Diehl},
  {Eifler}, {Evrard}, {Neto}, {Fernandez}, {Finley}, {Flaugher}, {Frieman},
  {Gerdes}, {Gruen}, {Gruendl}, {Gutierrez}, {Honscheid}, {James}, {Kuehn},
  {Kuropatkin}, {Lahav}, {Li}, {Lima}, {Maia}, {March}, {Marshall}, {Martini},
  {Melchior}, {Miller}, {Neilsen}, {Nichol}, {Nord}, {Ogando}, {Plazas},
  {Romer}, {Sako}, {Santiago}, {Schubnell}, {Smith}, {Soares-Santos},
  {Suchyta}, {Swanson}, {Tarle}, {Thaler}, {Thomas}, {Vikram}, {Walker},
  {Wechsler}, {Weller}, {Zuntz}, \& {DES Collaboration}}]{Crocce2016}
{Crocce} M. {et~al.}, 2016, MNRAS, 455, 4301

\bibitem[{{Croton} {et~al}\mbox{.}(2005){Croton}, {Farrar}, {Norberg},
  {Colless}, {Peacock}, {Baldry}, {Baugh}, {Bland-Hawthorn}, {Bridges},
  {Cannon}, {Cole}, {Collins}, {Couch}, {Dalton}, {De Propris}, {Driver},
  {Efstathiou}, {Ellis}, {Frenk}, {Glazebrook}, {Jackson}, {Lahav}, {Lewis},
  {Lumsden}, {Maddox}, {Madgwick}, {Peterson}, {Sutherland}, \&
  {Taylor}}]{Croton2005}
{Croton} D.~J. {et~al.}, 2005, MNRAS, 356, 1155

\bibitem[{{Dark Energy Survey Collaboration} {et~al}\mbox{.}(2016){Dark Energy
  Survey Collaboration}, {Abbott}, {Abdalla}, {Allam}, {Aleksi{\'c}}, {Amara},
  {Bacon}, {Balbinot}, {Banerji}, {Bechtol}, {Benoit-L{\'e}vy}, {Bernstein},
  {Bertin}, {Blazek}, {Dodelson}, {Bonnett}, {Brooks}, {Bridle}, {Brunner},
  {Buckley-Geer}, {Burke}, {Capozzi}, {Caminha}, {Carlsen}, {Carnero-Rosell},
  {Carollo}, {Carrasco-Kind}, {Carretero}, {Castander}, {Clerkin}, {Collett},
  {Conselice}, {Crocce}, {Cunha}, {D'Andrea}, {da Costa}, {Davis}, {Desai},
  {Diehl}, {Dietrich}, {Doel}, {Drlica-Wagner}, {Etherington}, {Estrada},
  {Evrard}, {Fabbri}, {Finley}, {Flaugher}, {Fosalba}, {Foley}, {Frieman},
  {Garc{\'{\i}}a-Bellido}, {Gaztanaga}, {Gerdes}, {Giannantonio}, {Goldstein},
  {Gruen}, {Gruendl}, {Guarnieri}, {Gutierrez}, {Hartley}, {Honscheid}, {Jain},
  {James}, {Jeltema}, {Jouvel}, {Kessler}, {King}, {Kirk}, {Kron}, {Kuehn},
  {Kuropatkin}, {Lahav}, {Li}, {Lima}, {Lin}, {Maia}, {Makler}, {Manera},
  {Maraston}, {Marshall}, {Martini}, {McMahon}, {Melchior}, {Merson}, {Miller},
  {Miquel}, {Mohr}, {Morice-Atkinson}, {Naidoo}, {Neilsen}, {Nichol}, {Nord},
  {Ogando}, {Ostrovski}, {Palmese}, {Papadopoulos}, {Peiris}, {Peoples},
  {Plazas}, {Percival}, {Reed}, {Romer}, {Roodman}, {Ross}, {Rozo}, {Rykoff},
  {Sadeh}, {Sako}, {S{\'a}nchez}, {Sanchez}, {Santiago}, {Scarpine},
  {Schubnell}, {Sevilla-Noarbe}, {Sheldon}, {Smith}, {Smith}, {Soares-Santos},
  {Sobreira}, {Soumagnac}, {Suchyta}, {Sullivan}, {Swanson}, {Tarle}, {Thaler},
  {Thomas}, {Thomas}, {Tucker}, {Vieira}, {Vikram}, {Walker}, {Wechsler},
  {Wester}, {Weller}, {Whiteway}, {Wilcox}, {Yanny}, {Zhang}, \&
  {Zuntz}}]{DES2016}
{Dark Energy Survey Collaboration} {et~al.}, 2016, MNRAS

\bibitem[{{Davidzon} {et~al}\mbox{.}(2016){Davidzon}, {Cucciati}, {Bolzonella},
  {De Lucia}, {Zamorani}, {Arnouts}, {Moutard}, {Ilbert}, {Garilli},
  {Scodeggio}, {Guzzo}, {Abbas}, {Adami}, {Bel}, {Bottini}, {Branchini},
  {Cappi}, {Coupon}, {de la Torre}, {Di Porto}, {Fritz}, {Franzetti}, {Fumana},
  {Granett}, {Guennou}, {Iovino}, {Krywult}, {Le Brun}, {Le F{\`e}vre},
  {Maccagni}, {Ma{\l}ek}, {Marulli}, {McCracken}, {Mellier}, {Moscardini},
  {Polletta}, {Pollo}, {Tasca}, {Tojeiro}, {Vergani}, \&
  {Zanichelli}}]{Davidzon2016}
{Davidzon} I. {et~al.}, 2016, A\&A, 586, A23

\bibitem[{{Desai} {et~al}\mbox{.}(2012){Desai}, {Armstrong}, {Mohr}, {Semler},
  {Liu}, {Bertin}, {Allam}, {Barkhouse}, {Bazin}, {Buckley-Geer}, {Cooper},
  {Hansen}, {High}, {Lin}, {Lin}, {Ngeow}, {Rest}, {Song}, {Tucker}, \&
  {Zenteno}}]{Desai2012}
{Desai} S. {et~al.}, 2012, ApJ, 757, 83

\bibitem[{{Eisenstein} {et~al}\mbox{.}(2001){Eisenstein}, {Annis}, {Gunn},
  {Szalay}, {Connolly}, {Nichol}, {Bahcall}, {Bernardi}, {Burles}, {Castander},
  {Fukugita}, {Hogg}, {Ivezi{\'c}}, {Knapp}, {Lupton}, {Narayanan}, {Postman},
  {Reichart}, {Richmond}, {Schneider}, {Schlegel}, {Strauss}, {SubbaRao},
  {Tucker}, {Vanden Berk}, {Vogeley}, {Weinberg}, \& {Yanny}}]{Eisenstein2001}
{Eisenstein} D.~J. {et~al.}, 2001, AJ, 122, 2267

\bibitem[{{Etherington} \& {Thomas}(2015)}]{Etherington2015}
{Etherington} J., {Thomas} D., 2015, MNRAS, 451, 5179

\bibitem[{Farouki \& Shapiro(1981)}]{Farouki1981}
Farouki R., Shapiro S.~L., 1981, AJ, 243, 32

\bibitem[{{Flaugher} {et~al}\mbox{.}(2015){Flaugher}, {Diehl}, {Honscheid},
  {Abbott}, {Alvarez}, {Angstadt}, {Annis}, {Antonik}, {Ballester}, {Beaufore},
  {Bernstein}, {Bernstein}, {Bigelow}, {Bonati}, {Boprie}, {Brooks},
  {Buckley-Geer}, {Campa}, {Cardiel-Sas}, {Castander}, {Castilla}, {Cease},
  {Cela-Ruiz}, {Chappa}, {Chi}, {Cooper}, {da Costa}, {Dede}, {Derylo},
  {DePoy}, {de Vicente}, {Doel}, {Drlica-Wagner}, {Eiting}, {Elliott}, {Emes},
  {Estrada}, {Fausti Neto}, {Finley}, {Flores}, {Frieman}, {Gerdes},
  {Gladders}, {Gregory}, {Gutierrez}, {Hao}, {Holland}, {Holm}, {Huffman},
  {Jackson}, {James}, {Jonas}, {Karcher}, {Karliner}, {Kent}, {Kessler},
  {Kozlovsky}, {Kron}, {Kubik}, {Kuehn}, {Kuhlmann}, {Kuk}, {Lahav}, {Lathrop},
  {Lee}, {Levi}, {Lewis}, {Li}, {Mandrichenko}, {Marshall}, {Martinez},
  {Merritt}, {Miquel}, {Mu{\~n}oz}, {Neilsen}, {Nichol}, {Nord}, {Ogando},
  {Olsen}, {Palaio}, {Patton}, {Peoples}, {Plazas}, {Rauch}, {Reil}, {Rheault},
  {Roe}, {Rogers}, {Roodman}, {Sanchez}, {Scarpine}, {Schindler}, {Schmidt},
  {Schmitt}, {Schubnell}, {Schultz}, {Schurter}, {Scott}, {Serrano}, {Shaw},
  {Smith}, {Soares-Santos}, {Stefanik}, {Stuermer}, {Suchyta}, {Sypniewski},
  {Tarle}, {Thaler}, {Tighe}, {Tran}, {Tucker}, {Walker}, {Wang}, {Watson},
  {Weaverdyck}, {Wester}, {Woods}, {Yanny}, \& {DES
  Collaboration}}]{Flaugher2015}
{Flaugher} B. {et~al.}, 2015, AJ, 150, 150

\bibitem[{{Fontanot} {et~al}\mbox{.}(2009){Fontanot}, {De Lucia}, {Monaco},
  {Somerville}, \& {Santini}}]{Fontanot2009}
{Fontanot} F., {De Lucia} G., {Monaco} P., {Somerville} R.~S., {Santini} P.,
  2009, MNRAS, 397, 1776

\bibitem[{{Fossati} {et~al}\mbox{.}(2015){Fossati}, {Wilman}, {Fontanot}, {De
  Lucia}, {Monaco}, {Hirschmann}, {Mendel}, {Beifiori}, \&
  {Contini}}]{Fossati2015}
{Fossati} M. {et~al.}, 2015, MNRAS, 446, 2582

\bibitem[{{Fumagalli} {et~al}\mbox{.}(2014){Fumagalli}, {Fossati}, {Hau},
  {Gavazzi}, {Bower}, {Sun}, \& {Boselli}}]{Fumagalli2014}
{Fumagalli} M., {Fossati} M., {Hau} G.~K.~T., {Gavazzi} G., {Bower} R., {Sun}
  M., {Boselli} A., 2014, MNRAS, 445, 4335

\bibitem[{Gallazzi {et~al}\mbox{.}(2009)Gallazzi, Bell, Wolf, Gray, Papovich,
  Barden, Peng, Meisenheimer, Heymans, van Kampen, Gilmour, Balogh, McIntosh,
  Bacon, Barazza, B\"{o}hm, Caldwell, H\"{a}u\ss~ler, Jahnke, Jogee, Lane,
  Robaina, Sanchez, Taylor, Wisotzki, \& Zheng}]{Gallazzi2009}
Gallazzi A. {et~al.}, 2009, ApJ, 690, 1883

\bibitem[{{Garilli} {et~al}\mbox{.}(2014){Garilli}, {Guzzo}, {Scodeggio},
  {Bolzonella}, {Abbas}, {Adami}, {Arnouts}, {Bel}, {Bottini}, {Branchini},
  {Cappi}, {Coupon}, {Cucciati}, {Davidzon}, {De Lucia}, {de la Torre},
  {Franzetti}, {Fritz}, {Fumana}, {Granett}, {Ilbert}, {Iovino}, {Krywult}, {Le
  Brun}, {Le F{\`e}vre}, {Maccagni}, {Ma{\l}ek}, {Marulli}, {McCracken},
  {Paioro}, {Polletta}, {Pollo}, {Schlagenhaufer}, {Tasca}, {Tojeiro},
  {Vergani}, {Zamorani}, {Zanichelli}, {Burden}, {Di Porto}, {Marchetti},
  {Marinoni}, {Mellier}, {Moscardini}, {Nichol}, {Peacock}, {Percival},
  {Phleps}, \& {Wolk}}]{Garilli2014}
{Garilli} B. {et~al.}, 2014, A\&A, 562, A23

\bibitem[{{Garilli} {et~al}\mbox{.}(2008){Garilli}, {Le F{\`e}vre}, {Guzzo},
  {Maccagni}, {Le Brun}, {de la Torre}, {Meneux}, {Tresse}, {Franzetti},
  {Zamorani}, {Zanichelli}, {Gregorini}, {Vergani}, {Bottini}, {Scaramella},
  {Scodeggio}, {Vettolani}, {Adami}, {Arnouts}, {Bardelli}, {Bolzonella},
  {Cappi}, {Charlot}, {Ciliegi}, {Contini}, {Foucaud}, {Gavignaud}, {Ilbert},
  {Iovino}, {Lamareille}, {McCracken}, {Marano}, {Marinoni}, {Mazure},
  {Merighi}, {Paltani}, {Pell{\`o}}, {Pollo}, {Pozzetti}, {Radovich}, {Zucca},
  {Blaizot}, {Bongiorno}, {Cucciati}, {Mellier}, {Moreau}, \&
  {Paioro}}]{Garilli2008}
{Garilli} B. {et~al.}, 2008, A\&A, 486, 683

\bibitem[{{Giannantonio} {et~al}\mbox{.}(2016){Giannantonio}, {Fosalba},
  {Cawthon}, {Omori}, {Crocce}, {Elsner}, {Leistedt}, {Dodelson},
  {Benoit-L{\'e}vy}, {Gazta{\~n}aga}, {Holder}, {Peiris}, {Percival}, {Kirk},
  {Bauer}, {Benson}, {Bernstein}, {Carretero}, {Crawford}, {Crittenden},
  {Huterer}, {Jain}, {Krause}, {Reichardt}, {Ross}, {Simard}, {Soergel},
  {Stark}, {Story}, {Vieira}, {Weller}, {Abbott}, {Abdalla}, {Allam},
  {Armstrong}, {Banerji}, {Bernstein}, {Bertin}, {Brooks}, {Buckley-Geer},
  {Burke}, {Capozzi}, {Carlstrom}, {Carnero Rosell}, {Carrasco Kind},
  {Castander}, {Chang}, {Cunha}, {da Costa}, {D'Andrea}, {DePoy}, {Desai},
  {Diehl}, {Dietrich}, {Doel}, {Eifler}, {Evrard}, {Neto}, {Fernandez},
  {Finley}, {Flaugher}, {Frieman}, {Gerdes}, {Gruen}, {Gruendl}, {Gutierrez},
  {Holzapfel}, {Honscheid}, {James}, {Kuehn}, {Kuropatkin}, {Lahav}, {Li},
  {Lima}, {March}, {Marshall}, {Martini}, {Melchior}, {Miquel}, {Mohr},
  {Nichol}, {Nord}, {Ogando}, {Plazas}, {Romer}, {Roodman}, {Rykoff}, {Sako},
  {Saliwanchik}, {Sanchez}, {Schubnell}, {Sevilla-Noarbe}, {Smith},
  {Soares-Santos}, {Sobreira}, {Suchyta}, {Swanson}, {Tarle}, {Thaler},
  {Thomas}, {Vikram}, {Walker}, {Wechsler}, \& {Zuntz}}]{Giannantonio2016}
{Giannantonio} T. {et~al.}, 2016, MNRAS, 456, 3213

\bibitem[{{G{\'o}rski} {et~al}\mbox{.}(2005){G{\'o}rski}, {Hivon}, {Banday},
  {Wandelt}, {Hansen}, {Reinecke}, \& {Bartelmann}}]{Gorski2005}
{G{\'o}rski} K.~M., {Hivon} E., {Banday} A.~J., {Wandelt} B.~D., {Hansen}
  F.~K., {Reinecke} M., {Bartelmann} M., 2005, ApJ, 622, 759

\bibitem[{Haas {et~al}\mbox{.}(2012)Haas, Schaye, \& Jeeson-Daniel}]{Haas2012}
Haas M.~R., Schaye J., Jeeson-Daniel A., 2012, MNRAS, 419, 2133

\bibitem[{{Hasinger} {et~al}\mbox{.}(2005){Hasinger}, {Miyaji}, \&
  {Schmidt}}]{Hasinger2005}
{Hasinger} G., {Miyaji} T., {Schmidt} M., 2005, A\&A, 441, 417

\bibitem[{{Hennig} {et~al}\mbox{.}(2016){Hennig}, {Mohr}, {Zenteno}, {Desai},
  {Dietrich}, {Bocquet}, {Strazzullo}, {Saro}, {Abbott}, {Abdalla}, {Bayliss},
  {Benoit-Levy}, {Bernstein}, {Bertin}, {Brooks}, {Capasso}, {Capozzi},
  {Carnero}, {Carrasco Kind}, {Carretero}, {Chiu}, {D'Andrea}, {daCosta},
  {Diehl}, {Doel}, {Eifler}, {Evrard}, {Fausti-Neto}, {Fosalba}, {Frieman},
  {Gangkofner}, {Gonzalez}, {Gruen}, {Gruendl}, {Gupta}, {Gutierrez},
  {Honscheid}, {Hlavacek-Larrondo}, {James}, {Kuehn}, {Kuropatkin}, {Lahav},
  {March}, {Marshall}, {Martini}, {McDonald}, {Melchior}, {Miller}, {Miquel},
  {Neilsen}, {Nord}, {Ogando}, {Plazas}, {Reichardt}, {Romer}, {Rozo},
  {Rykoff}, {Sanchez}, {Santiago}, {Schubnell}, {Sevilla-Noarbe}, {Smith},
  {Soares-Santos}, {Sobreira}, {Stalder}, {Stanford}, {Suchyta}, {Swanson},
  {Tarle}, {Thomas}, {Vikram}, {Walker}, \& {Zhang}}]{Hennig2016}
{Hennig} C. {et~al.}, 2016, arXiv:1604.00988

\bibitem[{{Hirschmann} {et~al}\mbox{.}(2012){Hirschmann}, {Somerville}, {Naab},
  \& {Burkert}}]{Hirschmann2012}
{Hirschmann} M., {Somerville} R.~S., {Naab} T., {Burkert} A., 2012, MNRAS, 426,
  237

\bibitem[{{Ilbert} {et~al}\mbox{.}(2013){Ilbert}, {McCracken}, {Le F{\`e}vre},
  {Capak}, {Dunlop}, {Karim}, {Renzini}, {Caputi}, {Boissier}, {Arnouts},
  {Aussel}, {Comparat}, {Guo}, {Hudelot}, {Kartaltepe}, {Kneib}, {Krogager},
  {Le Floc'h}, {Lilly}, {Mellier}, {Milvang-Jensen}, {Moutard}, {Onodera},
  {Richard}, {Salvato}, {Sanders}, {Scoville}, {Silverman}, {Taniguchi},
  {Tasca}, {Thomas}, {Toft}, {Tresse}, {Vergani}, {Wolk}, \&
  {Zirm}}]{Ilbert2013}
{Ilbert} O. {et~al.}, 2013, A\&A, 556, A55

\bibitem[{{Ilbert} {et~al}\mbox{.}(2010){Ilbert}, {Salvato}, {Le Floc'h},
  {Aussel}, {Capak}, {McCracken}, {Mobasher}, {Kartaltepe}, {Scoville},
  {Sanders}, {Arnouts}, {Bundy}, {Cassata}, {Kneib}, {Koekemoer}, {Le
  F{\`e}vre}, {Lilly}, {Surace}, {Taniguchi}, {Tasca}, {Thompson}, {Tresse},
  {Zamojski}, {Zamorani}, \& {Zucca}}]{Ilbert2010}
{Ilbert} O. {et~al.}, 2010, ApJ, 709, 644

\bibitem[{{Jarvis} {et~al}\mbox{.}(2016){Jarvis}, {Sheldon}, {Zuntz},
  {Kacprzak}, {Bridle}, {Amara}, {Armstrong}, {Becker}, {Bernstein}, {Bonnett},
  {Chang}, {Das}, {Dietrich}, {Drlica-Wagner}, {Eifler}, {Gangkofner}, {Gruen},
  {Hirsch}, {Huff}, {Jain}, {Kent}, {Kirk}, {MacCrann}, {Melchior}, {Plazas},
  {Refregier}, {Rowe}, {Rykoff}, {Samuroff}, {S{\'a}nchez}, {Suchyta},
  {Troxel}, {Vikram}, {Abbott}, {Abdalla}, {Allam}, {Annis}, {Benoit-L{\'e}vy},
  {Bertin}, {Brooks}, {Buckley-Geer}, {Burke}, {Capozzi}, {Carnero Rosell},
  {Carrasco Kind}, {Carretero}, {Castander}, {Clampitt}, {Crocce}, {Cunha},
  {D'Andrea}, {da Costa}, {DePoy}, {Desai}, {Diehl}, {Doel}, {Fausti Neto},
  {Flaugher}, {Fosalba}, {Frieman}, {Gaztanaga}, {Gerdes}, {Gruendl},
  {Gutierrez}, {Honscheid}, {James}, {Kuehn}, {Kuropatkin}, {Lahav}, {Li},
  {Lima}, {March}, {Martini}, {Miquel}, {Mohr}, {Neilsen}, {Nord}, {Ogando},
  {Reil}, {Romer}, {Roodman}, {Sako}, {Sanchez}, {Scarpine}, {Schubnell},
  {Sevilla-Noarbe}, {Smith}, {Soares-Santos}, {Sobreira}, {Swanson}, {Tarle},
  {Thaler}, {Thomas}, {Walker}, \& {Wechsler}}]{Jarvis2016}
{Jarvis} M. {et~al.}, 2016, MNRAS, 460, 2245

\bibitem[{{Kaiser}(1984)}]{Kaiser1984}
{Kaiser} N., 1984, ApJ, 284, L9

\bibitem[{{Kodama} \& {Bower}(2003)}]{Kodama2003}
{Kodama} T., {Bower} R., 2003, MNRAS, 346, 1

\bibitem[{Larson {et~al}\mbox{.}(1980)Larson, Tinsley, \&
  Caldwell}]{Larson1980}
Larson R.~B., Tinsley B.~M., Caldwell C.~N., 1980, AJ, 237, 692

\bibitem[{{Le F{\`e}vre} {et~al}\mbox{.}(2013){Le F{\`e}vre}, {Cassata},
  {Cucciati}, {Garilli}, {Ilbert}, {Le Brun}, {Maccagni}, {Moreau},
  {Scodeggio}, {Tresse}, {Zamorani}, {Adami}, {Arnouts}, {Bardelli},
  {Bolzonella}, {Bondi}, {Bongiorno}, {Bottini}, {Cappi}, {Charlot}, {Ciliegi},
  {Contini}, {de la Torre}, {Foucaud}, {Franzetti}, {Gavignaud}, {Guzzo},
  {Iovino}, {Lemaux}, {L{\'o}pez-Sanjuan}, {McCracken}, {Marano}, {Marinoni},
  {Mazure}, {Mellier}, {Merighi}, {Merluzzi}, {Paltani}, {Pell{\`o}}, {Pollo},
  {Pozzetti}, {Scaramella}, {Tasca}, {Vergani}, {Vettolani}, {Zanichelli}, \&
  {Zucca}}]{LeFevre2013}
{Le F{\`e}vre} O. {et~al.}, 2013, A\&A, 559, A14

\bibitem[{{Le F{\`e}vre} {et~al}\mbox{.}(2005){Le F{\`e}vre}, {Vettolani},
  {Garilli}, {Tresse}, {Bottini}, {Le Brun}, {Maccagni}, {Picat}, {Scaramella},
  {Scodeggio}, {Zanichelli}, {Adami}, {Arnaboldi}, {Arnouts}, {Bardelli},
  {Bolzonella}, {Cappi}, {Charlot}, {Ciliegi}, {Contini}, {Foucaud},
  {Franzetti}, {Gavignaud}, {Guzzo}, {Ilbert}, {Iovino}, {McCracken}, {Marano},
  {Marinoni}, {Mathez}, {Mazure}, {Meneux}, {Merighi}, {Paltani}, {Pell{\`o}},
  {Pollo}, {Pozzetti}, {Radovich}, {Zamorani}, {Zucca}, {Bondi}, {Bongiorno},
  {Busarello}, {Lamareille}, {Mellier}, {Merluzzi}, {Ripepi}, \&
  {Rizzo}}]{LeFevre2005}
{Le F{\`e}vre} O. {et~al.}, 2005, A\&A, 439, 845

\bibitem[{{Leistedt} {et~al}\mbox{.}(2015){Leistedt}, {Peiris}, {Elsner},
  {Benoit-L{\'e}vy}, {Amara}, {Bauer}, {Becker}, {Bonnett}, {Bruderer},
  {Busha}, {Carrasco Kind}, {Chang}, {Crocce}, {da Costa}, {Gaztanaga}, {Huff},
  {Lahav}, {Palmese}, {Percival}, {Refregier}, {Ross}, {Rozo}, {Rykoff},
  {S{\'a}nchez}, {Sadeh}, {Sevilla-Noarbe}, {Sobreira}, {Suchyta}, {Swanson},
  {Wechsler}, {Abdalla}, {Allam}, {Banerji}, {Bernstein}, {Bernstein},
  {Bertin}, {Bridle}, {Brooks}, {Buckley-Geer}, {Burke}, {Capozzi}, {Carnero
  Rosell}, {Carretero}, {Cunha}, {D'Andrea}, {DePoy}, {Desai}, {Diehl}, {Doel},
  {Eifler}, {Evrard}, {Fausti Neto}, {Flaugher}, {Fosalba}, {Frieman},
  {Gerdes}, {Gruen}, {Gruendl}, {Gutierrez}, {Honscheid}, {James}, {Jarvis},
  {Kent}, {Kuehn}, {Kuropatkin}, {Li}, {Lima}, {Maia}, {March}, {Marshall},
  {Martini}, {Melchior}, {Miller}, {Miquel}, {Nichol}, {Nord}, {Ogando},
  {Plazas}, {Reil}, {Romer}, {Roodman}, {Sanchez}, {Santiago}, {Scarpine},
  {Schubnell}, {Smith}, {Soares-Santos}, {Tarle}, {Thaler}, {Thomas}, {Vikram},
  {Walker}, {Wester}, {Zhang}, \& {Zuntz}}]{Leistedt2015}
{Leistedt} B. {et~al.}, 2015, arXiv:1507.05647

\bibitem[{{Li} \& {White}(2009)}]{Li2009}
{Li} C., {White} S.~D.~M., 2009, MNRAS, 398, 2177

\bibitem[{{Lilly} {et~al}\mbox{.}(2007){Lilly}, {Le F{\`e}vre}, {Renzini},
  {Zamorani}, {Scodeggio}, {Contini}, {Carollo}, {Hasinger}, {Kneib}, {Iovino},
  {Le Brun}, {Maier}, {Mainieri}, {Mignoli}, {Silverman}, {Tasca},
  {Bolzonella}, {Bongiorno}, {Bottini}, {Capak}, {Caputi}, {Cimatti},
  {Cucciati}, {Daddi}, {Feldmann}, {Franzetti}, {Garilli}, {Guzzo}, {Ilbert},
  {Kampczyk}, {Kovac}, {Lamareille}, {Leauthaud}, {Borgne}, {McCracken},
  {Marinoni}, {Pello}, {Ricciardelli}, {Scarlata}, {Vergani}, {Sanders},
  {Schinnerer}, {Scoville}, {Taniguchi}, {Arnouts}, {Aussel}, {Bardelli},
  {Brusa}, {Cappi}, {Ciliegi}, {Finoguenov}, {Foucaud}, {Franceschini},
  {Halliday}, {Impey}, {Knobel}, {Koekemoer}, {Kurk}, {Maccagni}, {Maddox},
  {Marano}, {Marconi}, {Meneux}, {Mobasher}, {Moreau}, {Peacock}, {Porciani},
  {Pozzetti}, {Scaramella}, {Schiminovich}, {Shopbell}, {Smail}, {Thompson},
  {Tresse}, {Vettolani}, {Zanichelli}, \& {Zucca}}]{Lilly2007}
{Lilly} S.~J. {et~al.}, 2007, ApJS, 172, 70

\bibitem[{{Maiolino} {et~al}\mbox{.}(2008){Maiolino}, {Nagao}, {Grazian},
  {Cocchia}, {Marconi}, {Mannucci}, {Cimatti}, {Pipino}, {Ballero}, {Calura},
  {Chiappini}, {Fontana}, {Granato}, {Matteucci}, {Pastorini}, {Pentericci},
  {Risaliti}, {Salvati}, \& {Silva}}]{Maiolino2008}
{Maiolino} R. {et~al.}, 2008, A\&A, 488, 463

\bibitem[{{Maraston}(2005)}]{Maraston2005}
{Maraston} C., 2005, MNRAS, 362, 799

\bibitem[{{Maraston} {et~al}\mbox{.}(2006){Maraston}, {Daddi}, {Renzini},
  {Cimatti}, {Dickinson}, {Papovich}, {Pasquali}, \& {Pirzkal}}]{Maraston2006}
{Maraston} C., {Daddi} E., {Renzini} A., {Cimatti} A., {Dickinson} M.,
  {Papovich} C., {Pasquali} A., {Pirzkal} N., 2006, ApJ, 652, 85

\bibitem[{Maraston {et~al}\mbox{.}(2013)Maraston, Pforr, Henriques, Thomas,
  Wake, Brownstein, Capozzi, Tinker, Bundy, Skibba, Beifiori, Nichol,
  Edmondson, Schneider, Chen, Masters, Steele, Bolton, York, Weaver, Higgs,
  Bizyaev, Brewington, Malanushenko, Malanushenko, Snedden, Oravetz, Pan,
  Shelden, \& Simmons}]{Maraston2013}
Maraston C. {et~al.}, 2013, MNRAS, 435, 2764

\bibitem[{{Maraston} {et~al}\mbox{.}(2009){Maraston}, {Str{\"o}mb{\"a}ck},
  {Thomas}, {Wake}, \& {Nichol}}]{Maraston2009}
{Maraston} C., {Str{\"o}mb{\"a}ck} G., {Thomas} D., {Wake} D.~A., {Nichol}
  R.~C., 2009, MNRAS, 394, L107

\bibitem[{{McNaught-Roberts} {et~al}\mbox{.}(2014){McNaught-Roberts},
  {Norberg}, {Baugh}, {Lacey}, {Loveday}, {Peacock}, {Baldry},
  {Bland-Hawthorn}, {Brough}, {Driver}, {Robotham}, \&
  {V{\'a}zquez-Mata}}]{McNaught-Roberts2014}
{McNaught-Roberts} T. {et~al.}, 2014, MNRAS, 445, 2125

\bibitem[{{Mohr} {et~al}\mbox{.}(2012){Mohr} {et~al.}}]{Mohr2012}
{Mohr} J.~J., {et~al.}, 2012, in Society of Photo-Optical Instrumentation
  Engineers (SPIE) Conference Series, Vol. 8451, Society of Photo-Optical
  Instrumentation Engineers (SPIE) Conference Series, p.~0

\bibitem[{{Mortlock} {et~al}\mbox{.}(2015){Mortlock}, {Conselice}, {Hartley},
  {Duncan}, {Lani}, {Ownsworth}, {Almaini}, {Wel}, {Huang}, {Ashby}, {Willner},
  {Fontana}, {Dekel}, {Koekemoer}, {Ferguson}, {Faber}, {Grogin}, \&
  {Kocevski}}]{Mortlock2015}
{Mortlock} A. {et~al.}, 2015, MNRAS, 447, 2

\bibitem[{Muldrew {et~al}\mbox{.}(2012)Muldrew, Croton, Skibba, Pearce, Ann,
  Baldry, Brough, Choi, Conselice, Cowan, Gallazzi, Gray, Gr\"{u}tzbauch, Li,
  Park, Pilipenko, Podgorzec, Robotham, Wilman, Yang, Zhang, \&
  Zibetti}]{Muldrew2012}
Muldrew S.~I. {et~al.}, 2012, MNRAS, 419, 2670

\bibitem[{Newman {et~al}\mbox{.}(2013)Newman, Cooper, Davis, Faber, Coil,
  Guhathakurta, Koo, Phillips, Conroy, Dutton, Finkbeiner, Gerke, Rosario,
  Weiner, Willmer, Yan, Harker, Kassin, Konidaris, Lai, Madgwick, Noeske,
  Wirth, Connolly, Kaiser, Kirby, Lemaux, Lin, Lotz, Luppino, Marinoni,
  Matthews, Metevier, \& Schiavon}]{Newman2013}
Newman J.~A. {et~al.}, 2013, ApJS, 208, 5

\bibitem[{{Nipoti} {et~al}\mbox{.}(2003){Nipoti}, {Stiavelli}, {Ciotti},
  {Treu}, \& {Rosati}}]{Nipoti2003}
{Nipoti} C., {Stiavelli} M., {Ciotti} L., {Treu} T., {Rosati} P., 2003, MNRAS,
  344, 748

\bibitem[{{Palmese} {et~al}\mbox{.}(2016){Palmese}, {Lahav}, {Banerji},
  {Gruen}, {Jouvel}, {Melchior}, {Aleksi{\'c}}, {Annis}, {Diehl}, {Jeltema},
  {Romer}, {Rozo}, {Rykoff}, {Seitz}, {Suchyta}, {Zhang}, {Abbott}, {Abdalla},
  {Allam}, {Benoit-L{\'e}vy}, {Bertin}, {Brooks}, {Buckley-Geer}, {Burke},
  {Capozzi}, {Carnero Rosell}, {Carrasco Kind}, {Carretero}, {Crocce}, {Cunha},
  {D'Andrea}, {da Costa}, {Desai}, {Dietrich}, {Doel}, {Estrada}, {Evrard},
  {Flaugher}, {Frieman}, {Gerdes}, {Goldstein}, {Gruendl}, {Gutierrez},
  {Honscheid}, {James}, {Kuehn}, {Kuropatkin}, {Li}, {Lima}, {Maia},
  {Marshall}, {Miller}, {Miquel}, {Nord}, {Ogando}, {Plazas}, {Roodman},
  {Sanchez}, {Scarpine}, {Sevilla-Noarbe}, {Smith}, {Soares-Santos},
  {Sobreira}, {Swanson}, {Tarle}, {Thomas}, {Tucker}, \&
  {Vikram}}]{Palmese2016}
{Palmese} A. {et~al.}, 2016, arXiv:1601.00589

\bibitem[{Peng {et~al}\mbox{.}(2010)Peng, Lilly, Kova\v{c}, Bolzonella,
  Pozzetti, Renzini, Zamorani, Ilbert, Knobel, Iovino, Maier, Cucciati, Tasca,
  Carollo, Silverman, Kampczyk, de~Ravel, Sanders, Scoville, Contini, Mainieri,
  Scodeggio, Kneib, {Le F\`{e}vre}, Bardelli, Bongiorno, Caputi, Coppa, de~la
  Torre, Franzetti, Garilli, Lamareille, {Le Borgne}, {Le Brun}, Mignoli,
  Montero, Pello, Ricciardelli, Tanaka, Tresse, Vergani, Welikala, Zucca,
  Oesch, Abbas, Barnes, Bordoloi, Bottini, Cappi, Cassata, Cimatti, Fumana,
  Hasinger, Koekemoer, Leauthaud, Maccagni, Marinoni, McCracken, Memeo, Meneux,
  Nair, Porciani, Presotto, \& Scaramella}]{Peng2010}
Peng Y.-j. {et~al.}, 2010, ApJ, 721, 193

\bibitem[{{Pipino} {et~al}\mbox{.}(2014){Pipino}, {Cibinel}, {Tacchella},
  {Carollo}, {Lilly}, {Miniati}, {Silverman}, {van Gorkom}, \&
  {Finoguenov}}]{Pipino2014}
{Pipino} A. {et~al.}, 2014, ApJ, 797, 127

\bibitem[{{Pozzetti} {et~al}\mbox{.}(2007){Pozzetti}, {Bolzonella},
  {Lamareille}, {Zamorani}, {Franzetti}, {Le F{\`e}vre}, {Iovino}, {Temporin},
  {Ilbert}, {Arnouts}, {Charlot}, {Brinchmann}, {Zucca}, {Tresse}, {Scodeggio},
  {Guzzo}, {Bottini}, {Garilli}, {Le Brun}, {Maccagni}, {Picat}, {Scaramella},
  {Vettolani}, {Zanichelli}, {Adami}, {Bardelli}, {Cappi}, {Ciliegi},
  {Contini}, {Foucaud}, {Gavignaud}, {McCracken}, {Marano}, {Marinoni},
  {Mazure}, {Meneux}, {Merighi}, {Paltani}, {Pell{\`o}}, {Pollo}, {Radovich},
  {Bondi}, {Bongiorno}, {Cucciati}, {de la Torre}, {Gregorini}, {Mellier},
  {Merluzzi}, {Vergani}, \& {Walcher}}]{Pozzetti2007}
{Pozzetti} L. {et~al.}, 2007, A\&A, 474, 443

\bibitem[{{Pozzetti} {et~al}\mbox{.}(2010){Pozzetti}, {Bolzonella}, {Zucca},
  {Zamorani}, {Lilly}, {Renzini}, {Moresco}, {Mignoli}, {Cassata}, {Tasca},
  {Lamareille}, {Maier}, {Meneux}, {Halliday}, {Oesch}, {Vergani}, {Caputi},
  {Kova{\v c}}, {Cimatti}, {Cucciati}, {Iovino}, {Peng}, {Carollo}, {Contini},
  {Kneib}, {Le F{\'e}vre}, {Mainieri}, {Scodeggio}, {Bardelli}, {Bongiorno},
  {Coppa}, {de la Torre}, {de Ravel}, {Franzetti}, {Garilli}, {Kampczyk},
  {Knobel}, {Le Borgne}, {Le Brun}, {Pell{\`o}}, {Perez Montero},
  {Ricciardelli}, {Silverman}, {Tanaka}, {Tresse}, {Abbas}, {Bottini}, {Cappi},
  {Guzzo}, {Koekemoer}, {Leauthaud}, {Maccagni}, {Marinoni}, {McCracken},
  {Memeo}, {Porciani}, {Scaramella}, {Scarlata}, \& {Scoville}}]{Pozzetti2010}
{Pozzetti} L. {et~al.}, 2010, A\&A, 523, A13

\bibitem[{{Rozo} {et~al}\mbox{.}(2016){Rozo}, {Rykoff}, {Abate}, {Bonnett},
  {Crocce}, {Davis}, {Hoyle}, {Leistedt}, {Peiris}, {Wechsler}, {Abbott},
  {Abdalla}, {Banerji}, {Bauer}, {Benoit-L{\'e}vy}, {Bernstein}, {Bertin},
  {Brooks}, {Buckley-Geer}, {Burke}, {Capozzi}, {Rosell}, {Carollo}, {Kind},
  {Carretero}, {Castander}, {Childress}, {Cunha}, {D'Andrea}, {Davis}, {DePoy},
  {Desai}, {Diehl}, {Dietrich}, {Doel}, {Eifler}, {Evrard}, {Neto}, {Flaugher},
  {Fosalba}, {Frieman}, {Gaztanaga}, {Gerdes}, {Glazebrook}, {Gruen},
  {Gruendl}, {Honscheid}, {James}, {Jarvis}, {Kim}, {Kuehn}, {Kuropatkin},
  {Lahav}, {Lidman}, {Lima}, {Maia}, {March}, {Martini}, {Melchior}, {Miller},
  {Miquel}, {Mohr}, {Nichol}, {Nord}, {O'Neill}, {Ogando}, {Plazas}, {Romer},
  {Roodman}, {Sako}, {Sanchez}, {Santiago}, {Schubnell}, {Sevilla-Noarbe},
  {Smith}, {Soares-Santos}, {Sobreira}, {Suchyta}, {Swanson}, {Thaler},
  {Thomas}, {Uddin}, {Vikram}, {Walker}, {Wester}, {Zhang}, \& {da
  Costa}}]{Rozo2016}
{Rozo} E. {et~al.}, 2016, MNRAS, 461, 1431

\bibitem[{{Salpeter}(1955)}]{Salpeter1955}
{Salpeter} E.~E., 1955, ApJ, 121, 161

\bibitem[{{S{\'a}nchez} {et~al}\mbox{.}(2014){S{\'a}nchez}, {Carrasco Kind},
  {Lin}, {Miquel}, {Abdalla}, {Amara}, {Banerji}, {Bonnett}, {Brunner},
  {Capozzi}, {Carnero}, {Castander}, {da Costa}, {Cunha}, {Fausti}, {Gerdes},
  {Greisel}, {Gschwend}, {Hartley}, {Jouvel}, {Lahav}, {Lima}, {Maia},
  {Mart{\'{\i}}}, {Ogando}, {Ostrovski}, {Pellegrini}, {Rau}, {Sadeh}, {Seitz},
  {Sevilla-Noarbe}, {Sypniewski}, {de Vicente}, {Abbot}, {Allam}, {Atlee},
  {Bernstein}, {Bernstein}, {Buckley-Geer}, {Burke}, {Childress}, {Davis},
  {DePoy}, {Dey}, {Desai}, {Diehl}, {Doel}, {Estrada}, {Evrard},
  {Fern{\'a}ndez}, {Finley}, {Flaugher}, {Frieman}, {Gaztanaga}, {Glazebrook},
  {Honscheid}, {Kim}, {Kuehn}, {Kuropatkin}, {Lidman}, {Makler}, {Marshall},
  {Nichol}, {Roodman}, {S{\'a}nchez}, {Santiago}, {Sako}, {Scalzo}, {Smith},
  {Swanson}, {Tarle}, {Thomas}, {Tucker}, {Uddin}, {Vald{\'e}s}, {Walker},
  {Yuan}, \& {Zuntz}}]{Sanchez2014}
{S{\'a}nchez} C. {et~al.}, 2014, MNRAS, 445, 1482

\bibitem[{Schawinski {et~al}\mbox{.}(2007)Schawinski, Kaviraj, Khochfar, Yoon,
  Yi, Deharveng, Boselli, Barlow, Conrow, Forster, Friedman, Martin, Morrissey,
  Neff, Schiminovich, Seibert, Small, Wyder, Bianchi, Donas, Heckman, Lee,
  Madore, Milliard, Rich, \& Szalay}]{Schawinski2007}
Schawinski K. {et~al.}, 2007, The Astrophysical Journal Supplement Series, 173,
  512

\bibitem[{{Schmidt}(1968)}]{Schmidt1968}
{Schmidt} M., 1968, ApJ, 151, 393

\bibitem[{{Sevilla} {et~al}\mbox{.}(2011){Sevilla}, {Armstrong}, {Bertin},
  {Carlson}, {Daues}, {Desai}, {Gower}, {Gruendl}, {Hanlon}, {Jarvis},
  {Kessler}, {Kuropatkin}, {Lin}, {Marriner}, {Mohr}, {Petravick}, {Sheldon},
  {Swanson}, {Tomashek}, {Tucker}, {Yang}, {Yanny}, \& {for the DES
  Collaboration}}]{Sevilla2011}
{Sevilla} I. {et~al.}, 2011, arXiv:1109.6741

\bibitem[{{Shattow} {et~al}\mbox{.}(2013){Shattow}, {Croton}, {Skibba},
  {Muldrew}, {Pearce}, \& {Abbas}}]{Shattow2013}
{Shattow} G.~M., {Croton} D.~J., {Skibba} R.~A., {Muldrew} S.~I., {Pearce}
  F.~R., {Abbas} U., 2013, MNRAS, 433, 3314

\bibitem[{{Springel} {et~al}\mbox{.}(2006){Springel}, {Frenk}, \&
  {White}}]{Springel2006}
{Springel} V., {Frenk} C.~S., {White} S.~D.~M., 2006, Nature, 440, 1137

\bibitem[{{Strauss} {et~al}\mbox{.}(2002){Strauss}, {Weinberg}, {Lupton},
  {Narayanan}, {Annis}, {Bernardi}, {Blanton}, {Burles}, {Connolly},
  {Dalcanton}, {Doi}, {Eisenstein}, {Frieman}, {Fukugita}, {Gunn},
  {Ivezi{\'c}}, {Kent}, {Kim}, {Knapp}, {Kron}, {Munn}, {Newberg}, {Nichol},
  {Okamura}, {Quinn}, {Richmond}, {Schlegel}, {Shimasaku}, {SubbaRao},
  {Szalay}, {Vanden Berk}, {Vogeley}, {Yanny}, {Yasuda}, {York}, \&
  {Zehavi}}]{Strauss2002}
{Strauss} M.~A. {et~al.}, 2002, AJ, 124, 1810

\bibitem[{{Taylor} {et~al}\mbox{.}(2009){Taylor}, {Franx}, {van Dokkum},
  {Bell}, {Brammer}, {Rudnick}, {Wuyts}, {Gawiser}, {Lira}, {Urry}, \&
  {Rix}}]{Taylor2009}
{Taylor} E.~N. {et~al.}, 2009, ApJ, 694, 1171

\bibitem[{{Thomas} {et~al}\mbox{.}(2005){Thomas}, {Maraston}, {Bender}, \&
  {Mendes de Oliveira}}]{Thomas2005}
{Thomas} D., {Maraston} C., {Bender} R., {Mendes de Oliveira} C., 2005, ApJ,
  621, 673

\bibitem[{Thomas {et~al}\mbox{.}(2010)Thomas, Maraston, Schawinski, Sarzi, \&
  Silk}]{Thomas2010}
Thomas D., Maraston C., Schawinski K., Sarzi M., Silk J., 2010, MNRAS, 1789,
  1775

\bibitem[{{Vikram} {et~al}\mbox{.}(2015){Vikram}, {Chang}, {Jain}, {Bacon},
  {Amara}, {Becker}, {Bernstein}, {Bonnett}, {Bridle}, {Brout}, {Busha},
  {Frieman}, {Gaztanaga}, {Hartley}, {Jarvis}, {Kacprzak}, {Kovacs}, {Lahav},
  {Leistedt}, {Lin}, {Melchior}, {Peiris}, {Rozo}, {Rykoff}, {Sanchez},
  {Sheldon}, {Troxel}, {Wechsler}, {Zuntz}, {Abbott}, {Abdalla}, {Armstrong},
  {Banerji}, {Bauer}, {Benoit-Levy}, {Bertin}, {Brooks}, {Buckley-Geer},
  {Burke}, {Capozzi}, {Carnero Rosell}, {Carrasco Kind}, {Castander}, {Crocce},
  {Cunha}, {D'Andrea}, {da Costa}, {DePoy}, {Desai}, {Diehl}, {Dietrich},
  {Estrada}, {Evrard}, {Fausti Neto}, {Fernandez}, {Flaugher}, {Fosalba},
  {Gerdes}, {Gruen}, {Gruendl}, {Honscheid}, {James}, {Kent}, {Kuehn},
  {Kuropatkin}, {Li}, {Maia}, {Makler}, {March}, {Marshall}, {Martini},
  {Merritt}, {Miller}, {Miquel}, {Neilsen}, {Nichol}, {Nord}, {Ogando},
  {Plazas}, {Romer}, {Roodman}, {Sanchez}, {Scarpine}, {Sevilla}, {Smith},
  {Soares-Santos}, {Sobreira}, {Suchyta}, {Swanson}, {Tarle}, {Thaler},
  {Thomas}, {Walker}, \& {Weller}}]{Vikram2015}
{Vikram} V. {et~al.}, 2015, arXiv:1504.03002

\bibitem[{{Vulcani} {et~al}\mbox{.}(2011){Vulcani}, {Poggianti},
  {Arag{\'o}n-Salamanca}, {Fasano}, {Rudnick}, {Valentinuzzi}, {Dressler},
  {Bettoni}, {Cava}, {D'Onofrio}, {Fritz}, {Moretti}, {Omizzolo}, \&
  {Varela}}]{Vulcani2011}
{Vulcani} B. {et~al.}, 2011, MNRAS, 412, 246

\bibitem[{{White} \& {Frenk}(1991)}]{White1991}
{White} S.~D.~M., {Frenk} C.~S., 1991, ApJ, 379, 52

\bibitem[{Wilman {et~al}\mbox{.}(2010)Wilman, Zibetti, \&
  Budav\'{a}ri}]{Wilman2010}
Wilman D.~J., Zibetti S., Budav\'{a}ri T., 2010, MNRAS, 406, 1701

\bibitem[{{Worthey} {et~al}\mbox{.}(1992){Worthey}, {Faber}, \&
  {Gonzalez}}]{Worthey1992}
{Worthey} G., {Faber} S.~M., {Gonzalez} J.~J., 1992, ApJ, 398, 69

\end{thebibliography}

\section*{Affiliations}

$^{1}$Institute of Cosmology and Gravitation, University of Portsmouth, Portsmouth, PO1 3FX, UK\\
$^{2}$SEPnet, South East Physics Network, (www.sepnet.ac.uk)\\
$^{3}$Centro de Investigaciones Energeticas, Medioambientales y Tecnologicas (CIEMAT), Madrid, Spain\\
$^{4}$Wisconsin IceCube Particle Astrophysics Center (WIPAC), Madison, WI 53703, USA\\
$^{5}$Department of Physics, University of Wisconsin–Madison, Madison, WI 53706, USA\\
$^{6}$Aix Marseille Université, CNRS, LAM (Laboratoire d'Astrophysique de Marseille) UMR 7326, 13388, Marseille, France\\
$^{7}$Observat{\'o}rio Nacional, Rua Gal. Jos{\'e} Cristino 77, Rio de Janeiro, RJ - 20921-400, Brazil\\
$^{8}$Laborat{\'o}rio Interinstitucional de e-Astronomia-LIneA, Rua Gal. Jos{\'e} Cristino 77, Rio de Janeiro, RJ 20921-400, Brazil\\
$^{9}$Sorbonne Universit{\'e}s, UPMC Univ Paris 6 et CNRS, UMR 7095, Institut d'Astrophysique de Paris, 98 bis bd Arago, 75014 Paris, France\\
$^{10}$Astrophysics Group, Department of Physics and Astronomy, University College London, 132 Hampstead Road, London, NW1 2PS, United Kingdom\\
$^{11}$National Center for Supercomputing Applications, 1205 West Clark St, Urbana, IL 61801, USA\\
$^{12}$ETH Z{\"u}rich, Institut f{\"u}r Astronomie, Wolfgang-Pauli-Str. 27, CH-8093 Z{\"u}rich, Switzerland\\
$^{13}$Cerro Tololo Inter-American Observatory, National Optical Astronomy Observatory, Casilla 603, La Serena, Chile\\                                
$^{14}$Department of Physics \& Astronomy, University College London, Gower Street, London, WC1E 6BT, UK\\                                              
$^{15}$Department of Physics and Electronics, Rhodes University, PO Box 94, Grahamstown, 6140, South Africa\\  
$^{16}$Fermi National Accelerator Laboratory, P. O. Box 500, Batavia, IL 60510, USA\\                                                                  
$^{17}$Carnegie Observatories, 813 Santa Barbara St., Pasadena, CA 91101, USA\\
$^{18}$CNRS, UMR 7095, Institut d'Astrophysique de Paris, F-75014, Paris, France\\                                                                        
$^{19}$Department of Astronomy, University of Illinois, 1002 W. Green Street, Urbana, IL 61801, USA\\
$^{20}$National Center for Supercomputing Applications, 1205 West Clark St., Urbana, IL 61801, USA\\                                                
$^{21}$Institut de Ci{\`e}ncies de l'Espai, IEEC-CSIC, Campus UAB, Carrer de Can Magrans, s/n,  08193 Bellaterra, Barcelona, Spain\\               
$^{22}$Institut de F{\'i}sica d'Altes Energies (IFAE), The Barcelona Institute of Science and Technology, Campus UAB, 08193 Bellaterra (Barcelona) Spain\\
$^{23}$Kavli Institute for Particle Astrophysics \& Cosmology, P. O. Box 2450, Stanford University, Stanford, CA 94305, USA\\                             
$^{24}$Excellence Cluster Universe, Boltzmannstr. 2, 85748 Garching, Germany\\                                                       
$^{25}$Faculty of Physics, Ludwig-Maximilians-Universit{\"a}t, Scheinerstr. 1, 81679 Munich, Germany\\                                                     
$^{26}$Department of Physics and Astronomy, University of Pennsylvania, Philadelphia, PA 19104, USA\\
$^{27}$Jet Propulsion Laboratory, California Institute of Technology, 4800 Oak Grove Dr., Pasadena, CA 91109, USA\\                                     
$^{28}$Department of Astronomy, University of Michigan, Ann Arbor, MI 48109, USA\\                                                                       
$^{29}$Department of Physics, University of Michigan, Ann Arbor, MI 48109, USA\\                                                                       
$^{30}$Kavli Institute for Cosmological Physics, University of Chicago, Chicago, IL 60637, USA\\
$^{31}$SLAC National Accelerator Laboratory, Menlo Park, CA 94025, USA\\                                                                                 
$^{32}$Center for Cosmology and Astro-Particle Physics, The Ohio State University, Columbus, OH 43210, USA\\                                             
$^{33}$Department of Physics, The Ohio State University, Columbus, OH 43210, USA\\                                                                       
$^{34}$Australian Astronomical Observatory, North Ryde, NSW 2113, Australia\\                                                                           
$^{35}$Departamento de F{\'i}sica Matem{\'a}tica,  Instituto de F{\'i}sica, Universidade de S{\~a}o Paulo,  CP 66318, CEP 05314-970, S{\~a}o Paulo, SP,  Brazil\\
$^{36}$Department of Astronomy, The Ohio State University, Columbus, OH 43210, USA\\                                                                     
$^{37}$Department of Astrophysical Sciences, Princeton University, Peyton Hall, Princeton, NJ 08544, USA\\                                              
$^{38}$Instituci{\'o} Catalana de Recerca i Estudis Avan{\c c}ats, E-08010 Barcelona, Spain\\                                                          
$^{39}$Max Planck Institute for Extraterrestrial Physics, Giessenbachstrasse, 85748 Garching, Germany\\                                                 
$^{40}$Department of Physics and Astronomy, Pevensey Building, University of Sussex, Brighton, BN1 9QH, UK\\                                             
$^{41}$Centro de Investigaciones Energ{\'e}ticas, Medioambientales y Tecnol{\'o}gicas (CIEMAT), Madrid, Spain\\                                        
$^{42}$ICTP South American Institute for Fundamental Research, Instituto de F{\'i}sica Te{\'o}rica, Universidade Estadual Paulista, S{\~a}o Paulo, Brazil\\
$^{43}$Argonne National Laboratory, 9700 South Cass Avenue, Lemont, IL 60439, USA\\                                                                       
$^{44}$Department of Physics, IIT Hyderabad, Kandi, Telangana-502285, India\\
 
\end{document}